%% file: main.tex
\documentclass{article}

\usepackage[margin=1in]{geometry} 
\usepackage{amsmath,amsthm,amssymb,amsfonts, fancyhdr, color, comment, environ, bm}
\numberwithin{equation}{section}
\usepackage{bbm}
\usepackage{lscape}

\usepackage{graphicx, subcaption}
\graphicspath{ {./} }

\usepackage[
    backend=biber, 
    style=numeric-comp,
    maxnames=5,
    minnames=1,
    sorting=nty,
    date=year,
    giveninits=true,
    url=false,
    isbn=false,
    doi=false,
]{biblatex}

\renewbibmacro*{issue+date}{%
  \setunit{\addcomma\space}
  \iffieldundef{issue}
    {\usebibmacro{date}}
    {\printfield{issue}%
     \setunit*{\addspace}%
     \usebibmacro{date}}%
  \newunit}
  
\DeclareFieldFormat[article, inbook]{citetitle}{#1}
\DeclareFieldFormat[article, inbook]{title}{#1} 
\DeclareFieldFormat{pages}{#1}

\renewbibmacro{in:}{%
  \ifentrytype{article}{}{\printtext{\bibstring{in}\intitlepunct}}}
\renewcommand{\intitlepunct}{\nopunct}

\AtEveryBibitem{%
  \clearfield{number}%
  \clearfield{issue}%
}

\usepackage[breaklinks=true]{hyperref}

\usepackage{url}
\usepackage{array}
\usepackage{tabularx}
\usepackage{booktabs}

\newcolumntype{C}[1]{>{\centering\arraybackslash}p{#1}}

\newcolumntype{L}[1]{>{\raggedright\arraybackslash}p{#1}}
\newcolumntype{E}{>{\centering\arraybackslash}X}
\newcolumntype{R}[1]{>{\centering\arraybackslash}p{#1}}

\usepackage{algorithm}
\usepackage{algpseudocode}
\usepackage{subcaption}
\usepackage{hyperref}

\newcommand{\inter}{\mathring{V}(T)}

\usepackage{chngcntr}

\newtheorem{theorem}{Theorem}[section]

\newtheorem{proposition}[theorem]{Proposition}

\newtheorem{lemma}[theorem]{Lemma}

\newtheorem{theoremA}{Theorem}[section] 
\newtheorem{propositionA}[theoremA]{Proposition}
\newtheorem{lemmaA}[theoremA]{Lemma}
 
\usepackage{etoc}

\title{Basis statistics for phylogenetic tree shape: \\ tree balance formulas beyond expectations}
\author{Daniel J.~Bauman\thanks{Institute for Computational and Mathematical Engineering, Stanford University, Stanford, CA 94305 USA. Email: djbauman@stanford.edu} \\ Noah A.~Rosenberg\thanks{Department of Biology, Stanford University, Stanford, CA 94305 USA.}}
\date{\today}

\begin{document}

\maketitle

\begin{abstract}
\noindent Summary statistics that measure the shapes of phylogenetic trees contributed to quantitatively understanding evolutionary processes. Among phylogenetic summary statistics, the \emph{tree balance indices} have been designed to capture intuitive notions of balance or imbalance in trees. Here, we characterize the tree balance indices in terms of \emph{basis statistics}, mapping the exponentially large space of rooted, binary, unlabeled trees to lower-dimensional spaces. We observe that many tree balance indices can be represented in terms of linear functions of these low-dimensional spaces, enabling us to compute exact formulas for the means, variances, and covariances of many balance indices under the Yule--Harding model. The new framework recapitulates existing formulas for the expectations of 8 indices: Sackin index, average leaf depth, total cophenetic index, $\hat{s}$-shape statistic, Colless index, corrected Colless index, quadratic Colless index, and rooted quartet index for binary trees. It produces new formulas for the expectations of 9 additional indices: equal weights Colless index, total $I$ index, total $I'$ index, Rogers $J$ index, total internal path length, total path length, average vertex depth, stairs1 index, and stairs2 index. Moreover, the framework also allows us to calculate the exact variance of all 17 indices that are represented by the bases as well as the covariance and correlation for all $\binom{17}{2} = 136$ pairs among the 17 indices. 
\end{abstract}

\section{Introduction}

The topology of a phylogenetic tree can reveal insights into the evolutionary process that produced the tree. Features of tree topologies have identified systematic patterns in the macroevolutionary processes of speciation and extinction~\cite{Slowinski1989, Heard1992, Kirkpatrick1993, Mooers1997, Aldous2001, Agapow2002,  Stadler2013, Pennell2025}. In viral phylogenies, they have been used to study selection and population structure in a setting of epidemics~\cite{Volz2013, Frost2013}. In population genetics, tree topologies have been used for testing population-genetic models through their influence on population-genetic statistics such as Tajima's $D$~\cite{Li2013, Ferretti2017}. 

To achieve quantitative summaries of phylogenetic tree topologies, many summary statistics have been devised. In particular, tree balance indices are summary statistics that measure the balance or imbalance of a given tree. Since the development of the first tree balance indices---the Sackin index, or the sum of the depths of leaves on a tree \cite{Sackin1972}, and the Colless index, the sum of an asymmetry measure across individual nodes \cite{Colless1982}---many statistics have been developed, measuring balance using features such as the balance of individual nodes in a tree, the height of a tree and its subtrees, the depths of the leaves, and the properties of subtrees \cite{Fischer2023, Kersting2025, Fischer2025}. 

In some instances, certain balance indices have been seen to be particularly correlated with one another \cite{Blum2006Mean, Cardona2013, Janzen2024}. For example, in the Yule--Harding model for tree shapes, Blum et al.~\cite{Blum2006Mean} studied the joint distribution of the Sackin and Colless indices, proving that these indices have a high correlation ($\approx 0.98$) in the limit of large tree sizes. Janzen \& Etienne~\cite{Janzen2024} studied tree summary statistics including balance indices in simulations of several models, uncovering clusters of highly correlated indices; these correlations between indices are highly model-dependent. Generally, however, the extent to which different balance indices are related to one another and the extent to which they capture distinct information about tree topologies are largely unknown. 

The Yule-Harding model or equal-rates Markov model (henceforth termed the \emph{Yule model}), first studied in the context of phylogenetic trees by Harding~\cite{Harding1971}, serves as a useful null model of evolution~\cite{Rosenberg2025}. In the Yule model, a branching process begins with one lineage. At a set of discrete future times, a lineage randomly chosen among extant lineages splits into two, and all lineages have an equal probability of being the next to split. This process forms a rooted, binary phylogenetic tree, as each lineage can be traced back to the starting lineage through a unique path of splits. The properties of many tree balance indices under the Yule model have been well-studied, as documented in the recent comprehensive survey by Fischer et al.~\cite{Fischer2023}.

The contribution of this paper is twofold. First, we propose to unify tree balance indices using a concept of \emph{basis statistics}. We introduce a perspective that unifies many existing balance indices in a common mathematical framework that identifies how distinct features of tree topologies are measured by each balance index, independently of a choice of phylogenetic model. Second, we obtain exact results on the mathematical properties of these basis statistics under the Yule model, enabling exact computation of the means, variances, and covariances between all statistics in a broad class of tree balance indices. 

In Section~\ref{section 2}, we introduce notation and terminology. Section~\ref{section 3} defines the concept of a \emph{basis} and introduces two bases that we call the \emph{clade-number basis} and the \emph{daughter-number basis}. In Section~\ref{section 4}, we evaluate whether each tree balance index surveyed by Fischer et al.~\cite{Fischer2023} is represented by the clade-number basis and the daughter-number basis. Section~\ref{section 5} synthesizes numerous results on trees under the Yule model to find mathematical formulas for the means and covariance matrices of the basis statistics. In Section \ref{section 6}, we apply the mathematical results obtained in Section~\ref{section 5} to find formulas for the means, variances, and covariances of all tree balance indices that we showed were represented by the bases in Section~\ref{section 4}. We conclude with a discussion of the results and further applications of the basis statistics framework.

\section{Notation and definitions} \label{section 2}

We rely largely on the notation of Fischer et al.~\cite{Fischer2023}. A \textit{rooted binary tree} $T = \big(V(T), E(T)\big)$ is a directed graph that has a vertex set $V(T)$ and a set of directed edges $E(T)$ and that satisfies three properties: (i) if the edges are considered to be undirected, then the graph is acyclic and connected; (ii) there exists exactly one vertex $\rho \in V(T)$ with in-degree 0, called the \textit{root} of the tree; (iii) each vertex has out-degree 0 (\textit{leaves}) or 2 (\textit{internal vertices}). For the rest of the paper, the term \textit{tree} refers to a rooted, binary, nonempty tree.

The set of leaves of $T$ is denoted $V_L(T)$. The set of internal vertices, also termed \emph{internal nodes}, is denoted $\inter = V(T) \setminus V_L(T)$. The \textit{size} of a tree is its number of leaves: $|T| = |V_L(T)|$. The set of rooted binary (unlabeled) tree shapes with $n$ leaves is denoted $\mathcal{T}_n$. 

If a path exists from vertex $u$ to vertex $v$, then we say that $v$ is \textit{descended} from $u$, and $u$ is an \textit{ancestor} of $v$; if this path is nontrivial and contains exactly one edge, then $v$ is a \textit{child} of $u$ and $u$ is the \textit{parent} of $v$. For all vertices $v \in V(T)$, there is a unique directed path from $\rho$ to $v$. The \textit{depth} of a vertex $v$ is the number of edges between $\rho$ and $v$ in this path and is denoted $\delta_T(v)$. The \textit{height} $H(T)$ of a tree is the maximal depth among its leaves, $H(T) = \max_{v \in V_L(T)} \delta_T(v)$. 

All internal nodes $u$ have out-degree 2, and hence an internal node $u$ has two descendants connected by exactly one edge from $u$. We call these descendants the \textit{daughters} of $u$. For an internal node $u$, we denote by $u_L$ and $u_R$ the left and right daughters of $u$; ``left'' and ``right'' are used for convenience, but an exchange of left and right does not change $T$. Any vertex $v \in V(T)$ induces a subtree $T_v$, which is the rooted, binary tree that has $v$ at the root and that contains all vertices $w \in V(T)$ that are descended from $v$ and the corresponding edges connecting $v$ to $w$. The leaves of $T_v$ form a \textit{clade} of size $n_v := |T_v|$. A subset of leaves $A \subset V_L(T)$ forms a clade if and only if $A = V_L(T_v)$ for some internal vertex $v \in V(T)$. For an internal node $v \in \inter$ with daughters $v_L, v_R$, its \textit{daughter subtrees} are the pair $(T_{v_L}, T_{v_R})$. Because tree topologies are equivalent under any left--right ordering of the daughters of a node, we assume by convention that $n_{v_L} \leq n_{v_R}$. Note that the daughter subtrees must satisfy $n_{v_L} + n_{v_R} = n_v$ because the leaf sets $T_{v_L}$ and $T_{v_R}$ are disjoint and contain all leaves in $T_v$.

Unless otherwise indicated, all vectors are column vectors, and they are denoted by bold letters. For example, to signify that $\mathbf{a}$ is a column vector, we write $\mathbf{a} = (a_1, a_2, \ldots, a_n)^t$.

\section{Bases of tree statistics} \label{section 3}

In this section, we describe the concept of a \emph{basis of statistics} (Section \ref{sec:3_1}). We explain the utility of such a basis (Section \ref{sec:3_2}), introduce two bases (Section \ref{sec:3_3}), and discuss their relationships (Section \ref{sec:3_4}), interpretations (Section \ref{sec:3_5}), and properties (Sections \ref{sec:3_6}).

\subsection{Definition of a basis of statistics}
\label{sec:3_1}

A summary statistic $A(T)$ of a tree is simply a function $A: \mathcal{T}_n \mapsto \mathbb{R}$. We define a \textit{basis of statistics} $\mathcal{B}$ on $\mathcal{T}_n$ as a set $\{B_i(T)\}_{i=1}^r$ of statistics $B_i: \mathcal{T}_n \mapsto \mathbb{R}$. We write this basis as a column vector $\mathcal{B}(T) = \big(B_1(T), B_2(T), \ldots, B_r(T)\big)^t$. A summary statistic $A$ can be \textit{represented} by the basis $\mathcal{B}$ if for all $T$ in $\mathcal{T}_n$,
\begin{equation}
    A(t) = \sum_{i=1}^r a_i B_i(T) = \mathbf{a}^t \, \mathcal{B}(T)
\end{equation}
for some coefficients $a_i\in \mathbb{R}$, where $\mathbf{a} = (a_1, a_2, \ldots, a_r)^t$. If a basis has been defined, then we can identify the statistic $A(T)$ with its coefficient vector $\mathbf{a}$. 

\subsection{Utility of representing statistics with a basis}
\label{sec:3_2}

To motivate the effort to write tree statistics in terms of linear combinations of sets of basis statistics, let $\mathcal{B}(T) = \big(B_1(T), B_2(T), \ldots, B_r(T)\big)^t$ be a basis of statistics. Let $\mathbb{P}$ be a probability distribution on $\mathcal{T}_n$, the trees of size $n$, such as the probability distribution generated by the Yule model~\cite{Steel2016}. If $T$ is a random tree sampled from the distribution $\mathbb{P}$, then $\mathcal{B}(T)$ is an associated random vector. The basis has an expectation vector $\boldsymbol{\mu} = \mathbb{E}[\mathcal{B}(T)]$ and a covariance matrix $\mathbf{\Sigma} = \mathbb{E}[(\mathcal{B}(T) - \boldsymbol{\mu})(\mathcal{B}(T) - \boldsymbol{\mu})^t]$, where the expectation of a function $f$ of the random variable $T$, $f: \mathcal{T}_n\to \mathbb{R}^k$, is calculated by $\mathbb{E}[f(T)] = \sum_{T \in \mathcal{T}_n} f(T) \, \mathbb{P}(T)$. 

Let $A_1(T)$ and $A_2(T)$ be two statistics that are represented by $\mathcal{B}$. We can identify them in terms of their (column) vectors of coefficients $\mathbf{a}_1$ and $\mathbf{a}_2$, such that $A_1(T) = \mathbf{a}_1^t \mathcal{B}(T)$ and $A_2(T) = \mathbf{a}^t_2 \mathcal{B}(T)$. Then we can find the mean of a statistic $A_1(T)$ by 
\begin{equation} \label{mean of statistic}
    \mathbb{E}[A_1(T)] = \mathbb{E} \big[\mathbf{a}^t_1 \mathcal{B}(T) \big] = \mathbf{a}^t_1 \mathbb{E}[\mathcal{B}(T)]= \mathbf{a}^t_1\boldsymbol{\mu}.
\end{equation}
Moreover, we can find the covariance between $A_1(T)$ and $A_2(T)$: 
\begin{align}
    \textrm{Cov} \big[A_1(T), A_2(T)  \big] &= \mathbb{E} \big[ \big(A_1(T)- \mathbb{E}[A_1(T)] \big) \big(A_2(T)-\mathbb{E}[A_2(T)] \big)^t \big] \nonumber \\
    &= \mathbb{E}\big[\mathbf{a}_1^t \big(\mathcal{B}(T) - \boldsymbol{\mu} \big) \big(\mathcal{B}(T) - \boldsymbol{\mu} \big)^t \mathbf{a}_2 \big] = \mathbf{a}_1^t \mathbf{\Sigma} \mathbf{a}_2. \label{covariance of statistics}
\end{align}
Notice that setting $A_2(T) = A_1(T)$ in eq.~\ref{covariance of statistics} recovers the variance of $A_1(T)$. Therefore, if we know the mean and covariance of the elements of a basis of statistics, we can immediately obtain the mean and variance of all statistics represented by the basis, as well as the covariance matrix between all pairs of these statistics.

\subsection{Two bases of statistics}
\label{sec:3_3}

We define two bases of statistics for trees $T \in \mathcal{T}_n$. 

\subsubsection{Clade-number basis}
\label{sec:3_3_1}

For $2 \leq r \leq n$, let $C_r(T)$ be the number of subtrees of size $r$ contained in $T$. These numbers have been explored numerous times in the phylogenetics literature with different names, such as the \textit{number of $r$-pronged nodes}~\cite{Rosenberg2006}. We define the \textit{clade-number basis} as the set $\{C_2(T), C_3(T), \ldots, C_n(T)\}$. We denote the clade-number basis by $\mathcal{C}$.

\subsubsection{Daughter-number basis}
\label{sec:3_3_2}

For $1 \leq i \leq \lfloor n/2 \rfloor$, $i \leq j \leq n-i$, let $D_{ij}(T)$ be the number of internal nodes of $T$ that have daughter subtrees of sizes $i$ and $j$, where $i \leq j$ by convention. We define the \textit{daughter-number basis} as the set 
$\{D_{ij}(T)\}_{\{(i,j) | 1 \leq i \leq \lfloor n/2\rfloor, \, i \leq j \leq n-i\}}$. By convention, we write the elements of the daughter-number basis in the order $D_{1,1}(T), D_{1,2}(T), \ldots, D_{1,n-1}(T), D_{2,2}(T), D_{2,3}(T), \ldots, \allowbreak D_{2,n-2}(T), \ldots, D_{\lfloor n/2\rfloor, \lceil n/2 \rceil}(T)$. We denote the daughter-number basis by $\mathcal{D}$. 

\subsection{Relationships between bases}
\label{sec:3_4}

One natural concept in this setting is that some bases are \emph{refinements} of others. By refinement, we mean that if $\mathcal{B} = (B_1, B_2, \ldots, B_r)$ and $\tilde{\mathcal{B}} = (\tilde{B}_1, \tilde{B}_2, \ldots, \tilde{B}_s)$ are two bases, then $\tilde{\mathcal{B}}$ is a \textit{refinement} of $\mathcal{B}$ if $B_i$ can be represented by $\tilde{\mathcal{B}}$ for all $i$. Because 
\begin{equation} \label{daughter to clade}
C_r(T) = \sum_{i=1}^{\lfloor r/2 \rfloor} D_{i,r-i}(T),
\end{equation}
the daughter-number basis $\mathcal{D}$ is a refinement of the clade-number basis $\mathcal{C}$. It follows that any statistic $A$ that can be represented by $\mathcal{C}$ can also be represented by $\mathcal{D}$. As we demonstrate later, many tree statistics can be represented by both bases or by only the daughter-number basis; however, a statistic cannot be represented by only the clade-number basis and not the daughter-number basis.

\subsection{Interpretation of bases}
\label{sec:3_5}

Summary statistics of trees map a tree $T \in \mathcal{T}_n$ to a real number. A basis of statistics simply maps a tree to a vector. By mapping the space of trees $\mathcal{T}_n \mapsto \mathbb{R}^k$ for some $k$ that is much smaller than the number of trees, we are embedding the tree space in a low-dimensional space. Studying the image of $\mathcal{T}_n$ in $\mathbb{R}^k$ under such a mapping enables quantitative analyses of the differences among trees. A single summary statistic that is represented by this basis corresponds to a projection of the image of $\mathcal{T}_n$ under this mapping onto a line in $\mathbb{R}^k$: because a statistic that can be represented by the basis is by definition equal to $\mathbf{a}^t \mathcal{B}(T)$ for a coefficient vector $\mathbf{a}$, the statistic corresponds to projection of $\mathcal{B}(T)$ onto the line in the direction of $\mathbf{a}$, followed by rescaling by the magnitude $\|\mathbf{a}\|_2$.

Trees cannot in general be recovered from their embedding under a given basis; multiple trees can map to the same basis vector. Moreover, even though the image of $\mathcal{T}_n$ under a given basis lies within $\mathbb{R}^k$, the underlying dimension of the set of points that correspond to images in $\mathbb{R}^k$ of trees in $\mathcal{T}_n$ may be less than $k$. We find the true dimension of the clade-number basis and the daughter-number basis below.  

\subsubsection{Dimensionality of the clade-number basis}
\label{sec:3_5_1}

The number of non-trivial values possible for the size of a clade is $n-1$; the clade-number basis has elements $C_2(T), C_3(T), \ldots, C_n(T)$. Hence, the clade-number basis is an embedding from $\mathcal{T}_n$ to $\mathbb{R}^{n-1}$. However, two identities reduce the dimension of the clade-number basis. A tree $T \in \mathcal{T}_n$ satisfies
\begin{align}
    C_n(T) &= 1, \label{1 tree size n} \\
    C_2(T) + C_3(T) + \ldots + C_{n-1}(T) &= n-2. 
\label{eq:Csum}
\end{align}
Eq.~\ref{1 tree size n} holds because a unique vertex, the root, has a subtree of size $n$ leaves. Eq.~\ref{eq:Csum} holds because in a rooted binary tree with $n$ leaves, the number of non-root internal vertices is $n-2$ \cite[Lemma 1.4]{Steel2016}. 

The identities \ref{1 tree size n} and \ref{eq:Csum} have two implications. First, $C_n(T)$ can be removed from the basis by writing the linear combination
\begin{equation} \label{remove Cn}
    C_n(T) = \frac{1}{n-2}\big[C_2(T) + C_3(T) + \ldots + C_{n-1}(T)\big].
\end{equation}
Second, the constant sum in eq.~\ref{eq:Csum} implies that one of $C_2(T), C_3(T), \ldots, C_{n-1}(T)$ can always be obtained from the other $n-2$ values, so that the underlying dimension of the embedding of $\mathcal{T}_n$ in the clade-number basis is no greater than $n-3$.

\subsubsection{Dimensionality of the daughter-number basis}
\label{sec:3_5_2}

The number of elements in the daughter-number basis is $\lfloor n/2\rfloor \, \lceil n/2 \rceil$. We notice that for each clade number $C_r(T)$ with $2 \leq r \leq n-1$, two distinct ways exist for writing this clade number in terms of the daughter numbers. The first is simply eq.~\ref{daughter to clade}, which counts each internal node of size $r$ by counting all daughter numbers of size $(i, r-i)$, $1 \leq i \leq \lfloor \frac{r}{2} \rfloor$. The second way to count the clade numbers of size $r$ is to sum the number of subtrees that have a left subtree of size $r$ and the number of subtrees that have a right subtree of size $r$. We can write this as
\begin{equation} \label{3.9}
    C_r(T) = \sum_{k=1}^{\lfloor n/2 \rfloor} \sum_{\ell = k}^{n-k} D_{k,\ell}(T) \,\big(\mathbbm{1}_{\{r=k\}} + \mathbbm{1}_{\{r=\ell\}}\big),
\end{equation}
where $\mathbbm{1}_{\{\cdot\}} = 1$ if the statement $\{\cdot\}$ is true and $0$ otherwise. This equation counts each daughter number $D_{i,j}(T)$ once if the left subtree is size $r$ and once if the right subtree is size $r$. By combining eqs.~\ref{daughter to clade} and~\ref{3.9}, we find that for each value of $r$, $2 \leq r \leq n-1$,
\begin{equation}
    \sum_{i=1}^{\lfloor r/2 \rfloor} D_{i,r-i}(T)  = \sum_{k=1}^{\lfloor n/2 \rfloor} \sum_{\ell = k}^{n-k} D_{k,\ell}(T) \,\big(\mathbbm{1}_{\{r=k\}} + \mathbbm{1}_{\{r=\ell\}}\big).
\label{eq:Dir}
\end{equation} 

In the case of $r = n$, we can combine eq.~\ref{daughter to clade} with eq.~\ref{1 tree size n} to observe that
\begin{equation}
    \sum_{i=1}^{\lfloor n/2 \rfloor} D_{i,n-i}(T) = 1.
\label{eq:Dirn}
\end{equation}
Eqs.~\ref{eq:Dir} and \ref{eq:Dirn} provide $n-1$ identities in the daughter-number basis, implying that it has underlying dimension at most $\lfloor n/2\rfloor \, \lceil n/2 \rceil - (n-1) = \lfloor (n-2)/2\rfloor \, \lceil (n-2)/2\rceil$. 

\subsection{Evaluating if statistics can be represented by a basis} \label{not represented by basis}
\label{sec:3_6}

Suppose we have a summary statistic $A(T)$ and a basis of statistics $\mathcal{B}(T)$. We wish to understand if the statistic can be represented by the basis---that is, if there exists a vector of coefficients $\mathbf{a}$ such that $A(T) = \mathbf{a}^t \mathcal{B}(T)$ for all trees $T \in \mathcal{T}_n$. 

From the definitions of the two bases $\mathcal{C}(T)$ and $\mathcal{D}(T)$ and the definitions of individual balance indices, it is often straightforward to exhibit a representation of a balance index statistic if it exists, as the balance index might be defined in terms of the basis elements. Later, in Section~\ref{section 4} and Appendix~\ref{Appendix A}, we systematically study tree balance indices and prove that many are represented by the daughter-number basis, or by both the clade-number basis and the daughter-number basis. 

For some balance index statistics, however, no straightforward correspondence exists between its definition and the clade-number basis or daughter-number basis. One approach to proving that a statistic \emph{cannot} be represented by a basis is to proceed by counterexample: if we can find two trees $T_1, T_2 \in \mathcal{T}_n$ such that $\mathcal{B}(T_1) = \mathcal{B}(T_2)$, but $A(T_1) \neq A(T_2)$, then the statistic $A(T)$ cannot be represented by the basis $\mathcal{B}$. We provide examples of such trees in Appendix~\ref{Appendix B}, sections~\ref{B.1} and~\ref{B.2}, to prove that specific indices are represented by $\mathcal{D}$ but not $\mathcal{C}$, or by neither basis. 

When a statistic $A(T)$ is equal to a \textit{nonlinear} function of a basis of statistics $\mathcal{B}(T)$, no such counterexample can be found since $A(T_1) = A(T_2)$ for any two trees $T_1, T_2$ satisfying $\mathcal{B}(T_1) = \mathcal{B}(T_2)$. To prove that $A(T)$ cannot be represented by a linear combination of the basis elements in this case, we introduce an approach based on linear algebra in Appendix~\ref{Appendix C}.

\section{Basis representations of tree balance indices} \label{section 4}

In this section, we demonstrate that the clade-number basis and the daughter-number basis represent many existing tree balance indices. We systematically examined all tree balance indices compiled by Fischer et al.~\cite[Chapters 5-23]{Fischer2023} to evaluate whether they can be represented by each basis. We did not include statistics listed in Chapter 24, ``Related tree shape statistics that are not (im)balance indices.'' Furthermore, one balance index, listed as the Colless-like indices~\cite[Chapter 16]{Fischer2023}, had a definition too general to be studied for binary trees and was hence excluded. For the rooted quartet index~\cite[Chapter 20]{Fischer2023}, we chose the definition that was specifically recommended for rooted binary trees. For indices with multiple definitions listed by Fischer et al., we picked the definition that was simplest to study in relation to the bases. In total, we analyzed 29 tree balance indices.

Based on the definition of an index, we evaluated if the given balance index can be represented by the clade-number basis and the daughter-number basis. In Appendix~\ref{Appendix A}, we state definitions for all indices, and we prove an explicit representation of the index by the clade-number basis or daughter-number basis if a representation exists. In Appendix~\ref{Appendix B} and Appendix~\ref{Appendix C}, for all indices that we claim are not represented by the clade-number basis or the daughter-number basis, we prove that such a representation does not exist. The results of whether each statistic is represented by these two bases are summarized in Table~\ref{tab:index_representations}.

\begin{table}[htbp]
    \centering
    \renewcommand{\arraystretch}{1.2}
    \setlength{\tabcolsep}{4pt}
    \begin{tabular}{|L{6.1cm}|C{1.7cm}|C{1.8cm}|C{1.6cm}|C{1.6cm}|C{1.8cm}|}
        \hline
        \textbf{Index name} & \textbf{Appendix number} & \textbf{Chapter in Fischer et al.~\cite{Fischer2023}} & \textbf{Notation} & \textbf{Clade-number basis} & \textbf{Daughter-number basis} \\
        \hline
        Sackin index & \ref{sackin} & 5 & $S(T)$ & \checkmark & \checkmark \\
        Average leaf depth & \ref{avg leaf depth} & 6 & $\bar{N}(T)$ & \checkmark & \checkmark \\
        Variance of leaf depths & \ref{var leaf depths} & 7& $\sigma^2_N(T)$ & $\times$ & $\times$ \\
        Total cophenetic index & \ref{total cophenetic} & 8 & $\Phi(T)$ & \checkmark & \checkmark \\
        $\hat{s}$-shape statistic & \ref{s-shape} & 9 & $\hat{s}(T)$ & \checkmark & \checkmark \\
        $B_1$ index & \ref{b1} &  10 & $B_1(T)$ & $\times$ & $\times$ \\
        $B_2$ index & \ref{b2} & 11 & $B_2(T)$ & $\times$ & $\times$ \\
        Colless index & \ref{colless} & 12 & $C(T)$ & $\times$ & \checkmark \\
        Corrected Colless index & \ref{corrected colless} & 13 & $I_C(T)$ & $\times$ & \checkmark \\
        Equal weights Colless ($I_2$) index & \ref{I2} & 14 & $I_2(T)$ & $\times$ & \checkmark \\
        Quadratic Colless index & \ref{quadratic colless} & 15 & $QC(T)$ & \checkmark & \checkmark \\
        Mean $I$ index & \ref{mean I index} & 17 & $\bar{I}(T)$ & $\times$ & $\times$ \\
        Total $I$ index & \ref{total I index} &  17 & $\Sigma I(T)$ & $\times$ & \checkmark \\
        Mean $I'$ index & \ref{mean I' index} & 17 & $\bar{I'}(T)$ & $\times$ & $\times$ \\
        Total $I'$ index & \ref{total I' index} &  17 & $\Sigma I'(T)$ & $\times$ & \checkmark \\
        Symmetry nodes index & \ref{symmetry nodes index} & 18 & $SNI(T)$ & $\times$ & $\times$ \\
        Rogers $J$ index & \ref{rogers j index} & 19 & $J(T)$ & $\times$ & \checkmark \\
        Rooted quartet index for binary trees & \ref{rqib} & 20 & $rQIB(T)$ & $\times$ & \checkmark \\
        Colijn-Plazzotta rank & \ref{CP rank} & 21 & $CP(T)$ & $\times$ & $\times$ \\
        Furnas rank & \ref{Furnas rank} &  22 & $F(T)$ & $\times$ & $\times$ \\
        Total internal path length & \ref{TIPL} & 23.1 & $TIP(T)$ & \checkmark & \checkmark \\
        Total path length & \ref{TPL} & 23.2 & $TPL(T)$ & \checkmark & \checkmark \\
        Average vertex depth & \ref{AVD} & 23.3 & $AVD(T)$ & \checkmark & \checkmark \\
        Maximum depth & \ref{maximum depth} & 23.4 & $mD(T)$ & $\times$ & $\times$ \\
        Maximum width & \ref{maximum width} & 23.5 & $mW(T)$ & $\times$ & $\times$ \\
        Modified maximum difference in widths & \ref{modified max dif widths} & 23.6 & $mdelW(T)$ & $\times$ & $\times$ \\
        Maximum width over maximum depths & \ref{max width over max depth} & 23.7 & $W/D(T)$ & $\times$ & $\times$ \\
        stairs1 index & \ref{stairs1} & 23.8 & $st1(T)$ & $\times$ & \checkmark \\
        stairs2 index & \ref{stairs2} & 23.9 & $st2(T)$ & $\times$ & \checkmark \\
        \hline
        \multicolumn{3}{|L{9.6cm}|}{\textbf{Total}} & \textbf{29} & \textbf{8} & \textbf{17} \\
        \hline
    \end{tabular}
    \caption{Tree balance indices and their representations by the clade-number basis and daughter-number basis. The clade-number basis and daughter-number basis are low-dimensional mappings from the space $\mathcal{T}_n$ of rooted binary tree shapes to a subset of $\mathbb{R}^k$, where $k$ is the number of statistics in the basis. The number of rooted binary tree shapes grows exponentially in the number of leaves $n$ according to the Wedderburn-Etherington sequence~\cite[eq.~2.8]{Harding1971}; meanwhile, the underlying dimensionality of the clade-number basis ($\sim n$) and the daughter-number basis ($\sim n^2 / 4$) grows polynomially (Section~\ref{sec:3_5}). Many, but not all, of the tree balance indices that we consider can be represented by a linear combination of the basis statistics. Checkmark (\checkmark) indicates that the index can be represented by the given basis, whereas cross ($\times$) indicates it cannot. Proofs of representations (\checkmark) appear in Appendix~\ref{Appendix A}, and proofs that representations do not exist ($\times$) appear in Appendix~\ref{Appendix B}.   }
    \label{tab:index_representations}
\end{table}

\subsection{Statistics represented by the clade-number basis} \label{sec:4_1}

Eight of the 29 balance indices can be represented by the clade-number basis: the Sackin index, average leaf depth, total cophenetic index, $\hat{s}$-shape statistic, quadratic Colless index, total internal path length, total path length, and average vertex depth. 

The Sackin index, total cophenetic index, and $\hat{s}$-shape statistic all have definitions that can be written in the form $\sum_{v \in \inter} f(n_v)$. Any statistic of this form can be represented by the clade-number basis as
\begin{equation}
    \sum_{v \in \inter} f(n_v) = \sum_{k = 2}^n f(k) \, C_k(T). \nonumber 
\end{equation}

The average leaf depth, total internal path length, total path length, and average vertex depth are rescaled or shifted versions of the Sackin index, when applied to binary trees. Hence, representations of these indices by the clade-number basis can be found using a representation of the Sackin index. 

The most surprising index to be represented by the clade-number basis is the quadratic Colless index. In Appendix~\ref{Appendix D}, we prove the following result.
\begin{proposition}  \label{cophenetic sackin qc theorem} 
    For a binary tree $T \in \mathcal{T}_n$, the Sackin index $S(T)$, quadratic Colless index $QC(T)$, and total cophenetic index $\Phi(T)$ satisfy the identity
    \begin{equation}
        QC(T) = 2\Phi(T) + S(T) - n(n-1). 
    \end{equation}
\end{proposition}    
This previously unreported identity establishes a new connection between these three balance indices. Because the Sackin index and total cophenetic index both are represented by the clade-number basis, Proposition~\ref{cophenetic sackin qc theorem} can be used to find a representation of the quadratic Colless index by the clade-number basis. 

\subsection{Statistics represented by the daughter-number basis} \label{sec:4_2}

All eight of the statistics represented by the clade-number basis are also represented by the daughter-number basis, as discussed in Section~\ref{sec:3_4}. Nine of the 21 remaining statistics are represented only by the daughter-number basis and not by the clade-number basis: the Colless index, corrected Colless index, $I_2$ index, total $I$ index, total $I'$ index, Rogers $J$ index, rooted quartet index for binary trees, stairs1, and stairs2. 

Each of these statistics has a definition that can be written in the form $\sum_{v \in \inter} f(n_{v_L}, n_{v_R})$. Any statistic of this type can be represented by the daughter-number basis as
\begin{equation}
    \sum_{v \in \inter} f(n_{v_L}, n_{v_R}) = \sum_{i = 1}^{\lfloor n/2 \rfloor} \sum_{j = i}^{n-i} f(i, j) \, D_{i, j}(T). \nonumber
\end{equation}

We present proofs that each of these statistics cannot be represented by the clade-number basis in Appendix~\ref{Appendix B}.

\subsection{Statistics represented by neither basis} \label{sec:4_3}

The remaining twelve balance indices are represented by neither basis: the variance of leaf depths, $B_1$ index, $B_2$ index, mean $I$ index, mean $I'$ index, symmetry nodes index, Colijn-Plazzotta rank, Furnas rank, maximum depth, maximum width, modified maximum difference in widths, and maximum width over maximum depths. 

The variance of leaf depths and the $B_2$ index are both functions of the depths of leaves, similar to the Sackin index, but their definitions involve quadratic and logarithmic terms (see Sections~\ref{var leaf depths} and \ref{b2} for definitions). These nonlinearities make it impossible to represent the statistics by a linear representation in terms of the basis statistics we have defined.

Neither of the rank indices, the Colijn-Plazzota rank and the Furnas rank, can be represented by either basis. The rank indices are both bijections between the integers and the tree shapes, and therefore each tree has a unique value. However, one can easily construct different trees that have identical clade or daughter numbers. Therefore, it is clear that these indices cannot be represented by either basis. 

The mean $I$ and mean $I'$ indices cannot be represented by either basis, but the closely-related total $I$ and total $I'$ indices can be. The explanation for this result is that the mean $I$ and mean $I'$ are formed from the total $I$ and total $I'$ indices by normalizing to obtain a nonlinear function. 

The symmetry nodes index cannot be represented by either basis. This result is unsurprising because the symmetry of a tree relies on the entire tree and is not just a property of the daughter splits. 

The remaining five indices are all related to the height or width of subtrees: the $B_1$ index, maximum depth, maximum width, modified maximum difference in widths, and maximum width over maximum depths. The height of a subtree is the maximum of leaf depths, and taking the maximum of a set is a nonlinear function. Therefore, it is unsurprising that none of these indices are able to be represented by either basis.

We present proofs in Appendix~\ref{Appendix B} that all twelve of these indices cannot be represented by either basis .

\section{Mean and covariance of basis elements under the Yule model} 
\label{section 5}

In this section, we find exact formulas under the Yule model for the mean and variance of the clade-number basis elements and the covariance of pairs of basis elements. We obtain similar results for the daughter-number basis.

\subsection{Preliminary results}
\label{sec:5_1}

For computations involving the clade-number and daughter-number bases, we rely on several previously known results, which we restate. 

\begin{theorem}[Group elimination property, \cite{Aldous1996}; p.~52 of \cite{Steel2016}] \label{Theorem 1}
    Consider a subset of the leaves of $T$, $A \subset V_L(T)$, $|A| = r$, where $T \in \mathcal{T}_n$ is a random tree sampled under the Yule model. If $A$ is a clade in $T$ with common ancestor $v$, then the subtree $T_v$ is distributed as $T_v \sim \mathcal{T}_r$, a random Yule-generated tree in $\mathcal{T}_r$. 
\end{theorem}
In other words, Theorem \ref{Theorem 1} states that for trees with $n$ leaves generated under the Yule model, the properties of their $r$-leaf clades follow the Yule model on trees with $r$ leaves.

Let $s(T)$ denote the number of \emph{symmetric nodes} in $T$, the number of internal nodes whose daughter subtrees have identical unlabeled topologies. The next theorem describes the probabilities of unlabeled tree shapes under the Yule model.

\begin{theorem}[Proposition 3.4 of \cite{Steel2016}; eq.~5 of \cite{Brown1994}] \label{Theorem 2}
    Let $T \in \mathcal{T}_n$ be a rooted binary unlabeled tree topology. The probability that a random tree on $n$ leaves generated under the Yule model has topology $T$ is equal to 
    \begin{equation*}
        P(T) = 2^{n - 1 - s(T)} \prod_{v \in \inter} \frac{1}{n_v - 1}.
    \end{equation*}
\end{theorem}

The next results concern features of subtree sizes under the Yule model.
\begin{theorem}[Theorem 4.4 of \cite{Rosenberg2006}] \label{Theorem 3}
    Consider a rooted, binary tree $T \in \mathcal{T}_n$ sampled under the Yule model. For $2 \leq r \leq n-1$, the clade number $C_r(T)$ has mean
    \begin{equation*} 
        M(n,r) := \mathbb{E}[C_r(T)] =\frac{2n}{r(r+1)},
    \end{equation*}
    and variance
    \begin{equation*} 
        V(n, r) := \emph{Var}[C_r(T)] =  \begin{cases}
            \frac{2(4r^2 - 3r - 4)(r-1)n}{r(r+1)^2(2r-1)(2r+1)}, & \text{if  } r < \frac{n}{2}, \\[1.4ex]
            \frac{(5r-7)(r-1)n}{r(r+1)^2(2r-1)}, & \text{if  } r = \frac{n}{2}, \\[1.4ex]
            \frac{2(r^2 + r - 2n)n}{r^2(r+1)^2}, & \text{if  } r > \frac{n}{2}.
        \end{cases}
    \end{equation*}
\end{theorem}
\noindent Note that trivially, because the root is the unique node with $n$ descendant leaves, $M(n,r)=\mathbb{E}[C_r(T)]=1$ and $V(n,r)=\textrm{Var}[C_r(T)]=0$ for the case of $r=n$.

\begin{theorem}[Eq.~2 of \cite{Tajima1983}; Theorem 3.7 of \cite{Rosenberg2006}] \label{Theorem 4}
    Consider a rooted, binary tree $T \in \mathcal{T}_n$ sampled under the Yule model. The probability $p_n(i)$ that the daughter subtrees of the root have $i$ and $n-i$ descendants, where $1 \leq i \leq n-i \leq n-1$, is 
    \begin{equation*}
        p_n(i) = \begin{cases}
            \frac{1}{n-1},  & \text{if } i = \frac{n}{2}, \\[1.4ex]
            \frac{2}{n-1}, &  \text{if } 1 \leq i < \frac{n}{2}.
        \end{cases}
    \end{equation*}
\end{theorem}

Let $[n]$ denote the set $\{1, 2, \ldots, n\}$.

\begin{theorem}[Theorem 4.5 of \cite{Zhu2011}] \label{Theorem 5}
Consider a rooted, binary labeled tree $T \in \mathcal{T}_n$ sampled under the Yule model, with labels $[n]$. Let $A, B \subset [n]$ be two subsets of size $|A| = r$, $|B| = s$, with $1 \leq r \leq n-1$ and $1 \leq s \leq n-1$. The probability $p_n(A, B)$ that both $A$ and $B$ form clades of $T$ is 
    \begin{equation*} 
        p_n(A, B) = \begin{cases}
            \frac{2n}{r(r+1)} \binom{n}{r}^{-1}, & \text{if } A = B,\\[1.4ex]
            \frac{4n}{r(r+1)(s+1)}\binom{n}{s}^{-1} \binom{s}{r}^{-1}, & \text{if } A \subsetneq B, \\[1.4ex]
            \frac{4n}{(r+1) s(s+1)}\binom{n}{r}^{-1}\binom{r}{s}^{-1}, & \text{if } B \subsetneq A, \\[1.4ex]
            \frac{2}{n-1} \binom{n}{r}^{-1}, & \text{if } A \cap B = \emptyset, \ A \cup B = [n], \\[1.4ex]
            4n \binom{r+s}{s}^{-1} \binom{n}{r+s}^{-1} G_n(r, s), & \text{if } A \cap B = \emptyset, \ A \cup B \subsetneq [n], \\[1.4ex]
            0, & \text{otherwise},
        \end{cases}
    \end{equation*}
    where 
    \begin{equation*} 
        G_n(r, s) = \frac{n}{rs(r+1)(s+1)} - \frac{r(r+1) + s(s+1) + rs}{rs(r+1)(s+1)(r+s+1)} + \frac{1}{(r+s)[(r+s)^2 - 1]}.
    \end{equation*}
    
\end{theorem}

\subsection{Lemmas}
\label{sec:5_2}

In addition to the previously reported results in Theorems~\ref{Theorem 1}-~\ref{Theorem 5}, to simplify the proofs of the main results, we provide a series of lemmas. For a rooted binary labeled tree $T \in \mathcal{T}_n$ with labels $[n]$, let $Z_A$ be the indicator variable that subset $A$ is a clade in $T$.
\begin{lemma} \label{lemma6}
    Consider a rooted binary labeled tree $T \in \mathcal{T}_n$ sampled under the Yule model, with labels $[n]$. Let $A, B \subset [n]$ be two subsets with sizes $|A| = r$, $|B| = s$, with $2 \leq r \leq s \leq n - 1$.  Then $\mathbb{E}[Z_A Z_B] = p_n(A, B)$.
\end{lemma}
\begin{proof}
As the expectation of the product of indicator variables, $\mathbb{E}[Z_A Z_B]$ is equivalent to the probability that $A$ and $B$ are both clades in $T$, which is given precisely by $p_n(A, B)$ from Theorem~\ref{Theorem 5}. 
\end{proof}
\begin{lemma} \label{lemma7}
    Consider a rooted binary labeled tree $T \in \mathcal{T}_n$ sampled under the Yule model, with labels $[n]$. Summing over all unordered pairs of disjoint subsets $A, B \subset [n]$ of size $|A| = |B| = r$, $2 \leq r \leq n$, we have
\begin{equation}  
\label{eq:lemma7}
\sum_{\big\{ \{A,B\} \, : \, A,B \subset [n], \, |A| = |B| = r, \, A\cap B = \emptyset \big\} } \mathbb{E}[Z_A Z_B] = V(n,r) - M(n,r) + M(n,r)^2.
\end{equation}
\end{lemma}
\begin{proof}
If $2 \leq r \leq n-1$, following eq.~4.10 of \cite{Rosenberg2006}, the variance $\textrm{Var}[C_r(T)] = V(n,r)$ can be written
    \begin{equation} \label{eq:lemma 7.1}
        V(n,r) = M(n,r) - M(n,r)^2 + \mathbb{E}\left[\sum_{\big\{ \{A,B\} \, : \, A,B \subset [n], \, |A| = |B| = r, \, A \neq B \big\} } Z_A Z_B \right].
    \end{equation}
    Because $A$ and $B$ cannot both be clades if $A$ and $B$ are not disjoint (and using the linearity of expectation),
    \begin{equation} \label{eq:lemma 7.2}
        \mathbb{E}\Bigg[\sum_{\substack{\big\{ \{A,B\} \, : \, A,B \subset [n], \\ |A| = |B| = r, \, A \neq B \big\} }} Z_A Z_B\Bigg] = \sum_{\substack{\big\{ \{A,B\} \, : \, A,B \subset [n], \\ |A| = |B| = r, \, A \neq B \big\} }} \mathbb{E}[Z_A Z_B] = \sum_{\substack{\big\{ \{A,B\} \, : \, A,B \subset [n], \\ |A| = |B| = r, \, A\cap B = \emptyset \big\} }} \mathbb{E}[Z_A Z_B].
    \end{equation}  
    Combining eqs.~\ref{eq:lemma 7.1} and \ref{eq:lemma 7.2} completes the proof.

    If $r=n$, then both sides of eq.~\ref{eq:lemma7} are 0 and the result is trivial.
\end{proof}

For a rooted binary labeled tree $T \in \mathcal{T}_n$ with labels $[n]$, let $W_A^{i,j}$ be the indicator that subset $A$ is a clade of size $r$ in $T$ with daughter subtrees of sizes $i, j$, with $1 \leq i \leq \lfloor r/2 \rfloor$, $j = r - i$.
\begin{lemma} \label{lemma8}
    Consider a rooted binary labeled tree $T \in \mathcal{T}_n$ sampled under the Yule model, with labels $[n]$. Let $A \subset [n]$ be a subset of size $r$, with $2 \leq r \leq n-1$, $1 \leq i \leq \lfloor \frac{r}{2} \rfloor$, and $j = r - i$. Then  
    $$\mathbb{E}[W^{i,j}_A] = p_r(i) \, \mathbb{E}[Z_A] = p_r(i) \, \binom{n}{r}^{-1} M(n,r).$$ 
\end{lemma}

\begin{proof} 
By the law of total probability,
    \begin{align}
        \mathbb{E}[W_A^{i,j}] &= \mathbb{E}[W_A^{i,j} | Z_A = 1] \, \mathbb{P}[Z_A = 1] + \mathbb{E}[W_A^{i,j}|Z_A = 0] \, \mathbb{P}[Z_A = 0] \nonumber \\
        &= \mathbb{E}[W_A^{i,j}|Z_A = 1] \, \mathbb{E}[Z_A] \nonumber \\
        &= p_r(i) \, \mathbb{E}[Z_A]. \label{eq:lemma 8.1}
    \end{align}
    For the last step, we use the group elimination property (Theorem \ref{Theorem 1}) to note that conditional on $A$ being a clade of $T$, $A$ is distributed as a Yule tree of size $|A|= r$ leaves, and hence its split probability at the root follows Theorem~\ref{Theorem 4}. Because each subset of size $r$ has the same probability of being a clade of $T$, and $C_r(T) = \sum_{|A| = r} Z_A$, we get
    \begin{equation} \label{eq:lemma 8.2}
        M(n,r) = \mathbb{E}[C_r(T)] = \sum_{\{A \, : \, A \subset [n], \, |A| = r\}} \mathbb{E}[z_A] = \binom{n}{r}  \, \mathbb{E}[Z_A].
    \end{equation}
    Combining eqs.~\ref{eq:lemma 8.1} and \ref{eq:lemma 8.2} completes the proof. 
\end{proof}

\begin{lemma} \label{lemma9}
Consider a rooted binary labeled tree $T \in \mathcal{T}_n$ sampled under the Yule model, with labels $[n]$. Let $A, B \subset [n]$ be two subsets with sizes $|A| = r$ and $|B| = s$, with $2 \leq r \leq s \leq n-1$. Suppose $i+j=r$ with $1 \leq i \leq \lfloor \frac{r}{2} \rfloor$, and suppose also  that $k+\ell = s$ with $1 \leq k \leq \lfloor \frac{s}{2} \rfloor$. Then
\begin{equation}
         \mathbb{E}[W^{i,j}_A W^{k,\ell}_B] := R_{ijk\ell}(A, B) =\begin{cases}
            p_r(i) \, \binom{n}{r}^{-1} M(n,r), & \text{case 1: } A = B, \ i = k,  \\[1ex]
            0, & \text{case 2: }A = B, \ i \neq k, \\[1ex]
            p_r(i) \, p_s(k) \, \binom{n}{s}^{-1}\binom{s}{r}^{-1}M(n,s) \, H(k, \ell, r, s) & \text{case 3: } A \subsetneq B, \\[1ex]
            0, & \text{case 4: }A \cap B \neq \emptyset, A \nsubseteq B, \\[1ex]
            p_r(i) \, p_s(k) \, p_n(A, B), & \text{case 5: }A \cap B = \emptyset,
        \end{cases}
    \end{equation}
    where $H(k, \ell, r, s) = \mathbbm{1}_{\{r \leq k\}} M(k, r) + \mathbbm{1}_{\{r \leq \ell\}}  M(\ell, r)$, and $\mathbbm{1}_{\{\cdot\}}$ equals 1 if the statement $\{\cdot\}$ is true and 0 otherwise. 
\end{lemma}

\begin{figure}[tbp]
    \centering
    \includegraphics[width=0.60\textwidth]{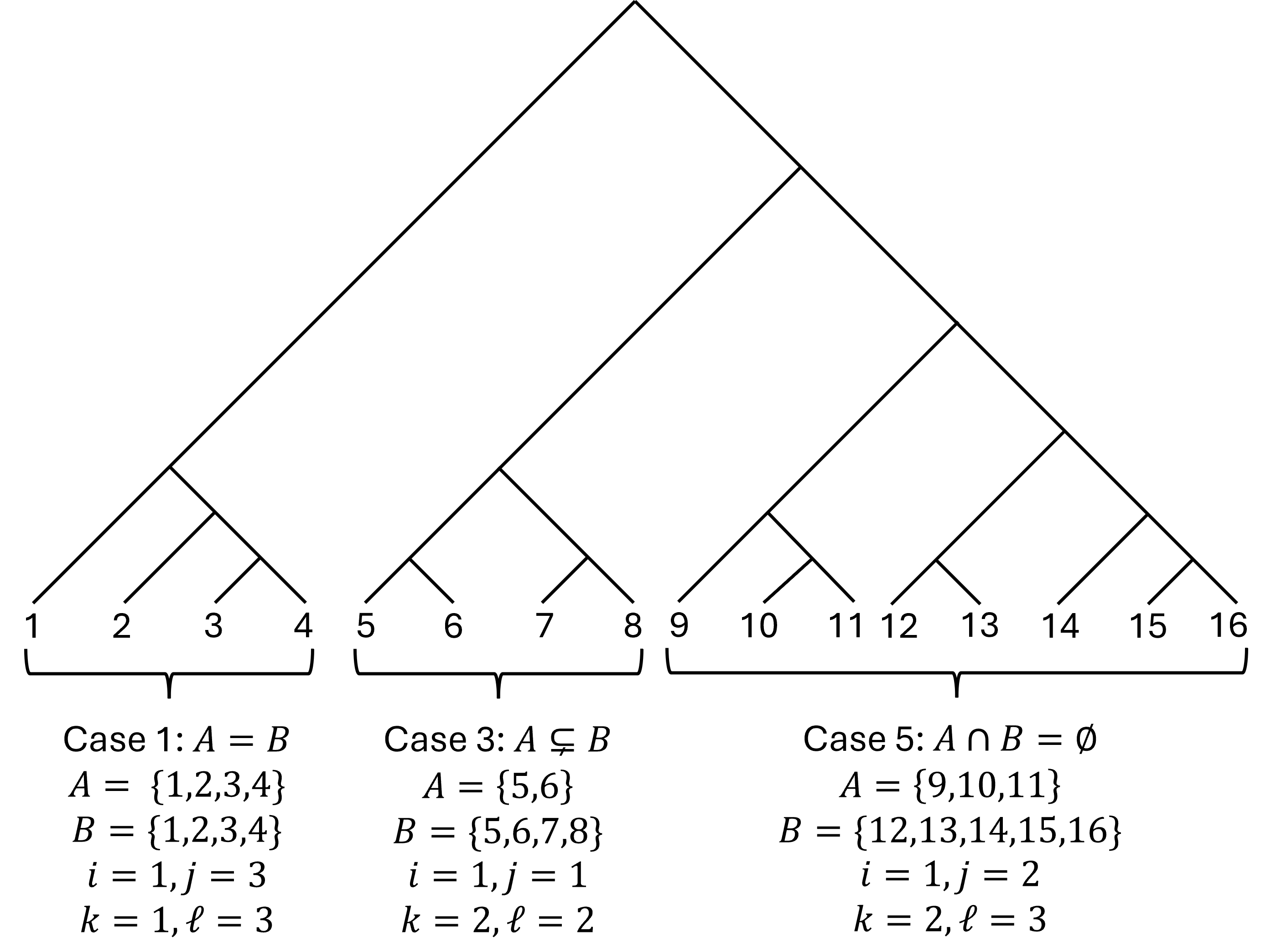}
    \caption{Schematic of the cases in Lemma~\ref{lemma9}. The three cases shown are the possible ways in which both subsets $A, B \subset [n]$ of the leaf set can be clades with a given daughter split $(i,j)$ of $A$ and $(k, \ell)$ of $B$.} 
    \label{fig:lemma 9 schematic}
\end{figure}

\begin{proof}
Case 1 (Figure \ref{fig:lemma 9 schematic}) is precisely the situation of Lemma~\ref{lemma8}, because $(W_A^{i,j})^2 = W_A^{i,j}$. 

Case 2 is 0 because a subset $A$ cannot both have a split $(i, j)$ and a split $(k, \ell)$ for the subtree sizes of its daughters, with $i \leq j$, $k \leq \ell$, and $i \neq k$. 

For Case 3 (Figure \ref{fig:lemma 9 schematic}), let $A \subsetneq B$. By the law of total expectation,
    \begin{align} 
        \mathbb{E}[W_A^{i,j} W_B^{k,\ell}] &= \mathbb{E}[W_A^{i,j} W_B^{k,\ell} | W_B^{k,\ell}= 1] \, \mathbb{P}[W_B^{k,\ell} = 1] +  \mathbb{E}[W_A^{i,j} W_B^{k,\ell} | W_B^{k,\ell}= 0] \, \mathbb{P}[W_B^{k,\ell} = 0] \nonumber \\
        &=  \mathbb{E}[W_A^{i,j} | W_B^{k,\ell} = 1] \, \mathbb{E}[W_B^{k,\ell}] \nonumber \\
        &= \mathbb{E}[W_A^{i,j} | W_B^{k,\ell} = 1]  \, p_s(k)\,\binom{n}{s}^{-1} M(n, s). \label{eq:lemma 9.1}
    \end{align}
    Here, we have used the fact that $W_B^{k,\ell}$ is an indicator random variable to equate $\mathbb{E}[W_B^{k,\ell}] = \mathbb{P}[W_B^{k,\ell} = 1]$, and for the final step we used Lemma \ref{lemma8}. 

    Next, we compute $\mathbb{E}[W_A^{i,j}|W_B^{k,\ell} = 1]$. If $W_B^{k,\ell} = 1$, then $B$ is a clade whose root has daughter clades of size $(k, \ell)$. Call these daughter clades $(L, R)$, where $|L| = k$ and $|R| = \ell$. The case of $k = \ell$ corresponds to arbitrarily assigning one subtree to be $L$ and the other to be $R$.
    
    For a fixed subset $C \subset B$ of size $|C| = k$, let $L_C$ be the indicator variable for the event that $C$ provides the leaf labels for the left subtree $L$. The number of possible subsets $C$ is $\binom{s}{k}$. Because we know that exactly one subset is equal to $L$, we have $\mathbb{P}[L_C = 1 | W_B^{k,\ell} = 1] = \binom{s}{k}^{-1}$. By the law of total expectation,
    \begin{align}
        \mathbb{E}[W_A^{i,j} | W_B^{k,\ell} = 1] &= \sum_{\{C \, : \, C \subset [n], \, |C| = k\} } \mathbb{E}[W_A^{i,j} | W_B^{k,\ell} L_C = 1] \, \mathbb{P}[L_C = 1 | W_B^{k,\ell} = 1] \nonumber \\
        &= \binom{s}{k}^{-1} \sum_{\{ C \, : \, C \subset [n], \, |C| = k \}} \mathbb{E}[W_A^{i,j} | W_B^{k,\ell} L_C = 1] \nonumber \\
        &= \binom{s}{k}^{-1} \left[ \sum_{\substack{\{C \, : \, C \subset [n], \, |C| = k, \\ A \subset C\} }} \mathbb{E}[W_A^{i,j} | W_B^{k,\ell}L_C = 1] + \sum_{\substack{\{C \, : \, C \subset [n], \, |C| = k, \\ A \subset B \setminus C\}}} \mathbb{E}[W_A^{i,j} | W_B^{k,\ell}L_C = 1] \right],\label{eq:lemma 9.2} 
    \end{align}
    where for the third equality we use the fact that $A \subsetneq B$ and $L_C=1$ imply that exactly one of the pair of statements ($A \subset C$, $A \subset B \setminus C$) must hold. 
    
    $A \subset C$ implies $r \leq k$. Conditional on $A \subset C$, with $C$ a clade, we use Theorem~\ref{Theorem 1} to deduce that the clade formed by $C$ is distributed as a Yule clade of size $k$. Applying Lemma \ref{lemma8} gives that, for $A \subset C$,
    \begin{equation} \label{eq:lemma 9.3}
        \mathbb{E}[W_A^{i,j} | W_B^{k,\ell}L_C = 1] = p_r(i) \, \binom{k}{r}^{-1}  \, M(k, r).
    \end{equation}
    The number of sets $C$ of size $k$ that contain the $r$ elements of $A$ is $\binom{s-r}{k-r}$, as $C$ contains the $r$ elements of $A$ and $k-r$ additional elements of $B \setminus A$, chosen from the $s-r$ elements of $B \setminus A$. 
    
    Similarly, $A \subset B \setminus C$ implies $r \leq \ell$. Conditional on $A \subset B \setminus C$, with $B \setminus C$ a clade, Lemma \ref{lemma8} implies that for $A \subset B \setminus C$, 
    \begin{equation} \label{eq:lemma 9.4}
        \mathbb{E}[W_A^{i,j} | W_B^{k,\ell}L_C = 1] = p_r(i) \, \binom{\ell}{r}^{-1} \, M(\ell, r).
    \end{equation}
     The number of sets $C$ of size $k$ such that $B \setminus C$, of size $\ell$, contains the $r$ elements of $A$ is $\binom{s-r}{\ell-r}$, as $B \setminus C$ contains the $r$ elements of $A$ and $\ell -r$ additional elements of $B \setminus A$, chosen from the $s-r$ elements of $B \setminus A$. 
        
    We can insert eqs.~\ref{eq:lemma 9.3} and \ref{eq:lemma 9.4} into eq.~\ref{eq:lemma 9.2} to obtain that
    \begin{align}
        \mathbb{E}[W_A^{i,j} | W_B^{k,\ell} = 1] &= \binom{s}{k}^{-1} \left[ \mathbbm{1}_{\{r \leq k\}} \, \binom{s-r}{k-r} \,  p_r(i) \, \binom{k}{r}^{-1} \, M(k, r) + \mathbbm{1}_{\{r \leq \ell\}} \, \binom{s-r}{\ell - r} p_r(i) \, \binom{\ell}{r}^{-1} \, M(\ell, r) \right] \nonumber \\
        &= p_r(i) \, \binom{s}{k}^{-1} \left[\mathbbm{1}_{\{r \leq k\}} \, \frac{\binom{s}{k}}{\binom{s}{r}} \, M(k, r) + \mathbbm{1}_{\{r \leq \ell\}} \, \frac{\binom{s}{\ell}}{\binom{s}{r}} \, M(\ell, r)\right] \nonumber \\
        &= p_r(i) \, \binom{s}{r}^{-1} \, H(k, \ell, r, s). \label{eq:lemma 9.5}
    \end{align}
    where in the last step, we use the binomial identity $\binom{s}{k} = \binom{s}{\ell}$ for $k + \ell = s$. Inserting eq.~\ref{eq:lemma 9.5} into eq.~\ref{eq:lemma 9.1} yields the final result for Case 3.

    Case 4 is 0 because if $A \cap B \neq \emptyset$ and $A \nsubseteq B$, then $A$ and $B$ cannot both be clades. 
    
    In case 5 (Figure \ref{fig:lemma 9 schematic}), we have $A \cap B = \emptyset$. Using the same notation from Lemma~\ref{lemma6}, where $Z_A$ is the indicator that $A$ is a clade, the law of total expectation gives
    \begin{equation}
    \label{eq:WW}
    \mathbb{E}[W_A^{i,j}W_B^{k,\ell}] = \mathbb{E}[W_A^{i,j}W_B^{k,\ell} | Z_AZ_B = 1] \, \mathbb{P}[Z_AZ_B = 1] = \mathbb{E}[W_A^{i,j}W_B^{k,\ell} | Z_AZ_B = 1] \, p_n(A, B)
    \end{equation}
    by Lemma~\ref{lemma6}. Because $A \cap B = \emptyset$, the subtree sizes for $A$ and $B$ are independent conditional on $Z_A Z_B = 1$:  
    \begin{equation*}
        \mathbb{E}[W_A^{i,j}W_B^{k,\ell} | Z_AZ_B = 1] = \mathbb{E}[W_A^{i,j}|Z_A = 1] \, \mathbb{E}[W_B^{k,\ell}|Z_B = 1].
    \end{equation*}
    In the proof of Lemma \ref{lemma8}, we used Theorem~\ref{Theorem 1} to find that $\mathbb{E}[W_A^{i,j} | Z_A = 1] = p_r(i)$ and similarly, $\mathbb{E}[W_B^{k, \ell} | Z_B = 1] = p_s(k)$. We insert these quantities into eq.~\ref{eq:WW}, concluding.
    \begin{equation*}
        \mathbb{E}[W_A^{i,j}W_B^{k,\ell}] = p_r(i)\, p_s(k) \, p_n(A, B).
    \end{equation*}

    This step concludes the proof of all five cases.
\end{proof}

With the lemmas now complete, describing features of clades and pairs of clades, we are now ready to study the vector of means $\boldsymbol{\mu}$ and the covariance matrix $\mathbf{\Sigma}$ for the two sets of basis statistics, $\mathcal{C}$ and $\mathcal{D}$.

\subsection{Clade-number basis}
\label{sec:5_3}

We use the results of Sections~\ref{sec:5_1} and \ref{sec:5_2} to state our main theorem concerning the clade-number basis.

\begin{theorem} 
\label{clade-number basis theorem}
    Let $C_r(T)$ denote the clade number of $T$ for $2 \leq r \leq n$, where $T \in\mathcal{T}_n$ is a random tree generated under the Yule model. 
    
\noindent (i) The mean of the clade number $C_r(T)$ is 
    \begin{equation} \label{eq:thm10i}
        M(n,r) := \mathbb{E}[C_r(T)] = \begin{cases}
            \frac{2n}{r(r+1)}, & 2 \leq r \leq n-1, \\[1.4ex]
            1, & r = n.
        \end{cases}
    \end{equation}

 \noindent (ii) The variance of the clade number $C_r(T)$ is  
    \begin{equation}
        V(n,r) := \emph{Var}[C_r(T)] = \begin{cases}
            \frac{2(4r^2 - 3r - 4)(r-1)n}{r(r+1)^2(2r-1)(2r+1)}, &  r < \frac{n}{2}, \\[1.4ex]
            \frac{(5r-7)(r-1)n}{r(r+1)^2(2r-1)}, & r = \frac{n}{2}, \\[1.4ex]
            \frac{2(r^2 + r - 2n)n}{r^2(r+1)^2}, & \frac{n}{2} < r \leq n-1, \\[1.4ex]
            0, & r = n.
        \end{cases}
    \end{equation}
    
\noindent (iii) For $2 \leq r < s \leq n-1$, the covariance between the clade numbers $C_r(T)$ and $C_s(T)$ is
    \begin{equation} \label{clade cov eq}
    C(n, r, s) := \emph{Cov}[C_r(T), C_s(T)] = \begin{cases}
        -\frac{4n(r-1)(r + 2s)}{s(s+1)(r+s)(r+s+1)(r+s-1)}, & r + s < n, \\[1.4ex]
        -\frac{2(r-1)\big(2r - s(s-3)\big)}{(r+1)s(s+1)(r+s-1)}, & r + s = n, \\[1.4ex]
        \frac{4n(s-n)}{r(r+1)s(s+1)}, & r + s > n.
    \end{cases}
\end{equation}
For $2 \leq r \leq n-1$, the covariance is $C(n, r, n) = \emph{Cov}[C_r(T), C_n(T)] = 0$. 
\end{theorem}
\begin{proof}

(i), (ii) These results simply augment Theorem~\ref{Theorem 3}, shown in \cite{Rosenberg2006}, with the case of $r=n$.

(iii) For the off-diagonal entries of the covariance matrix, we use Theorem~\ref{Theorem 5}. Let $2 \leq r < s \leq n-1$. Then  
\begin{align*}
    \textrm{Cov}[C_r(T), C_s(T)] &= \mathbb{E}[C_r(T) \, C_s(T)] - \mathbb{E}[C_r(T)]\, \mathbb{E}[C_s(T)]  \\
    &= \mathbb{E}[C_r(T)\,C_s(T)] - M(n, r) \, M(n,s).
\end{align*}
By definition of $C_r(T)$ as the count of clades in $T$ with precisely $r$ leaves, we can write
\begin{equation*}
    C_r(T) = \sum_{\substack{ \{A \, : \, A \subset [n], |A| = r\} }} Z_A,
\end{equation*}
recalling that $Z_A$ is the indicator variable that subset $A$ is a clade in $T$. 

With $|A| = r$ and $|B| = s$,
\begin{equation*}
    \mathbb{E}[C_r(T) \, C_s(T)] = \sum_{\substack{\{A \, : \, A \subset [n], \\ |A| = r\} }} \sum_{\substack{\{B \, : \, B \subset [n], \\ |B| = s\} }} \mathbb{E}[Z_AZ_B] = \sum_{\substack{\{A \, : \, A \subset [n], \\ |A| = r\} }} \sum_{\substack{\{B \, : \, B \subset [n], \\ |B| = s\} }} p_n(A, B),
\end{equation*}
by Lemma~\ref{lemma6}. Because $p_n(A, B) = 0$ if $A$ and $B$ contain shared elements but neither is a subset of the other ($A \cap B \neq \emptyset$, $A \nsubseteq B$, and $B \nsubseteq A$), we obtain
\begin{equation} \label{eq:theorem10.1}
    C(n,r,s) = \bigg[ 
        \sum_{\substack{\{A \, : \, A \subset [n], \\ |A| = r\} }} \sum_{\substack{\{B \, : \, B \subset [n], \\ |B| = s, \, A \subsetneq B\} }}
        p_n(A, B) \bigg] + \bigg[ 
         \sum_{\substack{\{A \, : \, A \subset [n], \\ |A| = r\} }} \sum_{\substack{\{B \, : \, B \subset [n], \\ |B| = s, \, A \cap B = \emptyset \} }}
        p_n(A, B) \bigg] - M(n,r) \, M(n,s).
\end{equation}
Using the second case of Theorem~\ref{Theorem 5} (for $A \subsetneq B$), we find that
\begin{equation} \label{eq:theorem10.2}
    \bigg[ 
        \sum_{\substack{\{A \, : \, A \subset [n], \\ |A| = r\} }} \sum_{\substack{\{B \, : \, B \subset [n], \\ |B| = s, \, A \subsetneq B\} }}
        p_n(A, B) \bigg] = \binom{n}{s}\binom{s}{r} \bigg[ \binom{n}{s}^{-1}\binom{s}{r}^{-1}\frac{4n}{r(r+1)(s+1)} \bigg] = M(n,s) \, M(s,r).
\end{equation}
Therefore, inserting eq.~\ref{eq:theorem10.2} into eq.~\ref{eq:theorem10.1} and simplifying using eq.~\ref{eq:thm10i},
\begin{align} 
    C(n, r, s) &= M(n,s) \, M(s,r) - M(n,r) \, M(n,s) + \bigg[ 
         \sum_{\substack{\{A \, : \, A \subset [n], \\ |A| = r\} }} \sum_{\substack{\{B \, : \, B \subset [n], \\ |B| = s, \, A \cap B = \emptyset \} }}
        p_n(A, B) \bigg] \nonumber \\
        &= \frac{4n(s-n)}{r(r+1)s(s+1)} + \bigg[ 
         \sum_{\substack{\{A \, : \, A \subset [n], \\ |A| = r\} }} \sum_{\substack{\{B \, : \, B \subset [n], \\ |B| = s, \, A \cap B = \emptyset \} }}
        p_n(A, B) \bigg]. \label{eq:theorem10.3}
\end{align}
We split the remaining sum in eq.~\ref{eq:theorem10.3} into three cases: $r + s < n$, $r + s = n$, and $r + s > n$. 

If $r + s < n$ and $A \cap B = \emptyset$, we know $p_n(A, B) = 4n\binom{r+s}{ r}^{-1} \binom{n}{r+s}^{-1} G_n(r, s)$ from the fifth case of Theorem~\ref{Theorem 5}. Simplifying and substituting for $G_n(r,s)$ gives
\begin{align}
    C(n,r,s) &= \frac{4n(s-n)}{r(r+1)s(s+1)} + \bigg[ 
         \sum_{\substack{\{A \, : \, A \subset [n], \\ |A| = r\} }} \sum_{\substack{\{B \, : \, B \subset [n], \\ |B| = s, \, A \cap B = \emptyset \} }}
       4n\binom{r+s}{ r}^{-1} \binom{n}{r+s}^{-1} G_n(r, s) \bigg] \nonumber \\
    &= \frac{4n(s-n)}{r(r+1)s(s+1)} + \binom{n}{r} \binom{n-r}{s} 4n\binom{r+s}{ r}^{-1} \binom{n}{r+s}^{-1} G_n(r, s) \nonumber \\
    &= \frac{4n(s-n)}{r(r+1)s(s+1)} + 4n \, G_n(r,s) \nonumber \\
    &= -\frac{4n(r-1)(r + 2s)}{s(s+1)(r+s)(r+s+1)(r+s-1)}. \label{eq:theorem10.4}
\end{align}

For the next case, if $r + s = n$ and $A \cap B = \emptyset$, we know that $p_n(A, B) = \frac{2}{n-1}\binom{n}{r}^{-1}$ from the fourth case of Theorem~\ref{Theorem 5}. Substituting into eq.~\ref{eq:theorem10.3} and simplifying gives
\begin{align}
    C(n,r,s) &= \frac{4n(s-n)}{r(r+1)s(s+1)} + \bigg[ 
         \sum_{\substack{\{A \, : \, A \subset [n], \\ |A| = r\} }} \sum_{\substack{\{B \, : \, B \subset [n], \\ |B| = s, \, A \cap B = \emptyset \} }}
        \frac{2}{n-1} \binom{n}{r}^{-1} \bigg] \nonumber \\
        &= \frac{4n(s-n)}{r(r+1)s(s+1)} + \binom{n}{r} \binom{n-r}{n-r} \frac{2}{n-1} \binom{n}{r}^{-1} \nonumber \\
        &= -\frac{2(r-1)[2r - s(s-3)]}{(r+1)s(s+1)(r+s-1)}, \label{eq:theorem10.5}
\end{align}
where the last step makes use of $r+s=n$.

For the final case, if $r + s > n$, then it is impossible to have $A \cap B = \emptyset$, so the final terms in eq.~\ref{eq:theorem10.3} vanish. Therefore, in this case,
\begin{equation} \label{eq:theorem10.6}
    C(n,r,s) = \frac{4n(s-n)}{r(r+1)s(s+1)}. 
\end{equation}

Finally, note that for $s=n$, because $C_n(T)$ is a constant $C_n(T)=1$, $\textrm{Cov}[C_r(T), C_s(T)] = \mathbb{E}[C_r(T) \times 1] - \mathbb{E}[C_r(T)] \times 1 = 0$.
\end{proof}

\subsection{Daughter-number basis}
\label{sec:5_4}

In a similar manner to our computation for the clade-number basis, we use the results of Sections~\ref{sec:5_1} and \ref{sec:5_2} to state our main theorem for the daughter-number basis. 

\begin{theorem} \label{daughter-number basis theorem}
    Let $D_{i,j}(T)$ denote the daughter number of $T$ for $1 \leq i \leq j$, $i + j = r \leq n$, and let $D_{k,\ell}(T)$ denote the daughter number of $T$ for $1 \leq k \leq \ell$, $k + \ell = s \leq n$, where $T \in \mathcal{T}_n$ is a random tree generated under the Yule model. 
    
\noindent (i) The mean of the daughter number $D_{i,j}(T)$ is 
    \begin{align} \label{daughter mean eq}
        M_D(n, i, j) := \mathbb{E}[D_{i,j}(T)] &= p_{r}(i) \, M(n, r). 
    \end{align}

\noindent (ii) The variance of the daughter number $D_{i,j}(T)$ is 
    \begin{align} \label{daughter var eq}
        V_D(n, i, j) := \emph{Var}[D_{i,j}(T)] &= p_{r}(i) \, [1 - p_{r}(i)] \, M(n, r) + p_{r}(i)^2 \, V(n, r). 
    \end{align}

\noindent (iii) If $r=s$ and $i \neq k$, then the covariance between the daughter numbers $D_{i,j}(T)$ and $D_{k ,\ell}(T)$ is 
    \begin{align}
        C_D(n, i, j, k, \ell) := \emph{Cov}[D_{i,r-i}, D_{k,r-k}] = p_r(i) \, p_r(k) \, [V(n,r) - M(n,r)].
    \end{align}

\noindent (iv) If $r < s$, then the covariance between the daughter numbers $D_{i,j}(T)$ and $D_{k,\ell}(T)$ is 
    \begin{align}
        C_D(n, i, j, k, \ell) := \emph{Cov}[D_{i,j}(T), D_{k,\ell}(T)] &= p_r(i) \, p_s(k) \, \bigg[C(n, r, s) \nonumber \\
        & \quad + M(n, s) [\mathbbm{1}_{\{k \geq r\}} M(k,r) + \mathbbm{1}_{\{\ell \geq r\}} M(\ell, r) - M(s, r)]\bigg],
    \end{align}
    where $\mathbbm{1}_{\{\cdot\}}$ equals 1 if the statement $\{\cdot\}$ is true and 0 otherwise. 
\end{theorem}

\begin{proof} (i) For a subset $A\subset [n]$, let $W^{i,j}_A$ be the indicator that $A$ is a clade in $T$ with daughter subtrees of sizes $(i, j)$. By definition of the daughter number,
    \begin{equation*}
        D_{i,j}(T) = \sum_{ \{ A \, : \, A \subset [n], \, |A| = r \} } W^{i,j}_A.
    \end{equation*}
    Taking expectations and applying Lemma~\ref{lemma8} gives
    \begin{align*}
        \mathbb{E}[D_{i,j}(T)] &= \mathbb{E} \bigg[\sum_{ \substack{\{A \, : \, A \subset [n], \\ |A| = r\} } } W_A^{i,j} \bigg] = \binom{n}{r} \, p_r(i) \, \binom{n}{r}^{-1}  M(n,r) = p_r(i) \, M(n,r).
    \end{align*}

To prove parts (ii), (iii), and (iv) of the theorem, we first derive a general formula for the covariance between any two daughter numbers $D_{i,j}(T)$ and $D_{k, \ell}(T)$; the main results all follow from special cases of this formula. For any $i, j, k, \ell$, we can write the covariance as
\begin{align}
    \textrm{Cov}[D_{i,j}(T), D_{k,\ell}(T)] &= \mathbb{E}[D_{i,j}(T) \, D_{k,\ell}(T)] - \mathbb{E}[D_{i,j}(T)] \, \mathbb{E}[D_{k,\ell}(T)] \nonumber \\
    &= \mathbb{E} \bigg[ \sum_{\substack{\{A \, : \, A \subset [n], \\ |A| = r\} } } \sum_{ \substack{\{B \, : \, B \subset [n], \\ |B| = s\} }} W_A^{i,j}W_B^{k,\ell} \bigg] - p_r(i) \, p_s(k) \, M(n,r) \, M(n,s) \nonumber \\
    &= \bigg[ \sum_{\substack{\{A \, : \, A \subset [n], \\ |A| = r\} } } \sum_{ \substack{\{B \, : \, B \subset [n], \\ |B| = s\} }} \mathbb{E}[W_A^{i,j}W_B^{k,\ell}] \bigg] - p_r(i) \, p_s(k) \, M(n,r) \, M(n,s). \label{eq:theorem11.1}
\end{align}

(ii) By setting $i = k$ and $j = \ell$ in eq.~\ref{eq:theorem11.1}, we write the variance of the daughter number $D_{i,j}(T)$ as 
\begin{align}
    \textrm{Var}[D_{i,j}(T)] &= \bigg[\sum_{ \substack{\{A \, : \, A \subset [n], \\ |A| = r\} } } \sum_{\substack{\{B \, : \, B \subset [n], \\ |B| = r\} } } \mathbb{E}[W_A^{i,j}W_B^{i,j}]\bigg] - p_r(i)^2 \, M(n,r)^2. \label{eq:theorem11.2}
\end{align}
Using Lemma~\ref{lemma9}, we know that $\mathbb{E}[W_A^{i,j}W_B^{i,j}] = R_{ijk\ell}(A,B)$ is nonzero only if $A = B$ or $A \cap B = \emptyset$, so eq.~\ref{eq:theorem11.2} becomes
\begin{align}
    \textrm{Var}[D_{i,j}(T)] &= \bigg[\sum_{ \substack{\{A \, : \, A \subset [n], \\ |A| = r\} } } \mathbb{E}[W_A^{i,j}W_A^{i,j}]\bigg] + \bigg[\sum_{ \substack{\{A \, : \, A \subset [n], \\ |A| = r\} } } \sum_{\substack{\{B \, : \, B \subset [n], \\ |B| = r, \, A \cap B = \emptyset\} } } \mathbb{E}[W_A^{i,j}W_B^{i,j}]\bigg] - p_r(i)^2 \, M(n,r)^2 \nonumber \\
    &= \binom{n}{r}  p_r(i) \, \binom{n}{r}^{-1}\, M(n,r) + \bigg[\sum_{ \substack{\{A \, : \, A \subset [n], \\ |A| = r\} } } \sum_{\substack{\{B \, : \, B \subset [n], \\ |B| = r, \, A \cap B = \emptyset\} } } p_r(i)^2 p_n(A, B)\bigg] - p_r(i)^2 \, M(n,r)^2 \nonumber \\
    &= p_r(i) \, M(n,r) + p_r(i)^2 \bigg[\sum_{ \substack{\{A \, : \, A \subset [n], \\ |A| = r\} } } \sum_{\substack{\{B \, : \, B \subset [n], \\ |B| = r, \, A \cap B = \emptyset\} } } p_n(A, B)\bigg] - p_r(i)^2 \, M(n,r)^2. \label{eq:theorem11.3}
\end{align}
Next, we apply Lemma~\ref{lemma6} and Lemma~\ref{lemma7} to the sum in eq.~\ref{eq:theorem11.3} to obtain
\begin{align}
    \textrm{Var}[D_{i,j}(T)] &= p_r(i) \, M(n,r) + p_r(i)^2 \big[V(n,r) - M(n,r) + M(n,r)^2\big] - p_r(i)^2 \, M(n,r)^2 \nonumber \\
    &= p_{r}(i) \, [1 - p_{r}(i)] \, M(n, r) + p_{r}(i)^2 \, V(n, r). 
\end{align}

(iii) Similarly, if $r = s$ and $i \neq k$, we can use eq.~\ref{eq:theorem11.1} to write the covariance of the daughter numbers $D_{i,r-i}(T)$, $D_{k,r-k}(T)$ as
\begin{align}
    \textrm{Cov}[D_{i,r-i}(T), D_{k,r-k}(T)] &= \bigg[ \sum_{\substack{\{A \, : \, A \subset [n], \\ |A| = r\} } } \sum_{ \substack{\{B \, : \, B \subset [n], \\ |B| = r\} }} \mathbb{E}[W_A^{i,r-i} \, W_B^{k,r-k}] \bigg] - p_r(i) \, p_r(k) \, M(n,r)^2. \label{eq:theorem11.4}
\end{align}
Using Lemma~\ref{lemma9}, eq.~\ref{eq:theorem11.4} becomes
\begin{align}
    \textrm{Cov}[D_{i,r-i}(T), D_{k,r-k}(T)] &= \bigg[ \sum_{\substack{\{A \, : \, A \subset [n], \\ |A| = r\} } } \sum_{ \substack{\{B \, : \, B \subset [n], \\ |B| = r, \, A \cap B = \emptyset\} }} p_r(i) \, p_r(k) \, p_n(A, B) \bigg] - p_r(i) \, p_r(k) \, M(n,r)^2 \nonumber \\
    &= p_r(i) \, p_r(k) \bigg[ \sum_{\substack{\{A \, : \, A \subset [n], \\ |A| = r\} } } \sum_{ \substack{\{B \, : \, B \subset [n], \\ |B| = r, \, A \cap B = \emptyset\} }} p_n(A, B) \bigg] - p_r(i) \, p_r(k) \, M(n,r)^2. \label{eq:theorem11.5}
\end{align}
By applying  Lemma~\ref{lemma6} and Lemma~\ref{lemma7} to the sum in eq.~\ref{eq:theorem11.5}, we obtain
\begin{align}
    \textrm{Cov}[D_{i,r-i}(T), D_{k,r-k}(T)] &= p_r(i) \, p_r(k) \big[ V(n,r) - M(n,r) + M(n,r)^2 \big] - p_r(i) \, p_r(k) \, M(n,r)^2 \nonumber \\
    &= p_r(i) \, p_r(k) \, [V(n,r) - M(n,r)]. 
\end{align}

(iv) Finally, if $r < s$, we begin from eq.~\ref{eq:theorem11.1}. By applying Lemma~\ref{lemma9} to the sum in eq.~\ref{eq:theorem11.1}, we get
\begin{align}
    \textrm{Cov}[D_{i,j}(T), D_{k,\ell}(T)] &= \bigg[ \sum_{\substack{\{A \, : \, A \subset [n], \\ |A| = r\} } } \sum_{ \substack{\{B \, : \, B \subset [n], \\ |B| = s, \, A \subsetneq B\} }} R_{ijk\ell}(A, B) \bigg] + \bigg[ \sum_{\substack{\{A \, : \, A \subset [n], \\ |A| = r\} } } \sum_{ \substack{\{B \, : \, B \subset [n], \\ |B| = s, \, A \cap B = \emptyset\} }} R_{ijk\ell}(A,B) \bigg] \nonumber \\
    & \, - p_r(i) \, p_r(k) \, M(n,r) \, M(n,s). \label{eq:theorem11.6}
\end{align}
 By case 3 of Lemma~\ref{lemma9}, we find that
\begin{align}
     \sum_{\substack{\{A \, : \, A \subset [n], \\ |A| = r\} } } \sum_{ \substack{\{B \, : \, B \subset [n], \\ |B| = s, \, A \subsetneq B\} }} R_{ijk\ell}(A, B)  &=  \binom{n}{s} \, \binom{s}{r} \, p_r(i) \, p_s(k) \, \binom{n}{s}^{-1} \binom{s}{r}^{-1} M(n,s) \,  H(k,\ell, r,n) \nonumber \\
    &=p_r(i) \, p_s(k) \, M(n,s) \big[ \mathbbm{1}_{\{r \leq k\}} \, M(k, r) + \mathbbm{1}_{\{r \leq \ell\}} \, M(\ell, r)  \big], \label{eq:theorem11.7}
\end{align}
where we use the definition of $H$ from Lemma~\ref{lemma9}. 

Next, applying case 5 of Lemma~\ref{lemma9}, we find that
\begin{equation} \label{eq:theorem11.8}
    \sum_{\substack{\{A \, : \, A \subset [n], \\ |A| = r\} } } \sum_{ \substack{\{B \, : \, B \subset [n], \\ |B| = s, \, A \cap B = \emptyset\} }} R_{ijk\ell}(A,B) = p_r(i) \, p_s(k)  \bigg[ \sum_{\substack{\{A \, : \, A \subset [n], \\ |A| = r\} } } \sum_{ \substack{\{B \, : \, B \subset [n], \\ |B| = s, \, A \cap B = \emptyset\} }} p_n(A, B) \bigg].
\end{equation}
In eq.~\ref{eq:theorem10.3}, we previously observed that
\begin{equation}
    C(n, r, s) = M(n,s) \, M(s,r) - M(n,r) \, M(n,s) + \bigg[ 
         \sum_{\substack{\{A \, : \, A \subset [n], \\ |A| = r\} }} \sum_{\substack{\{B \, : \, B \subset [n], \\ |B| = s, \, A \cap B = \emptyset \} }}
        p_n(A, B) \bigg], \label{eq:theorem11.9}
\end{equation}
where in the context of the proof it was assumed that $r \leq s \leq n-1$. However, the equation still holds if $r = s = n$ because both sides are 0. Therefore, for any $(r, s)$, we have
\begin{equation}
    \sum_{\substack{\{A \, : \, A \subset [n], \\ |A| = r\} }} \sum_{\substack{\{B \, : \, B \subset [n], \\ |B| = s, \, A \cap B = \emptyset \} }}
        p_n(A, B) = C(n, r, s) - M(n, s) \, M(s,r) + M(n,r) \, M(n,s). \label{eq:theorem11.10}
\end{equation}
Substituting eq.~\ref{eq:theorem11.10} into eq.~\ref{eq:theorem11.8} gives

\begin{equation} \label{eq:theorem11.11}
    \sum_{\substack{\{A \, : \, A \subset [n], \\ |A| = r\} } } \sum_{ \substack{\{B \, : \, B \subset [n], \\ |B| = s, \, A \cap B = \emptyset\} }} R_{ijk\ell}(A,B) = p_r(i) \, p_s(k) \big[ C(n,r,s) - M(n,s) \, M(s,r)  + M(n,r) \, M(n,s) \big].
\end{equation}

By inserting eqs.~\ref{eq:theorem11.7} and \ref{eq:theorem11.11} into eq.~\ref{eq:theorem11.6}, we obtain
\begin{align}
    \textrm{Cov}[D_{ij}(T), D_{k\ell}(T)] &= p_r(i) \, p_s(k) \, M(n,s) \big[ \mathbbm{1}_{\{r \leq k\}} \, M(k, r) + \mathbbm{1}_{\{r \leq \ell\}} \, M(\ell, r)  \big] \nonumber \\
    & \quad + p_r(i) \, p_s(k) \big[C(n,r,s) + M(n,r) \, M(n,s) - M(n,s) \, M(s,r)] \nonumber \\
    & \quad - p_r(i) \, p_s(k) \, M(n,r) \, M(n,s) \nonumber \\
    &=  p_r(i) \, p_s(k) \, \big[C(n, r, s) + M(n, s) [ \mathbbm{1}_{\{k \geq r\}} M(k,r) + \mathbbm{1}_{\{\ell \geq r\}} M(\ell, r) - M(s, r)]\big]. 
\end{align}
This proves the theorem. 
\end{proof}

\section{New mathematical results for balance indices under the Yule model: mean, variance, covariance} \label{section 6}

Exact formulas for the mean and variance under the Yule model for many of the balance indices listed are already known (i.e., Table 4.5 of Fischer et al.~\cite{Fischer2023}). In addition, the covariances are known for a small number of pairs of indices: Cardona et al.~previously found exact, simplified formulas for the covariance between the Sackin index and the total cophenetic index \cite{Cardona2013}, and Blum et al.~described the asymptotic large-$n$ limit of the covariance between the Sackin index and Colless index \cite{Blum2006Mean}.

However, many gaps exist in knowledge of the properties of tree balance indices under the Yule model. In this section, we apply the results obtained in Section~\ref{section 5} to find exact formulas for the mean and variance of all 17 balance indices that can be represented by the clade-number basis or the daughter-number basis, and the covariance of all $\binom{17}{2}$ pairs of indices. We combine specific basis representations of indices from Appendix~\ref{Appendix A} with Theorems~\ref{clade-number basis theorem} and \ref{daughter-number basis theorem} to obtain the mean (eq.~\ref{mean of statistic}) and variance and covariance (eq.~\ref{covariance of statistics}) of indices. Using this method, we derive 9 new formulas for means of indices. Moreover, we obtain exact covariance results for all $\binom{17}{2} = 136$ pairs of indices. 

\subsection{Exact results for the mean of 17 balance indices under the Yule model} \label{sec:6_1}

Of the 17 balance indices represented by the clade-number basis or the daughter-number basis, seven were listed by Fischer et al.~\cite[Table 4.5 and Chapter 23]{Fischer2023} as having a known mean under the Yule model, and one was derived more recently~\cite{Cleary2025}. In this section, we obtain the means of all 17 indices under the Yule model, including 9 new results. 

Consider a statistic $A(T)$ that is represented by the clade-number basis with coefficients $a_n(r)$ for $2 \leq r \leq n$. From eq.~\ref{mean of statistic}, its expected value under the Yule model is given by
\begin{align} \label{clade represented mean}
    \mathbb{E}\big[ A(T) \big] &= \mathbb{E} \bigg[ \sum_{r=2}^n a_n(r) \, C_r(T) \bigg] = \sum_{r=2}^n a_n(r) \, M(n,r) = \bigg[\sum_{r=2}^{n-1} a_n(r) \, \frac{2n}{r(r+1)} \bigg] + a_n(n),
\end{align}
where we use Theorem~\ref{Theorem 3} for the exact expression of the mean $M(n,r)$. 

Similarly, consider a statistic $B(T)$ that can be represented by the daughter-number basis with coefficients $b_n(i,j)$, for $1 \leq i \leq \lfloor n/2 \rfloor$, $i \leq j \leq n-i$. From eq.~\ref{mean of statistic}, its expected value under the Yule model is 
\begin{align}
    \mathbb{E}\big[B(T) \big] &= \mathbb{E} \bigg[ \sum_{i = 1}^{\lfloor n/2 \rfloor} \sum_{j = i}^{n-i} b_n(i, j) \, D_{i, j}(T).  \bigg] \nonumber \\
    &= \sum_{i = 1}^{\lfloor n/2 \rfloor} \sum_{j = i}^{n-i} b_n(i, j) \, M_D(n, i, j) \nonumber \\
    &= \sum_{i = 1}^{\lfloor n/2 \rfloor} \sum_{j = i}^{n-i} b_n(i, j) \, p_{i+j}(i) \, M(n, i+j),\label{eq:6.2}
\end{align}
where $M_D(n,i,j)$ is the mean of the daughter number $D_{i,j}(T)$ and we substituted for this quantity using Theorem~\ref{daughter-number basis theorem} in the last step. Transforming the indices of this double summation by $(i, j) \mapsto (r, i)$ with $r = i + j$, we can rewrite eq.~\ref{eq:6.2} as
\begin{align}
    \mathbb{E}\big[B(T) \big] &= \sum_{r=2}^n \sum_{i=1}^{\lfloor r/2 \rfloor} b_n(i, r-i) \, p_r(i) \, M(n, r) \nonumber \\
    &= \sum_{r=2}^n M(n, r) \sum_{i=1}^{\lfloor r/2 \rfloor} b_n(i, r-i) \, p_r(i). \label{eq:6.3}
\end{align}
This transformation can make simplification easier in many cases. 

For all 17 indices, we can combine the coefficients in the clade-number basis or the daughter-number basis found in Appendix~\ref{Appendix A} with eqs.~\ref{clade represented mean} and \ref{eq:6.3} to obtain exact expressions for the expected value of the index under the Yule model. In many cases, the corresponding equations can be simplified considerably. As an example, we derive a known result: the expected value of the Sackin index under the Yule model.  

\begin{theorem}
    Consider a rooted, binary tree $T \in \mathcal{T}_n$ sampled under the Yule model. The expected value of the Sackin index $S(T)$ satisfies $\mathbb{E}[S(T)] = 2n(H_n - 1)$, where $H_n = \sum_{i=1}^n 1/i$ denotes the $n$th harmonic number. 
\end{theorem}

\begin{proof}

As shown in Section~\ref{sackin}, the Sackin index is represented by the clade-number basis as $S(T) = \sum_{r=2}^n r \, C_r(T)$, so it has coefficients $a_n(r) = r$. Therefore, substituting these coefficients into eq.~\ref{clade represented mean} gives
\begin{align}
    \mathbb{E}\big[S(T)\big] &= \bigg[ \sum_{r=2}^{n-1} \frac{2nr}{r(r+1)} \bigg] + n \nonumber \\
    &= 2n \bigg(\sum_{r=2}^{n-1} \frac{1}{r+1}\bigg) + n \nonumber \\
    &= 2n \bigg(\sum_{r=3}^n \frac{1}{r}  + \frac{1}{2}\bigg) \nonumber \\
    &= 2n \, (H_n - 1), \nonumber
\end{align}
proving the theorem.
\end{proof}

As a second example, we can use eq.~\ref{clade represented mean} to quickly obtain the expected value of the $\hat{s}$-shape statistic under the Yule model.  
\begin{theorem}
    Consider a rooted, binary tree $T \in \mathcal{T}_n$ sampled under the Yule model. The expected value of the $\hat{s}$-shape statistic $\hat{s}(T)$ satisfies $\mathbb{E}[\hat{s}(T)] = \log(n-1) + 2n \sum_{r=2}^{n-1} [{\log(r-1)}]/[{r(r+1)}]$.
\end{theorem}

\begin{proof}

As shown in Section~\ref{s-shape}, the $\hat{s}$-shape statistic is represented by the clade-number basis as $\hat{s}(T) = \sum_{r=2}^n \log(r-1) \, C_r(T)$, so it has coefficients $a_n(r) = \log(r-1)$. Therefore, substituting these coefficients into eq.~\ref{clade represented mean} gives that
\begin{align}
    \mathbb{E}\big[\hat{s}(T)\big] &= \bigg[ \sum_{r=2}^{n-1} \frac{2n \log(r-1)}{r(r+1)} \bigg] + \log(n-1), \nonumber
\end{align}
proving the theorem.
\end{proof}
Note that Proposition 10 of Cleary et al.~\cite{Cleary2025} reported this expectation, but did not include a term $\log(n-1)$ corresponding to the root vertex.

In Table~\ref{tab:new_mean_results}, we list formulas for the expected value under the Yule model of all 17 balance indices that are represented by the clade-number or daughter-number basis. We simplified each expression as much as possible using standard combinatorial identities. We plot the resulting expressions for all 17 indices in Figure~\ref{fig:mean var and cv}a  to observe how the expected value of each statistic behaves as a function of $n$. 

\begin{table}[hp]
    \centering
    \renewcommand{\arraystretch}{1.2}
    \setlength{\tabcolsep}{4pt}
    \begin{tabularx}{\linewidth}{|L{0.22\linewidth} | E | R{1.45cm} |}
        \hline
        
        \textbf{Index name} & \textbf{Expectation under Yule model} & \textbf{New result}\\ \hline
        
        Sackin index & $2n (H_n - 1)$ & No \\ \hline
        
        Average leaf depth & $2 (H_n - 1)$ & No \\ \hline
        
        Total cophenetic index & $n (n+1 - 2H_n)$ & No \\ \hline
        
        $\hat{s}$-shape statistic & 
        $\displaystyle
        \log(n-1) + 2n \sum_{r=2}^{n-1} \frac{\log(r - 1)}{r(r+1)}
        $ 
        & No \\ \hline
        
        Colless index & 
        $\displaystyle
        \begin{cases}
            n \big(H_{n/2} - 1\big), & n \text{ even} \\
            n \big(H_{(n-1)/2} - 1 \big) + 1, & n \text{ odd}
        \end{cases}
        $ 
        & No \\ \hline
        
        Corrected Colless index &  
        $\displaystyle
        \begin{cases}
            \dfrac{2n \big(H_{n/2} - 1 \big)}{(n-2)(n-1)}, &  n \text{ even} \\
            \dfrac{2n \big(H_{(n-1)/2} - 1 \big) + 2}{(n-2)(n-1)}, &  n \text{ odd}
        \end{cases}
        $
        & No \\ \hline
        
        Equal weights Colless ($I_2$) index & 
        $ \displaystyle
       \begin{cases}
            \dfrac{4n(n-1)\big(H_n - H_{n/2} \big) - n}{6(n-1)(n-2)},  &  n \text{ even} \\
            \dfrac{4n(n-2)\big(H_{n-1} - H_{(n-1)/2} \big) - n + 3}{6(n-2)^2},  &  n \text{ odd}
        \end{cases} 
        $ 
        & Yes \\ \hline
        
        Quadratic Colless index & $n \big(n+1- 2H_n\big)$ & No \\ \hline
        
        Total $I$ index & 
        $\displaystyle
        \begin{cases}
            n\big(H_{n} - H_{n/2} - \frac{5}{12} \big), &  n \text{ even} \\
            n\big(H_{n-1} - H_{(n-1)/2} - \frac{5}{12}\big), &  n \text{ odd}
        \end{cases}
        $ 
        & Yes \\ \hline
        
        Total $I'$ index & 
        $ \displaystyle
        \frac{n-2}{4}
        $ 
        & Yes \\ \hline
        
        Rogers $J$ index & 
        $ \displaystyle
        \begin{cases}
            2n\big(H_{n/2 - 1} - H_{n-1} + 1\big), &  n \text{ even} \\
            2n\big(H_{(n-1)/2} - H_{n-1} + 1\big) - 2, &  n \text{ odd}
        \end{cases}
        $ 
        & Yes \\ \hline
        
        Rooted quartet index for binary trees  & $ \displaystyle \frac{1}{3}\binom{n}{4}$ & No\\ \hline
        
        Total internal path length & $2 - 4n + 2n H_n $ & Yes \\ \hline

        Total path length & $2 - 6n + 4n H_n$ & Yes \\ \hline
        
        Average vertex depth & $\displaystyle \dfrac{2 - 6n + 4n H_n}{2n - 1}$ & Yes \\ \hline
        
        stairs1 index& 
        $ \displaystyle
        \begin{cases}
            \frac{2n}{n-1}\big(H_{n/2 - 1} - H_{n-1} + 1\big), &  n \text{ even} \\
            \frac{2n}{n-1}\big(H_{(n-1)/2} - H_{n-1} + 1\big) - 2, &  n \text{ odd}
        \end{cases}
        $ 
        & Yes \\ \hline

        stairs2 index & 
        $
            \displaystyle
            \begin{cases}
                \dfrac{2n H^{(2)}_{n-1} - n H^{(2)}_{n/2} - 2\big(H_{n-1} - H_{n/2} \big) - (n-1)}{n-1}, & n \text{ even} \\
                \dfrac{2n H^{(2)}_{n-1} - n H^{(2)}_{(n-1)/2} - 2\big(H_{n-1} - H_{(n-1)/2} \big) - (n-1)}{n-1}, & n \text{ odd}
            \end{cases}
        $
        & Yes \\ \hline
    \end{tabularx}

    \caption{The expected value under the Yule model of the 17 tree balance indices that are represented by the clade-number basis or the daughter-number basis. ``No'' indicates that the quantity was previously known~\cite[Table 4.5 and Chapter 23]{Fischer2023}; in the case of the $\hat{s}$-shape statistic, the formula was unknown in~\cite{Fischer2023} but was reported later in~\cite[Proposition 10]{Cleary2025} (with a small error). ``Yes'' indicates that the results obtained through this work are new. The harmonic numbers $H_k = \sum_{i=1}^k \frac{1}{i}$ appear in many formulas, and the second-order harmonic numbers $H^{(2)}_k = \sum_{i=1}^k {1}/{i^2}$ are used for the stairs2 index.}
    \label{tab:new_mean_results}
\end{table}

\subsection{Exact results for the variance of balance indices and covariance between balance indices under the Yule model}

Of the 17 balance indices represented by the clade-number basis or the daughter-number basis, seven were listed by Fischer et al.~\cite{Fischer2023} as having a known variance under the Yule model. Using the results in Section~\ref{section 5}, we obtain the variances of all 17 indices. Moreover, the same calculation allows us to find the covariances between all pairs of these indices.  

Consider two statistics $A(T), B(T)$ that are represented by the clade-number basis with coefficients $a_n(r), b_n(r)$ for $2 \leq r \leq n$. Using eq.~\ref{covariance of statistics}, the covariance between $A(T)$ and $B(T)$ under the Yule model is
\begin{equation} \label{clade covariance eq}
    \textrm{Cov}\big[ A(T), B(T) \big] = \bigg[\sum_{r=2}^n a_n(r) \,  b_n(r) \, V(n, r) \bigg] +  \bigg[2 \sum_{r=2}^{n-1} \sum_{s = r + 1}^{n} a_n(r) \, b_n(s) \, C(n, r, s) \bigg],
\end{equation}
where $V(n,r)$ and $C(n,r,s)$ are the variance and covariance of the clade-number basis from Theorem~\ref{clade-number basis theorem}. Notice that eq.~\ref{clade covariance eq} also gives the variance by setting $A(T) = B(T)$. Although the sums in eq.~\ref{clade covariance eq} could in principle be simplified into more compact expressions for different statistics, due to the complicated equations for $V(n,r)$ and $C(n,r,s)$, we were unable to recover any closed-form expressions for the variances and covariances we studied. 

Similarly, consider two statistics $A(T), B(T)$ that are represented by the daughter-number basis with coefficients $a_n(i,j)$, $b_n(i,j)$ for $1 \leq i \leq \lfloor n/2 \rfloor$, $i \leq j \leq n - i$. Their covariance under the Yule model is
\begin{align} 
    \textrm{Cov}\big[ A(T), B(T) \big] =& \bigg[ \sum_{i=1}^{\lfloor n/2 \rfloor} \sum_{j=i}^{n-i} a_n(i,j) \, b_n(i,j) \, V_D(n,i,j) \bigg] \nonumber \\
    &+ \bigg[ 2 \sum_{i=1}^{\lfloor n/2 \rfloor} \sum_{j=i}^{n-i} \sum_{k = i}^{\lfloor n/2 \rfloor} \sum_{\ell = k + 1}^{n - k} a_n(i,j) \, b_n(k, \ell) \, C_D(n, i, j, k, \ell) \bigg], \label{daughter covariance eq}
\end{align}
where $V_D(n,i,j)$ and $C_d(n,i,j,k,\ell)$ are the variance and covariance of the daughter-number basis from Theorem~\ref{daughter-number basis theorem}. As with the clade-number basis, this equation gives the variance of a statistic $A(T)$ by setting $A(T) = B(T)$. 

\begin{figure}
    \centering
    \includegraphics[width=\linewidth]{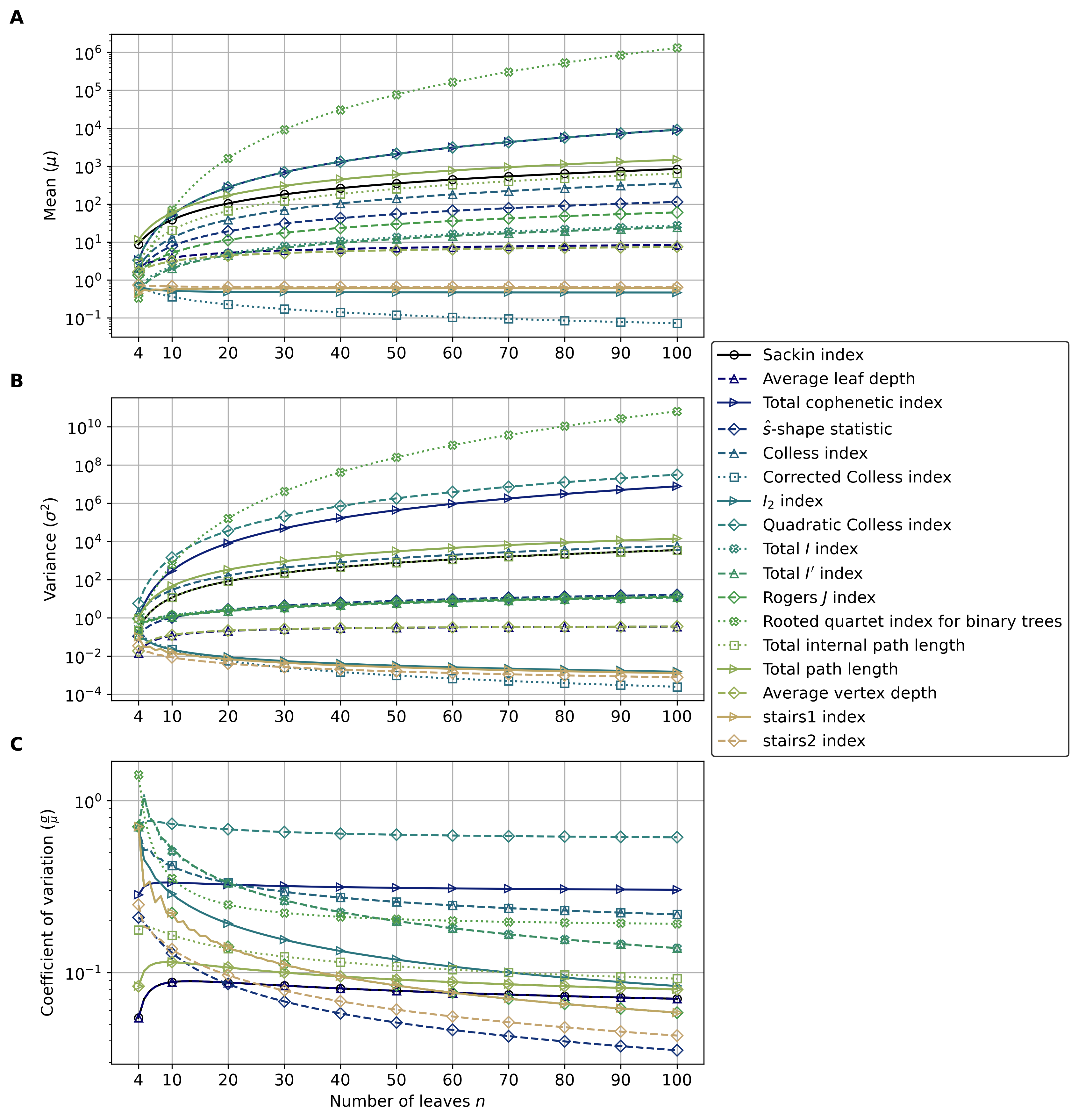}
    \caption{The mean, variance, and coefficient of variation under the Yule model of all 17 balance indices represented by the clade-number basis or the daughter-number basis for $4 \leq n \leq 100$. (A) The mean of each balance index was computed using the formulas in Table~\ref{tab:new_mean_results}. (B) The variance of each balance index was calculated by combining eq.~\ref{daughter covariance eq} with the specific coefficients of the daughter-number basis representation of a statistic found in Appendix~\ref{Appendix A}. (C) The coefficient of variation, $CV$. Low values $CV < 1$ correspond to distributions that are concentrated tightly around the mean, whereas high values $CV > 1$ correspond to high-variance distributions. }
    \label{fig:mean var and cv}
\end{figure}

We numerically evaluated eqs.~\ref{clade covariance eq} and eq.\ref{daughter covariance eq} to find the exact variances of all 17 balance indices represented by either basis and the covariances of all pairs. In Figure~\ref{fig:mean var and cv}b, we plot the variance of each index as a function of $n$, the number of leaves. Furthermore, we characterize the distribution of each index by calculating the coefficient of variation $CV = {\sqrt{\textrm{Var}\big(A(T)\big)}}/{\mathbb{E}[A(T)]}$ for each index in Figure~\ref{fig:mean var and cv}c. We observe from the figure that for large $n$, all balance indices studied have a coefficient of variation that is much smaller than $1$, indicating that the distribution of the balance index under the Yule model is tightly concentrated around the mean. 

\begin{figure}
    \centering
    \includegraphics[width=0.8\linewidth]{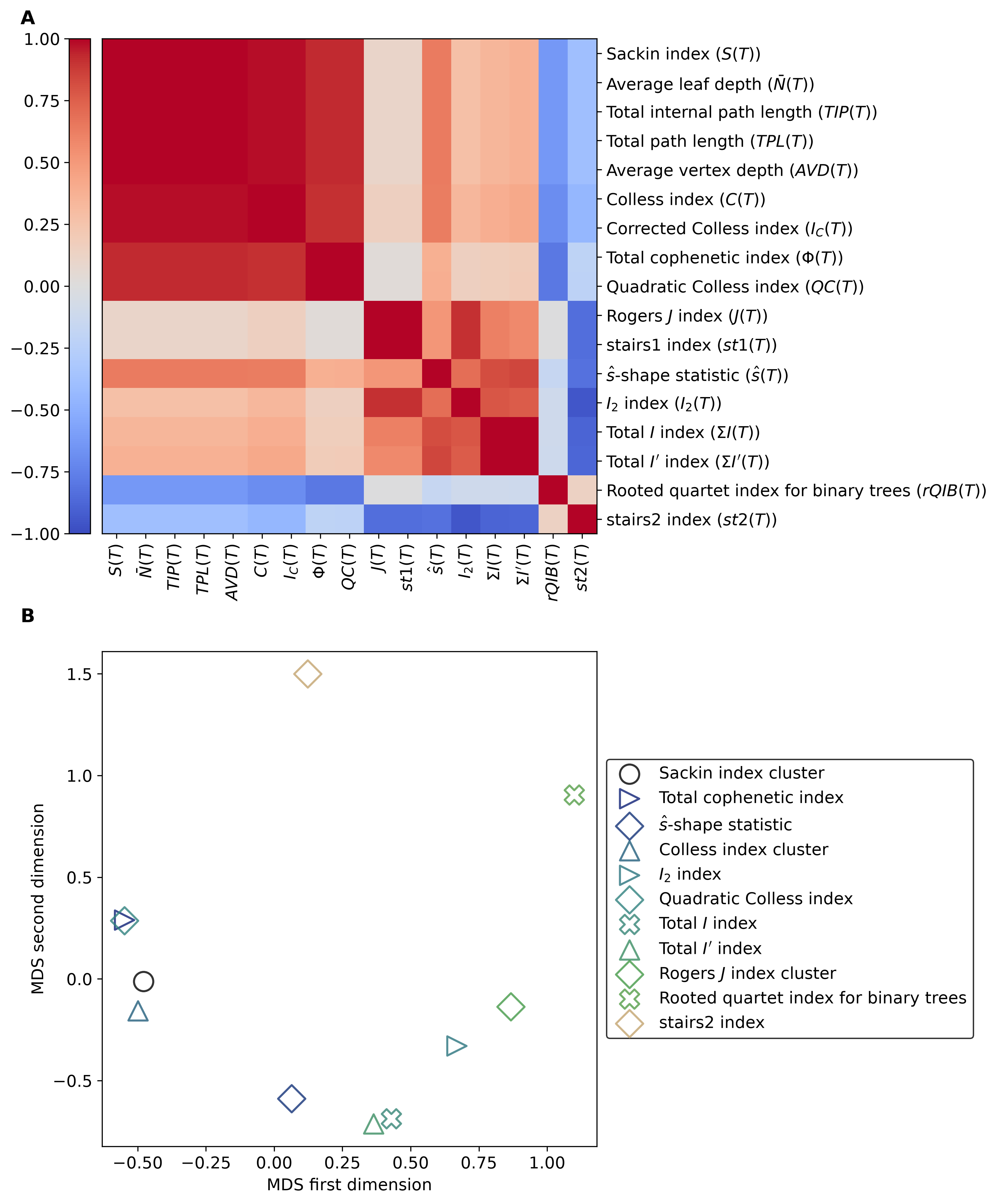}
    \caption{Correlations between pairs of balance statistics. (A) Correlation coefficients between all pairs among 17 balance indices represented by the daughter-number basis for $n = 100$, as computed from the ratio of the covariance (eq.~\ref{daughter covariance eq}) to the square root of the products of the variances (eq.~\ref{daughter covariance eq}). (B) Metric multidimensional scaling plot for the set of 17 indices represented by the daughter-number basis, treating the correlation matrix in panel (A) as a similarity matrix $S$, with accompanying distance matrix $D$ found by the transformation $d_{i,j} = \sqrt{2(1-s_{i,j})}$~\cite[eq. 14.2.14]{Mardia1979}. Several sets of indices are perfectly correlated, causing their associated points in this panel to be coincident. We indicate this overlap by plotting representative points for the Sackin index cluster (Sackin index, average leaf depth, total internal path length, total path length, and average vertex depth), Colless index cluster (Colless index and corrected Colless index), and Rogers $J$ index cluster (Rogers $J$ index and stairs1 index). The total cophenetic index $\Phi(T)$ and quadratic Colless index $QC(T)$ have points that appear incident in the plot, but their correlation is not exactly 1.}
    \label{fig:correlation matrix and mds}
\end{figure}

In Figure~\ref{fig:correlation matrix and mds}, we study correlations between balance indices. We calculated the exact correlation between each pair of indices at the fixed value $n = 100$, and we plot this matrix in Figure~\ref{fig:correlation matrix and mds}A. Additionally, we conducted metric multidimensional scaling (MDS) on the correlation matrix to perform a low-dimensional clustering of indices~\cite[Chapter 14]{Mardia1979}. Due to the fact that some indices are rescaled or shifted versions of other indices, we reduced the number of indices considered to a single representative index in each group; hence, Figure~\ref{fig:correlation matrix and mds}B considers only 11 indices. 

We observe several highly correlated clusters in Figure~\ref{fig:correlation matrix and mds}B. As observed previously, the Sackin index and Colless index are highly correlated, with correlation coefficient $\approx 0.98$~\cite[Theorem 2]{Blum2006Mean}. Also clustered with these indices are the quadratic Colless index and total cophenetic index. Next, a group of indices consisting of the $\hat{s}$-shape statistic, total $I$ index, total $I'$ index, $I_2$ index, and Rogers $J$ index form a cluster. Two remaining indices are not highly correlated with any of the other indices: the rooted quartet index for binary trees and the stairs2 index. 

\clearpage

\section{Discussion and future directions} \label{section 7}

In this paper, we have proposed a framework to study tree balance indices using sets of basis statistics. We provided examples of two bases, the clade-number basis and the daughter-number basis, that are able to represent many existing tree balance indices. We calculated the expected value and covariance for elements of both bases under the Yule model, and we used these results to find formulas for the expected value and variance of 17 tree balance indices under the Yule model, and formulas for the covariances of pairs of indices. 

\subsection{Connections to prior work} \label{sec:7_1}
Tree balance indices have often been studied individually; examples of recent studies include~\cite{Mir2013, Coronado2019, Coronado2020, Bartoszek2021, Kersting2021, King2021, Bienvenu2021, Goh2022, Currie2024}. However, several authors have attempted to study tree balance in a more unified way that simultaneously considers multiple statistics designed to measure balance. Aldous proposed a method to study tree balance based on performing median regression on the scatterplot of pairs $(\log{m}, \log{i})$ of a tree, where $m$ is the size of a subtree and $i$ is the size of that subtree's smaller daughter node~\cite{Aldous2001}. In this method, the slope of the regression line corresponds to a notion of the balance of the tree. We observe that this regression can be computed exactly with the knowledge of the tree's daughter numbers, and therefore can be interpreted as a nonlinear map from the daughter-number basis to a regression slope $\beta$ and intercept $\hat{y}$. 

Matsen's geometric approach to tree shape proposes embedding the space of trees $\mathcal{T}_n$ in a lower-dimensional subspace by using multidimensional scaling with respect to a particular choice of metric over $\mathcal{T}_n$~\cite{Matsen2006}. This approach studied statistics that can best differentiate trees that are furthest apart according to the metric. Matsen also used linear algebra techniques to find sets of statistics that give complementary information. Similarly, our basis statistics concept embeds the space of trees in a lower-dimensional subspace. Unlike an embedding from multidimensional scaling, the embeddings provided by the clade-number basis and daughter-number basis are interpretable in relation to relatively simple properties of trees.

In a subsequent paper, Matsen also considered classes of statistics he called \textit{binary recursive tree shape statistics} that provide a general framework for defining many existing tree shape statistics, including balance indices~\cite{Matsen2007}. Our basis statistics approach relies on characterizing tree statistics in terms of relatively simple explicit features of tree shapes rather than a recursive formulation. 

Recently, Fischer et al.~\cite{Fischer2025} have introduced an approach for generalizing tree balance indices by defining families of statistics using what they call balance ``metaconcepts''. This framework, which is similar to our basis statistics approach in considering functions computed at the level of tree vertices---and even examines functions that amount to considering clade sizes---has been valuable for identifying the trees that minimize and maximize tree-based statistics~\cite{Cleary2025, Fischer2025, Fischer2027}. Our focus here has been on probabilistic results, but some potential may exist to combine the metaconcept and basis statistics frameworks to obtain additional mathematical results on tree balance statistics.

Many of the results in Section \ref{section 4} have analogs in studies of binary search trees, by a correspondence between models of random binary search trees and the Yule--Harding model~\cite{Fuchs2025}. The covariance matrix of the clade number basis has been previously obtained in the literature on binary search trees. Indeed, Dennert et al.~calculated the covariance between $C_r(T)$ and $C_s(T)$ when $r + s < n$~\cite[Proposition 2.1]{Dennert2010}, along with a corresponding limiting law for the joint distribution of the clade numbers of small subtree sizes, and Holmgren \& Janson obtained expressions for the full covariance matrix~\cite[Lemma 3.3]{Holmgren2015}, along with a limiting law for functions of the subtrees. It may be possible to extend the results of Holmgren \& Janson to find the covariance matrix of the daughter-number basis, and to prove limiting results for statistics that are represented by the clade-number or daughter-number basis.

Finally, several authors have examined properties of the clade numbers of a Yule-generated tree beyond the mean and variance, including distributional properties of the clade numbers. This work includes asymptotic results for the distributions of the numbers of clades of different sizes under the Yule model~\cite{McKenzie2000, Blum2005, Chang2009}, and numerous results on the joint distribution of the number of clades of size 2 (cherries) and 3 (pitchforks)~\cite{Disanto2013, Wu2016, Choi2020, Choi2021}. Extending this work to explore the joint distribution of clades of larger sizes could provide stronger results about tree balance indices than the results we obtained here using the mean, variance, and covariance of the clade numbers.

\subsection{Further applications of the basis statistics approach} \label{sec:7_2}

\subsubsection{Interpreting how tree balance indices measure balance} \label{sec:7_2_1}

Many tree shape properties have been used to define balance indices,. These include the leaf depth, subtree size, node balance, rank, height, and width of trees. For statistics that are represented by the clade-number basis or the daughter-number basis (Table~\ref{tab:index_representations}), a representation identifies how a particular balance index measures nodes of a given size or daughter split. This perspective allows direct comparisons between the ways that different statistics measure these tree properties. 

The 12 balance indices we studied that are not represented by either basis (Table~\ref{tab:index_representations}) are primarily based on the rank, height, or width, or on nonlinear functions of leaf depths of trees. Very few properties of these statistics, such as their behavior under the Yule model, are known~\cite{Fischer2023}. This fact indicates that they are more challenging to mathematically characterize. At the same time, they capture novel information about tree shapes that is not contained in either the clade-number basis or the daughter-number basis. 

In Appendix~\ref{Appendix C}, we use a computational approach to find approximate representations of these balance indices in our two bases. This analysis produces several results. First, the residual error of the approximation provides some measure of how close each statistic is to being represented by the basis (Figures~\ref{fig:clade projection}, \ref{fig:daughter projection}), with the interpretation that the error in the approximation can serve as a proxy for the amount of information about tree shape encoded in a statistic and not captured by the basis. Second, the coefficients of the approximate representation show how important each feature in the basis is to that statistic (Figure~\ref{fig:coefficients}). Therefore, using this computational approach can allow insights even into indices that are not exactly represented by the bases. 

\subsubsection{Defining new summary statistics} \label{sec:7_2_2}

The clade-number basis and daughter-number basis can be used to define new summary statistics of trees. Any coefficient vector $\mathbf{a}$ with the same length as the basis $\mathcal{B}(T)$ defines a summary statistic $A(T) := \mathbf{a}^t \mathcal{B}(T)$. Therefore, each basis is associated with an entire class of statistics corresponding to different coefficient vectors.

We can use this class and the properties of the basis to find new summary statistics that can be designed for particular tasks. As one example, we can find the statistic that, among statistics represented by the clade-number basis, is maximal for the Yule model in expectation among trees of size $n$. More precisely, for some $n$, consider the clade-number basis $\mathcal{C}(T) \in \mathbb{R}^{n-1}$ and the set of all coefficient vectors $\mathbf{a} \in \mathbb{R}^{n-1}$. Let $m_{\mathcal{C}}$ be the maximal statistic,
\begin{align}
    m_{\mathcal{C}} &= \arg\max_{ \substack{\mathbf{a} \in \mathbb{R}^{n-1} \\ \|a\|_2 = 1} } \mathbb{E} \big[ \mathbf{a}^t \mathcal{C}(T) \big] \nonumber \\
    &= \arg\max_{ \substack{\mathbf{a} \in \mathbb{R}^{n-1} \\ \|a\|_2 = 1}} \mathbf{a}^t \mathbb{E}\big[\mathcal{C}(T) \big] \nonumber \\
    &= \arg\max_{ \substack{\mathbf{a} \in \mathbb{R}^{n-1} \\ \|a\|_2 = 1}} \mathbf{a}^t \boldsymbol{\mu}_{\mathcal{C}}. \label{optimal problem}
\end{align}
Here, $\boldsymbol{\mu}_{\mathcal{C}}= \mathbb{E}[\mathcal{C}(T)]$ is the expected value of the clade-number basis under the Yule model. We take the maximum over all coefficient vectors of a given magnitude. By elementary linear algebra, eq.~\ref{optimal problem} is solved by choosing $\mathbf{a} \propto \boldsymbol{\mu}_{\mathcal{C}}$, or $m_{\mathcal{C}} = \boldsymbol{\mu}_{\mathcal{C}} / \|\boldsymbol{\mu}_{\mathcal{C}}\|_2$. Because we can multiply this coefficient vector by any scalar and recover a statistic with the same ability to distinguish trees, we can ignore all normalization constants. Additionally, because $C_n(T) = 1$ for all trees, we can set the coefficient of $C_n(T)$ to 0 without loss of generality. Then using Theorem~\ref{clade-number basis theorem}, we can write this new statistic as
\begin{equation} \label{maximal yule clade}
    M_{\mathcal{C}}(T) = \sum_{k=2}^{n-1} \frac{1}{k(k+1)} C_k(T).
\end{equation}

This statistic is explicitly designed for a hypothesis test of whether a tree is consistent with the Yule model, similar in spirit to the $D$-index proposed by Blum and Fran\c{c}ois~\cite{Blum2005} and Matsen's effort to computationally choose a statistic from a class of statistics that is optimal for a particular application~\cite{Matsen2007}. Because it has been designed to be maximal in expectation for trees generated under the Yule model, a hypothesis test based on this statistic is naturally one-sided, because lower values would correspond to rejecting the test; it is not obvious whether a one-sided test or two-sided test should be used when many balance indices are selected as test statistics. The methods of Section~\ref{section 6} immediately apply to recover the expectation and variance of $M_{\mathcal{C}}(T)$ under the Yule model. Analogous summary statistics that come from the daughter-number basis or are designed to test other probability models of trees could be defined similarly. 

\subsubsection{Defining a distance between a tree and the Yule model} \label{sec:7_2_3}

We studied how the clade-number basis $\mathcal{C}(T)$ and daughter-number basis $\mathcal{D}(T)$ behave under the Yule model by explicitly computing the mean and covariance matrices $\boldsymbol{\mu}_{\mathcal{C}}$, $\mathbf{\Sigma}_{\mathcal{C}}$, $\boldsymbol{\mu}_{\mathcal{D}}$, $\mathbf{\Sigma}_{\mathcal{D}}$ of these two bases. Although the focus of this paper is to use this information to study balance indices, these results may have further applications. In particular, the two bases provide simple low-dimensional embeddings of the tree space $\mathcal{T}_n \mapsto \mathbb{R}^k$, such that the mean and covariance of the image of the embedding in $\mathbb{R}^k$ have been found for the Yule model. Therefore, one can study the distribution of trees $\mathcal{T}_n$ under the Yule model by directly studying the image of trees under the basis statistics mapping, utilizing techniques from multivariate statistical analysis. 

We propose that these bases can be used to define distances from a tree $T \in \mathcal{T}_n$ to the distribution $\mathcal{Y}_n$ of trees under the Yule model, by calculating the Mahalanobis distance of the image of $T$ under the basis $\mathcal{B}(T)$ to the distribution of the basis. The Mahalanobis distance between a vector $\mathbf{x} \in \mathbb{R}^n$ and a multivariate distribution $\mathcal{Q}$ on $\mathbb{R}^n$ with mean $\boldsymbol{\mu} \in \mathbb{R}^n$ and covariance matrix $\mathbf{\Sigma} \in \mathbb{R}^{n \times n}$ is defined as
\begin{equation} \label{mahalanobis}
    d(\mathbf{x}, \mathcal{Q}) = \sqrt{\big(\mathbf{x} - \boldsymbol{\mu}\big)^t \, \mathbf{\Sigma}^{-1} \,  \big(\mathbf{x} - \boldsymbol{\mu}\big)}. 
\end{equation}
The Mahalanobis distance has several properties that are particularly useful. If $\mathbf{x}$ is randomly sampled from the distribution $\mathcal{Q}$, and $\mathcal{Q} \sim \mathcal{N}(\mu, \Sigma)$ is a multivariate normal distribution, then $d(\mathbf{x}, \mathcal{Q})^2$ follows a $\chi^2$ distribution with $n$ degrees of freedom~\cite{Ghorbani2019}. This property, widely used in outlier detection, enables the distance to provide a parametric test for whether $\mathbf{x}$ follows the distribution $\mathcal{Q}$~\cite{Ghorbani2019}. Moreover, if the covariance matrix is not invertible because the underlying dimension of the distribution $\mathcal{Q}$ is less than $n$, then the distance can be defined by replacing $\Sigma^{-1}$ in eq.~\ref{mahalanobis} by the pseudoinverse $\Sigma^{\dagger}$ of the covariance  (cite). 

In particular, we can use the Mahalanobis distance to define a distance between a tree and the Yule model as
\begin{equation} \label{tree mahalanobis}
    D(T, \mathcal{Y}_n) := D\big(\mathcal{B}(T), \mathcal{B}(\mathcal{Y}_n)\big) = \sqrt{\big(\mathcal{B}(T) - \boldsymbol{\mu}_{\mathcal{B}}\big)^t \, \mathbf{\Sigma_B}^{\dagger} \, \big(\mathcal{B}(T) - \mathbf{\mu_B}\big)}.
\end{equation}
Note that the fact that the clade-number basis and daughter-number basis have underlying dimension less than the number of basis elements (Section~\ref{sec:3_5}) implies that we must take the pseudoinverse in eq.~\ref{tree mahalanobis}. 

Using this distance, we can reinterpret the classic problem of testing whether a phylogenetic tree is plausibly generated under the Yule model by restating it as an outlier detection problem~\cite{Ghorbani2019}. This approach has several features that may make it particularly useful. First, it is a naturally one-sided test because a higher value indicates that the tree is more likely to be an outlier under the Yule model (see also Blum \& Fran\c{c}ois~\cite{Blum2005}). Second, results by Dennert and Gr{\"u}bel~\cite[Theorem 2.2]{Dennert2010} show that as $n \to \infty$, the joint distribution of the number of small subtrees with at most $m$ leaves, for fixed $m$, converges to a multivariate Gaussian. Therefore, if the distance in eq.~\ref{tree mahalanobis} is calculated using only basis elements corresponding to small subtrees of a tree, there is promise that the Mahalanobis distance will resemble a $\chi^2$ distribution and allow for parametric hypothesis testing. 

\subsubsection{Extensions to other phylogenetic models} \label{sec:7_2_4}
One of the limitations of this work is that we studied the stochastic properties of the two bases only under the Yule model, despite the fact that many empirical trees have been observed to be more imbalanced than the Yule model (e.g.~\cite{Blum2006Random, Guyer1991, Mooers1997}). Understanding our basis statistics under alternative evolutionary models could help better quantify the shapes of trees these models generate.  

While we established the mean and covariance of the clade-number basis and the daughter-number basis for trees under the Yule model, the mean and covariance could in principle be similarly found for any other probability distribution on the space of trees $\mathcal{T}_n$. For another distribution, if results analogous to Theorem~\ref{clade-number basis theorem} and Theorem~\ref{daughter-number basis theorem} were obtained, then corresponding results to Section~\ref{section 6} regarding means, variances, and covariances would follow as a corollary. Therefore, calculating the mean and covariance of the two bases in various phylogenetic models will greatly extend knowledge of how phylogenetic tree balance indices behave in other models beyond the Yule and next-most common model, the Uniform model. Our recent progress in this direction, in a study in which we discovered a formula for the expected value of the Sackin index in a class of phylogenetic models known as biased speciation models~\cite{Bauman2026}, could be extended to focus on studying the clade-number basis and daughter-number basis in this class of models. 

\subsubsection{Alternative choices of bases} \label{sec:7_2_5}

The two bases we have defined, the clade-number basis and the daughter-number basis, were chosen because they are intuitive and low-dimensional, and because they represent many tree shape statistics. However, this concept of a basis could be applied to statistics that measure other features of tree shape. A different basis could be designed to represent some of the balance indices that are not possible to represent by the clade-number basis or the daughter-number basis. 

As a first example, define the ``height-number basis'' to be the set of statistics $H_{m,k}(T)$ for the number of subtrees of size $m$ with height $k$, where the \textit{height} of a subtree is defined as the maximum depth of a vertex ($2 \leq m \leq n$, $\lceil \log_2{m}\rceil  \leq k \leq m-1$). Several statistics such as the $B_1$ index and maximum depth can be represented by the height-number basis. However, the possibility of obtaining exact results comparable to Theorem~\ref{clade-number basis theorem} might be limited, because properties such as the expected height of a random tree are known only in a limiting sense~\cite{Devroye1986, Reed2003, Drmota2001, Drmota2003}. 

Second, let the ``leaf-depth basis'' be the set of statistics $L_k(T)$ that count the number of leaves in $T$ with a given depth $k$ to the root ($1 \leq k \leq n-1$). This basis could potentially represent balance indices that are defined in terms of the depths of leaves, such as the $B_2$ index or the variance of leaf depths.

Finally, consider the ``tree-number basis'' defined as follows. For a tree $S$, let $Q_{S}(T)$ count the number of subtrees of $T$ that are equal to $S$. The mean and variance of these quantities $Q_S(T)$ have been previously discovered by Rosenberg~\cite[Theorem 5.1]{Rosenberg2006}, and the covariance of this basis may be possible to establish by using similar ideas to the proofs of Theorems~\ref{clade-number basis theorem} and \ref{daughter-number basis theorem} from Section~\ref{section 5}. The tree-number basis can trivially represent any tree shape statistic $A(T)$ by setting the coefficient of $Q_S(T)$ equal to $A(S)$ for all trees $S \in \mathcal{T}_n$, and setting the coefficient of $Q_S(T)$ to 0 for all $S \in \mathcal{T}$ with $|S| < n$. Beyond this trivial representation, it would be interesting to explore what different representations are possible. Although this basis offers no dimensionality reduction to the space of trees, it may serve as a useful theoretical tool to understand the relationship between different tree shape statistics. 

\subsubsection{Connections to similar approaches in population genetics} \label{sec:7_2_6}

Just as many tree balance indices have been proposed to measure phylogenetic trees and test trees against the Yule model, numerous population-genetic statistics are designed to summarize population-genetic samples and to test the neutral coalescent model of population genetics~\cite{Wakeley2009, Hahn2019, Rosenberg2025Mathematical}. Achaz recognized that many population-genetic statistics can be represented as a linear function of the ``site-frequency spectrum'' $\xi$, a vector with entry $\xi_i$ that counts the number of polymorphic sites that appear in exactly $i$ individuals among a genetic sample of $n$ individuals $(1 \leq i \leq n-1)$~\cite{Achaz2009}. This fundamental observation created a unified framework for understanding the mathematical properties of these statistics, suggesting designs of new statistics for novel applications~\cite{Achaz2009, Ferretti2010, Ronen2013}.

We view our work as performing a similar role in the decades-long history of tree balance computations under the Yule model as that of the work of Achaz for the decades of site-frequency statistics under the neutral coalescent. First, the frameworks are  similar in that they first decompose a complex dataset into a low-dimensional representation: the set of sequences is reduced to its site-frequency spectrum in the population-genetic context, and a phylogenetic tree is reduced to a basis of statistics in the phylogenetic context. Second, the summarized representations are valuable because a broad variety of statistical approaches to examine the data are simple linear combinations of the elements of the low-dimensional representations. This fact enables mathematical results for an entire class of statistics to be established in both cases. Third, examining the class of statistics that are linear combinations of these low dimensional spaces can help generate novel statistics. We have proposed two potential approaches for doing so in Sections~\ref{sec:7_2_2} and \ref{sec:7_2_3}.

The analogy suggests that the clade numbers or daughter numbers of a phylogenetic tree can be regarded as a natural ``site-frequency spectrum'' for summarizing and interpreting phylogenetic trees. The analogy, together with extensive analysis of site-frequency statistics under a variety of population-genetic models, may provide a new mechanism for obtaining deeper understanding in mathematical phylogenetics. 

\bigskip
\noindent {\bf Funding.} We acknowledge support from NIH grant R01 HG005855 and NSF grant DMS-2450005.

\bigskip
\noindent {\bf Data availability statement.} The paper has no associated data. A public Github repository containing code to perform all calculations in this paper and reproduce all figures can be found at \url{https://github.com/dbauman18/phylogenetic_basis_statistics}.

\bigskip
\noindent {\bf Ethical statement.} The authors have no relevant financial or non-financial interests to disclose.
 
\printbibliography

\newpage
\appendix
\counterwithin{table}{section}
\counterwithin{figure}{section}
\renewcommand{\thetable}
{\thesection\arabic{table}}
\renewcommand{\thefigure}
{\thesection\arabic{figure}}

\input{Appendix_A}
\clearpage
\input{Appendix_B}

\clearpage
\input{Appendix_C}

\clearpage
\input{Appendix_D}

\end{document}

\typeout{get arXiv to do 4 passes: Label(s) may have changed. Rerun}

%% file: Appendix_A.tex
\section{Representations of specific balance indices} \label{Appendix A}

In this section, we study tree balance indices to discover representations of them by the clade-number basis or the daughter-number basis, should such a representation exist. We describe the tree balance indices in the order listed in Fischer et al.~\cite{Fischer2023}. To find representations for specific balance indices, most of the results are immediate consequences of applying the following three propositions to the definition of that index. To clarify notation, $n_v$ refers to the number of leaves in a subtree induced by $v$, while the appearance of $n$ as a subscript, such as $\mathcal{T}_n$, indicates that $n$ is the size of the tree. 


\begin{propositionA} \label{clade basis representations}
    Let $A(T)$ be a summary statistic of a tree $T \in \mathcal{T}_n$ such that $A(T) = \sum_{v \in \inter} a_n(n_v)$ for a function $a_n: \mathbb{N} \to \mathbb{R}$, where $n_v$ is the size of the subtree of vertex $v$. Then $A(T)$ can be represented by the clade-number basis as 
    \begin{equation}
        A(T) = \sum_{r=2}^n a_n(r) \, C_r(T). \nonumber
    \end{equation}
\end{propositionA}

\begin{proof}
    This result follows immediately from grouping the sum into internal vertices with identical sizes $n_v$:
    \begin{align}
        A(T) = \sum_{v \in \inter} a_n(n_v) &= \sum_{r=2}^n \sum_{\substack{v \in \inter \,| \\ n_v = r}} a_n(n_v) \nonumber \\
        &= \sum_{r=2}^n a_n(r) \sum_{\substack{v \in \inter \,| \\ n_v = r }} 1 \nonumber \\
        &= \sum_{r=2}^n a_n(r) \, C_r(T). \nonumber
    \end{align}
\end{proof}

Similarly, we can use the following proposition to prove that a balance index of a particular form is represented by the daughter-number basis. 

\begin{propositionA} \label{daughter basis representations}
    Let $B(T)$ be a summary statistic of a tree $T \in \mathcal{T}_n$ such that $B(T) = \sum_{v \in \inter} b_n(n_{v_L}, n_{v_R})$ for a function $b_n: \mathbb{N} \times \mathbb{N} \to \mathbb{R}$, where $n_{v_L} \leq n_{v_R}$ are the sizes of the left and right subtrees of $v$. Then $B(T)$ can be represented by the daughter-number basis as 
    \begin{equation}
    B(T) = \sum_{i = 1}^{\lfloor n/2 \rfloor} \sum_{j = i}^{n-i} b_n(i, j) \, D_{i, j}(T). \nonumber
\end{equation}
\end{propositionA}

\begin{proof}
    This result follows immediately from grouping the sum into internal vertices with identical daughter sizes $(n_{v_L}, n_{v_R})$:
    \begin{align}
        B(T) = \sum_{v \in \inter} b_n(n_{v_L}, n_{v_R}) &= \sum_{i = 1}^{\lfloor n/2 \rfloor} \sum_{j = i}^{n-i} \sum_{ \substack{v \in \inter \,| \\ (n_{v_L}, n_{v_R}) = (i,j)}} b_n(n_{v_L}, n_{v_R}) \nonumber \\
        &= \sum_{i = 1}^{\lfloor n/2 \rfloor} \sum_{j = i}^{n-i} b_n(i, j) \sum_{ \substack{ v \in \inter \,| \\ (n_{v_L}, n_{v_R}) = (i,j) }} 1 \nonumber \\
        &= \sum_{i = 1}^{\lfloor n/2 \rfloor} \sum_{j = i}^{n-i} b_n(i, j) \, D_{i, j}(T). \nonumber
    \end{align}
\end{proof}

Finally, we show that, as observed in Section~\ref{sec:3_4}, any index that is represented by the clade-number basis can also be represented by the daughter-number basis. 

\begin{propositionA} \label{clade to daughter}
    Let $A(T)$ be a summary statistic of a tree $T \in \mathcal{T}_n$ that is represented by the clade-number basis as $A(T) = \sum_{r=2}^n a_n(r) \, C_r(T)$. Then $A(T)$ can be represented by the daughter-number basis as
    \begin{equation}
        A(T) = \sum_{i = 1}^{\lfloor n/2 \rfloor} \sum_{j = i}^{n-i} a_n(i + j) \, D_{i,j}(T). \nonumber
    \end{equation}
\end{propositionA}

\begin{proof}
    Notice that we can write $A(T) = \sum_{v \in \inter} a_n(n_v)$. Now define $b_n(i, j) := a_n(i + j)$ for $(i, j) \in \mathbb{N} \times \mathbb{N}$. Because $n_v = n_{v_L} + n_{v_R}$, we have that $A(T) = \sum_{v \in \inter} b_n(n_{v_L}, n_{v_R})$. By Proposition~\ref{daughter basis representations}, $A(T)$ can be represented in the daughter-number basis by $A(T) = \sum_{i = 1}^{\lfloor n/2 \rfloor} \sum_{j = i}^{n-i} b_n(i, j) \, D_{i,j}(T)$. Substituting $a_n(i+j)$ for $b_n(i,j)$ completes the proof. 
\end{proof}

We now turn to studying basis representations for the following list of tree balance indices. For each index, we state whether a representation in each basis exists. If a representation exists, we find a representation as an application of Propositions~\ref{clade basis representations}, \ref{daughter basis representations}, and \ref{clade to daughter}. If it does not exist, we state this fact, and prove the statement in Appendix~\ref{Appendix B} using the methods outlined in Section~\ref{sec:3_6}. 

Table~\ref{tab:appendix notation} lists and describes the terms necessary to define the indices considered, following the notation used by Fischer et al.~\cite{Fischer2023}.

\begin{table}[tb]
    \centering
    \renewcommand{\arraystretch}{1.4}
    \begin{tabular}{|c|c|}
    \hline
        \textbf{Notation} & \textbf{Meaning} \\
    \hline
        $V(T)$ & Set of vertices \\
    \hline
        $\inter$ & Set of interior vertices \\
    \hline
        $V_L(T)$ & Set of leaves \\
    \hline
        $\rho$ & Root of the tree \\
    \hline
        $T_v$ & Subtree induced by $v$ \\
    \hline
        $n_v$ & Size of subtree induced by $v$ \\
    \hline
        $v_L, v_R$ & Daughters of $v$ (for $v \in \inter$; $n_{v_L} \leq n_{v_R}$) \\
    \hline
        $C_r(T)$ & Number of vertices with subtrees of size $r$, $2 \leq r \leq n$ \\
    \hline
        $D_{i,j}(T)$ & Number of vertices with daughter subtrees of sizes $(i,j)$, $1 \leq i \leq \lfloor \frac{n}{2} \rfloor$, $i \leq j \leq n-i$ \\
    \hline
        $\delta_T(v)$ & Depth of vertex $v$ (path length between root and $v$) \\
    \hline
        $h(T)$ & Height of tree (maximum depth of a leaf in $T$) \\
    \hline
        $s(T)$ & Number of symmetric nodes (nodes $v \in \inter$ with $T_{v_{L}} = T_{v_R}$) \\
    \hline
        $w_T(i)$ & Width of tree at level $i$, $0 \leq i \leq h(T)$ (number of vertices $v$ of depth $i$) \\
    \hline
    \end{tabular}
    \caption{Necessary notation to define the tree balance indices studied in Appendix~\ref{Appendix A}, following the conventions of Fischer et al.~\cite{Fischer2023}. }
    \label{tab:appendix notation}
\end{table}

\subsection{Sackin index} \label{sackin}
The Sackin index $S(T)$ is defined as~\cite[Chapter 5]{Fischer2023}
\begin{equation}
    S(T) = \sum_{v \in V_L(T)} \delta_T(v) = \sum_{v \in \inter} n_v. \nonumber
\end{equation}
The Sackin index can be represented by the clade-number basis as 
\begin{equation}
    S(T) = \sum_{r=2}^n r \, C_r(T). \nonumber
\end{equation}
It can be represented by the daughter-number basis as
\begin{equation}
    S(T) = \sum_{i = 1}^{\lfloor n/2 \rfloor} \sum_{j = i}^{n-i} (i + j) \, D_{i, j}(T). \nonumber
\end{equation}

\subsection{Average leaf depth} \label{avg leaf depth}
The average leaf depth is defined as~\cite[Chapter 6]{Fischer2023}
\begin{equation}
    \bar{N}(T) =\frac{1}{n} \sum_{v \in V_L(T)} \delta_T(v) =  \frac{1}{n} S(T). \nonumber
\end{equation}
The average leaf depth can be represented by the clade-number basis as
\begin{equation}
    \bar{N}(T) = \sum_{r=2}^n \frac{r}{n} \, C_r(T) \nonumber. 
\end{equation}
It can be represented by the daughter-number basis as
\begin{equation} \nonumber
    \bar{N}(T) = \sum_{i = 1}^{\lfloor n/2 \rfloor} \sum_{j = i}^{n-i} \frac{i + j}{n}\, D_{i, j}(T).
\end{equation}

\subsection{Variance of leaf depths} \label{var leaf depths}
The variance of leaf depths is defined as~\cite[Chapter 7]{Fischer2023}
\begin{equation}
    \sigma^2_N(T) = \frac{1}{n} \sum_{v \in V_L(T)}\big[ \delta_T(v) - \bar{N}(T) \big]^2. \nonumber
\end{equation}
The variance of leaf depths cannot be represented by the clade-number basis or the daughter-number basis; see Appendix~\ref{B.2}. 

\subsection{Total cophenetic index} \label{total cophenetic}
The total cophenetic index is defined as~\cite[Chapter 8]{Fischer2023}
\begin{equation}
    \Phi(T) = \sum_{v \in \inter \setminus \{\rho\}} \binom{n_v}{2}, \nonumber
\end{equation}
where $\rho$ is the root of the tree. The total cophenetic index can be represented by the clade-number basis as 
\begin{equation}
    \Phi(T) = \sum_{r=2}^{n-1} \binom{r}{2} \, C_r(T). \nonumber
\end{equation}
It can be represented by the daughter-number basis as
\begin{equation}
    \Phi(T) = \sum_{i = 1}^{\lfloor n/2 \rfloor} \sum_{j = i}^{n-i} b_n(i, j) \, D_{i, j}(T), \nonumber
\end{equation}
where 
$$
b_n(i,j) = \begin{cases}
    \binom{i+j}{2} & \text{if } i + j < n, \\
    0 & \text{if } i + j = n.
\end{cases}
$$

\subsection{\texorpdfstring{$\hat{s}$-shape statistic}{s-shape statistic}} \label{s-shape}
The $\hat{s}$-shape statistic is defined as~\cite[Chapter 9]{Fischer2023}
\begin{equation}
    \hat{s}(T) = \sum_{v \in \inter} \log(n_v - 1). \nonumber
\end{equation}
The $\hat{s}$-shape statistic can be represented by the clade-number basis as 
\begin{equation}
    \hat{s}(T) = \sum_{r=2}^n \log(r-1) \, C_r(T). \nonumber
\end{equation}
It can be represented by the daughter-number basis as
\begin{equation}
    \hat{s}(T) = \sum_{i = 1}^{\lfloor n/2 \rfloor} \sum_{j = i}^{n-i} \log(i + j - 1)  \, D_{i, j}(T). \nonumber
\end{equation}

\subsection{\texorpdfstring{$B_1$ index}{B1 index}} \label{b1}
The $B_1$ index is defined as~\cite[Chapter 10]{Fischer2023}
\begin{equation}
    B_1(T) = \sum_{v \in \inter \setminus \{\rho\}} \frac{1}{h(T_v)}. \nonumber
\end{equation}
The $B_1$ index cannot be represented by the clade-number basis or the daughter-number basis; see Appendix~\ref{B.2}. 

\subsection{\texorpdfstring{$B_2$ index}{B2 index}} \label{b2}

For a rooted binary tree, the $B_2$ index is defined as~\cite[Chapter 11]{Fischer2023}
\begin{equation}
    B_2(T) = \sum_{v \in V_L(T)} \delta_L(v) \left(\frac{1}{2}\right)^{\delta_L(v)}. \nonumber
\end{equation}
The $B_2$ index cannot be represented by the clade-number basis or the daughter-number basis; see Appendix~\ref{B.2}.

\subsection{Colless index} \label{colless}
The Colless index is defined as~\cite[Chapter 12]{Fischer2023}
\begin{equation}
    C(T) = \sum_{v \in \inter} |n_{v_L} - n_{v_R}|, \nonumber
\end{equation}
where $v_L, v_R$ are the daughters of $v$. The Colless index cannot be represented by the clade-number basis; see Appendix~\ref{B.1}. It can be represented by the daughter-number basis as 
\begin{equation}
    C(T) = \sum_{i = 1}^{\lfloor n/2 \rfloor} \sum_{j = i}^{n-i} (j-i) \, D_{i,j}(T). \nonumber
\end{equation}

\subsection{Corrected Colless index} \label{corrected colless}
The corrected Colless index is defined as~\cite[Chapter 13]{Fischer2023}
\begin{equation}
    I_C(T) = \frac{2 C(T)}{(n-1)(n-2)} = \frac{2}{(n-1)(n-2)}\sum_{v \in \inter} |n_{v_L} - n_{v_R}|. \nonumber
\end{equation}
Th corrected Colless index cannot be represented by the clade-number basis; see Appendix~\ref{B.1}. It can be represented by the daughter-number basis as 
\begin{equation}
    I_C(T) = \sum_{i = 1}^{\lfloor n/2 \rfloor} \sum_{j = i}^{n-i} \frac{2(j-i)}{(n-1)(n-2)} \, D_{i,j}(T). \nonumber
\end{equation}

\subsection{\texorpdfstring{Equal weights Colless ($I_2$) index}{Equal weights Colless index}} \label{I2}
The $I_2$ index is defined as~\cite[Chapter 14]{Fischer2023}
\begin{equation}
    I_2(T) = \frac{1}{n-2} \sum_{\substack{v \in \inter \,| \\ n_v > 2}} \frac{|n_{v_L} - n_{v_R}|}{n_v - 2}. \nonumber
\end{equation}
The $I_2$ index cannot be represented by the clade-number basis; see Appendix~\ref{B.1}. It can be represented by the daughter-number basis as 
\begin{equation}
    I_2(T) = \sum_{i = 1}^{\lfloor n/2 \rfloor} \sum_{j = i}^{n-i} b_n(i,j) \, D_{i,j}(T), \nonumber
\end{equation}
where 
$$
b_n(i,j) = \begin{cases}
    0 & \text{if } i + j = 2, \\
    \frac{j-i}{(n-2)(i+j - 2)} & \text{if } i + j > 2.
\end{cases}
$$

\subsection{Quadratic Colless index} \label{quadratic colless}
The quadratic Colless index is defined as~\cite[Chapter 15]{Fischer2023}
\begin{equation}
    QC(T) = \sum_{v \in \inter} (n_{v_L} - n_{v_R})^2. \nonumber
\end{equation}
Although it does not immediately follow from this definition and Proposition~\ref{clade basis representations}, the quadratic Colless index can be represented by the clade-number basis. To find a representation, we use Proposition~\ref{cophenetic sackin qc theorem} (see Appendix~\ref{Appendix D} for proof) to write $QC(T) = 2\Phi(T) + S(T) - n(n-1)$, and combine the representations of the total cophenetic index (Section~\ref{total cophenetic}) and Sackin index (Section~\ref{sackin}) with the fact that $C_n(T) = 1$. We obtain that the quadratic Colless index is represented by the clade-number basis as
\begin{align}
    QC(T) &= \bigg[ \sum_{r=2}^{n-1} r^2 \, C_r(T) \bigg] + (2n-n^2) \, C_n(T). \nonumber
\end{align} 
It can be represented by the daughter-number basis as 
\begin{equation}
    QC(T) = \sum_{i = 1}^{\lfloor n/2 \rfloor} \sum_{j = i + 1}^{n-i} (i-j)^2 D_{i,j}(T). \nonumber
\end{equation}

\subsection{\texorpdfstring{Mean $I$ index}{Mean I index}} \label{mean I index}
The mean $I$ index is defined as~\cite[Chapter 17]{Fischer2023}
\begin{equation}
    \Bar{I}(T) = \frac{1}{|\{v \in \inter, \, n_v \geq 4\}|} \sum_{\substack{v \in \inter \,| \\ n_v \geq 4}} I_v, \nonumber \\
\end{equation}
where 
$$I_v = \frac{n_{v_R} - \lceil \frac{n_v}{2}\rceil}{(n_v - 1) - \lceil \frac{n_v}{2}\rceil}.$$
The mean $I$ index cannot be represented by the clade-number basis or the daughter-number basis; see Appendix~\ref{B.2}.

\subsection{\texorpdfstring{Total $I$ index}{Total I index}} \label{total I index}
The total $I$ index is defined as~\cite[Chapter 17]{Fischer2023}
\begin{equation}
    \Sigma I(T) = \sum_{\substack{v \in \inter \,| \\ n_v \geq 4}} I_v, \nonumber
\end{equation}
where $I_v$ is defined in Section~\ref{mean I index}. The total $I$ index cannot be represented by the clade-number basis; see Appendix~\ref{B.1}. It can be represented by the daughter-number basis as
\begin{equation}
    \Sigma I(T) = \sum_{i = 1}^{\lceil n/2 \rceil} \sum_{j = i}^{n-i}  b_n(i,j)\, D_{i,j}(T) \nonumber,
\end{equation}
where
$$
    b_n(i,j) = \begin{cases}
        0 & \text{if } i + j < 4, \\
        \frac{j - \lceil \frac{i+j}{2}\rceil}{(i+j - 1) - \lceil \frac{i+j}{2}\rceil} & \text{if } i +j \geq 4.
    \end{cases}
$$

\subsection{\texorpdfstring{Mean $I'$ index}{Mean I' index}} \label{mean I' index}
The mean $I'$ index is defined as~\cite[Chapter 17]{Fischer2023}
\begin{equation}
    \Bar{I'}(T) = \frac{1}{|\{v \in \inter, \, n_v \geq 4\}|} \sum_{\substack{v \in \inter \,| \\ n_v \geq 4}} I'_v, \nonumber 
\end{equation}
where 
$$
I'_v = \begin{cases}
            I_v & \text{if } n_v \text{ odd}, \\
            \frac{n_v - 1}{n_v} I_v & \text{if } n_v \text{ even}.
        \end{cases}
$$    
The mean $I'$ index cannot be represented by the clade-number basis or the daughter-number basis; see Appendix~\ref{B.2}.

\subsection{\texorpdfstring{Total $I'$ index}{Total I' index}} \label{total I' index}
The total $I'$ index is defined as~\cite[Chapter 17]{Fischer2023}
\begin{equation}
    \Sigma I'(T) = \sum_{\substack{v \in \inter \,| \\ n_v \geq 4}} I'_v, \nonumber
\end{equation}
where $I'_v$ is defined in Section~\ref{mean I' index}. The total $I'$ index cannot be represented by the clade-number basis; see Appendix~\ref{B.1}. It can be represented by the daughter-number basis as
\begin{equation}
    \Sigma I'(T) = \sum_{i = 1}^{\lceil n/2 \rceil} \sum_{j = i}^{n-i} b_n(i,j) \, D_{i,j}(T), \nonumber
\end{equation}
where 
$$
b_n(i,j) = 
    \begin{cases}
        0 & \text{if } i + j < 4, \\
        \frac{j - \lceil \frac{i+j}{2}\rceil}{(i+j - 1) - \lceil \frac{i+j}{2}\rceil} & \text{if } i + j \geq 4, \ i + j \text{ odd}, \\
        \frac{i + j - 1}{i + j} \cdot \frac{j - \lceil \frac{i+j}{2}\rceil}{(i+j - 1) - \lceil \frac{i+j}{2}\rceil} & \text{if } i + j \geq 4, \ i + j \text{ even}.
    \end{cases}
$$

\subsection{Symmetry nodes index} \label{symmetry nodes index}
The symmetry nodes index is defined as~\cite[Chapter 18]{Fischer2023}
\begin{equation}
    SNI(T) = n - 1 - s(T) \nonumber.
\end{equation}
The symmetry nodes index cannot be represented by the clade-number basis or the daughter-number basis; see Appendix~\ref{B.2}.

\subsection{\texorpdfstring{Rogers $J$ index}{Rogers J index}} \label{rogers j index}
The Rogers $J$ index is defined as~\cite[Chapter 19]{Fischer2023}
\begin{equation}
    J(T) = \sum_{v \in \inter} \big(1 - \mathbbm{1}_{\{n_{v_L} = n_{v_R} \}} \big) \nonumber,
\end{equation}
where $\mathbbm{1}_{\{\cdot\}}$ equals 1 if the statement $\{\cdot\}$ is true and 0 otherwise. The Rogers $J$ index cannot be represented by the clade-number basis; see Appendix~\ref{B.1}. It can be represented by the daughter-number basis as 
\begin{equation}
    J(T) = \sum_{i = 1}^{\lfloor n/2 \rfloor} \sum_{j = i}^{n-i} \mathbbm{1}_{\{i < j\}} \, D_{i,j}(T) \nonumber. 
\end{equation}

\subsection{Rooted quartet index for binary trees} \label{rqib}
The rooted quartet index for binary trees $rQIB(T)$ is defined as~\cite[Chapter 20]{Fischer2023}
\begin{equation}
    rQIB(T) = \sum_{v \in \inter}\binom{n_{v_L}}{2} \binom{n_{v_R}}{2}. \nonumber
\end{equation}
The rooted quartet index for binary trees cannot be represented by the clade-number basis; see Appendix~\ref{B.1}. It can be represented by the daughter-number basis as 
\begin{equation}
    rQIB(T) = \sum_{i = 1}^{\lfloor n/2 \rfloor} \sum_{j = i}^{n-i} \binom{i}{2}\binom{j}{2} \, D_{i, j}(T) \nonumber, 
\end{equation}
where $\binom{1}{2} = 0$ by convention.

\subsection{Colijn-Plazzotta rank} \label{CP rank}
The Colijn-Plazzotta rank $CP(T)$ is a bijection from binary trees to the postive integers which orders trees in an ordering that preserves the criteria for a balance index~\cite{Rosenberg2021}. The Colijn-Plazzotta rank is defined recursively~\cite[Chapter 21]{Fischer2023}. For the base case, $CP(T) = 1$ if $T$ has only one leaf. Otherwise,
\begin{equation}
    CP(T) = \frac{1}{2} \,  CP(T_1) \, \big[CP(T_1) - 1\big] + CP(T_2) + 1, \nonumber
\end{equation}
where the root of $T$ has daughter subtrees $T_1, T_2$ with $CP(T_1) \geq CP(T_2)$. The Colijn-Plazzotta rank cannot be represented by the clade-number basis or the daughter-number basis; see Appendix~\ref{B.2}.

\subsection{Furnas rank} \label{Furnas rank}
The Furnas rank $F(T)$ is a bijection from binary trees to the positive integers, defined as~\cite[Chapter 22]{Fischer2023} 
\begin{equation}
    F(T) = r_n(T), \nonumber
\end{equation}
where $r_n(T)$ is the order of $T$ in a particular ordering among trees of size $n$ defined by Furnas~\cite{Furnas1984}. The Furnas rank cannot be represented by the clade-number basis or the daughter-number basis; see Appendix~\ref{B.2}.

\subsection{Total internal path length} \label{TIPL}

The total internal path length $TIP(T)$ is defined as~\cite[Chapter 23.1]{Fischer2023}
\begin{equation}
    TIP(T) = \sum_{v \in \inter} \delta_T(v). \nonumber
\end{equation}
For a binary tree $T$, the total internal path length is related to the Sackin index~\cite[Lemma 23.1]{Fischer2023} by
\begin{equation}
    S(T) = TIP(T) + 2(n-1). \nonumber
\end{equation}
Therefore, by using the representation of the Sackin index by the clade-number basis from Section~\ref{sackin} and writing $2(n-1) = 2(n-1) C_n(T)$, the total internal path length can be represented by the clade-number basis as
\begin{equation}
    TIP(T) =  \bigg[\sum_{r=2}^{n-1} r \, C_r(T) \bigg] + (2-n) C_n(T). \nonumber
\end{equation}
It can be represented by the daughter-number basis as
\begin{equation}
    TIP(T) = \sum_{i = 1}^{\lfloor n/2 \rfloor} \sum_{j = i}^{n-i} b_n(i,j) \, D_{i,j}(T), \nonumber
\end{equation}
where 
$$
b_n(i,j) = 
\begin{cases}
    i + j & \text{if } i + j < n, \\
    2 - n & \text{if } i +j = n.
\end{cases}
$$

\subsection{Total path length} \label{TPL}

The total path length $TPL(T)$ is defined as~\cite[Chapter 23.2]{Fischer2023}
\begin{equation}
    TPL(T) = \sum_{v \in V(T)} \delta_T(v) = S(T) + TIP(T). \nonumber
\end{equation}
By summing the representations of the Sackin index (Section~\ref{sackin}) and total internal path length (Section~\ref{TIPL}), the total path length can be represented by the clade-number basis as
\begin{equation}
    TPL(T) = \bigg[\sum_{r=2}^{n-1} 2r \, C_r(T)  \bigg] + 2 C_n(T). \nonumber
\end{equation}
It can be represented by the daughter-number basis as
\begin{equation}
    TPL(T) = \sum_{i = 1}^{\lfloor n/2 \rfloor} \sum_{j = i}^{n-i} b_n(i,j) \, D_{i,j}(T), \nonumber
\end{equation}
where 
$$
b_n(i,j)= 
\begin{cases}
    2(i + j) & \text{if } i + j < n, \\
    2 & \text{if } i +j = n.
\end{cases}
$$

\subsection{Average vertex depth} \label{AVD}
The average vertex depth $AVD(T)$ is defined as~\cite[Chapter 23.3]{Fischer2023}
\begin{equation}
    AVD(T) = \frac{1}{|V(T)|} \sum_{v \in V(T)} \delta_T(v) = \frac{1}{|V(T)|} TPL(T). \nonumber
\end{equation}
Because a binary tree has $2n - 1$ total vertices, $|V(T)| = 2n-1$. Therefore, using the prior result for the total path length $TPL(T)$ (Section~\ref{TPL}), the average vertex depth can be represented by the clade-number basis as
\begin{equation} \nonumber
    AVD(T) = \bigg[\sum_{r = 2}^{n-1} \frac{2r}{2n - 1} \, C_r(T) \bigg] + \frac{2}{2n - 1} C_n(T).
\end{equation}
It can be represented by the daughter-number basis as
\begin{equation}
    AVD(T) = \sum_{i = 1}^{\lfloor n/2 \rfloor} \sum_{j = i}^{n-i} b_n(i,j) \, D_{i,j}(T), \nonumber
\end{equation}
where 
$$
b_n(i,j) = 
\begin{cases}
    \frac{2(i+j)}{2n - 1} & \text{if } i + j < n, \\
    \frac{2}{2n - 1} & \text{if } i +j = n.
\end{cases}
$$

\subsection{Maximum depth} \label{maximum depth}
The maximum depth $mD(T)$ is defined as~\cite[Chapter 23.4]{Fischer2023}
\begin{equation}
    mD(T) = h(T). \nonumber
\end{equation}
The maximum depth cannot be represented by the clade-number basis or the daughter-number basis; see Appendix~\ref{B.2}.

\subsection{Maximum width} \label{maximum width}
The maximum width $mW(T)$ is defined as~\cite[Chapter 23.5]{Fischer2023}
\begin{equation}
    mW(T) = \max_{i = 0, 1, \ldots, h(T)} w_T(i). \nonumber
\end{equation}
The maximum width cannot be represented by the clade-number basis or the daughter-number basis; see Appendix~\ref{B.2}.

\subsection{Modified maximum difference in widths} \label{modified max dif widths}
The modified maximum difference in widths $mdelW(T)$ is defined as~\cite[Chapter 23.6]{Fischer2023}
\begin{equation}
    mdelW(T) = \max_{i = 0, 1, \ldots, h(T) - 1} \big[ w_T(i + 1) - w_T(i) \big]. \nonumber
\end{equation}
The modified maximum difference in widths cannot be represented by the clade-number basis or the daughter-number basis; see Appendix~\ref{B.2}. 

\subsection{Maximum width over maximum depth} \label{max width over max depth}
The maximum width over maximum depth $W/D(T)$ is defined as~\cite[Chapter 23.7]{Fischer2023}
\begin{equation}
    W/D(T) = \frac{mW(T)}{mD(T)}. \nonumber
\end{equation}
The maximum width over maximum depth cannot be represented by the clade-number basis or the daughter-number basis; see Appendix~\ref{B.2}.

\subsection{stairs1 index} \label{stairs1}
The stairs1 index is defined as~\cite[Chapter 23.8]{Fischer2023}
\begin{equation} \nonumber
    st1(T) = \frac{1}{n-1} \sum_{v \in \inter} \big(1 - \mathbbm{1}_{\{n_{v_L} = n_{v_R}\}} \big) = \frac{J(T)}{n-1}, 
\end{equation}
where $J(T)$ is the Rogers $J$ index. The stairs1 index cannot be represented by the clade-number basis; see Appendix~\ref{B.1}. It can be represented by the daughter-number basis as 
\begin{equation}
    st1(T) = \sum_{i = 1}^{\lfloor n/2 \rfloor} \sum_{j = i}^{n-i} \frac{\mathbbm{1}_{\{i < j\}}}{n - 1} \, D_{i,j}(T) \nonumber. 
\end{equation}

\subsection{stairs2 index} \label{stairs2}
The stairs2 index is defined as~\cite[Chapter 23.9]{Fischer2023}
\begin{equation}
    st2(T) = \frac{1}{n-1} \sum_{v \in \inter} \frac{n_{v_L}}{n_{v_R}}, \nonumber
\end{equation}
where we recall that by convention, $n_{v_L} \leq n_{v_R}$ for the daughters $v_L, v_R$ of $v$. The stairs2 index cannot be represented by the clade-number basis; see Appendix~\ref{B.1}. It can be represented by the daughter-number basis as 
\begin{equation}
    st2(T) = \sum_{i = 1}^{\lfloor n/2 \rfloor} \sum_{j = i}^{n-i} \frac{i}{j(n-1)} \, D_{ij}(T) \nonumber. 
\end{equation}

%% file: Appendix_B.tex
\section{Proving balance indices are not represented by bases} \label{Appendix B}

\subsection{Balance indices not represented by the clade number basis} \label{B.1}
\begin{figure}[htbp]
    \centering
    \begin{subfigure}[t]{0.48\textwidth}
        \centering
        \includegraphics[height=1.2in]{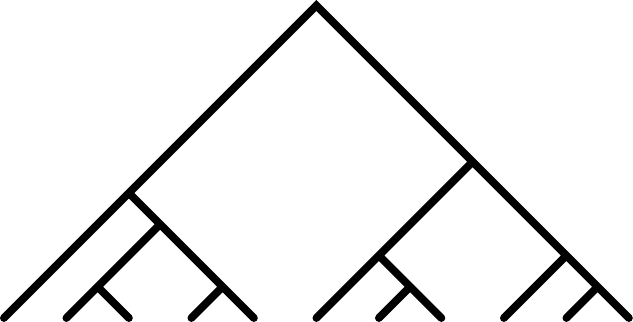}
        \caption{}\label{fig:B_1a}
    \end{subfigure}%
    ~ 
    \begin{subfigure}[t]{0.48\textwidth}
        \centering
        \includegraphics[height=1.2in]{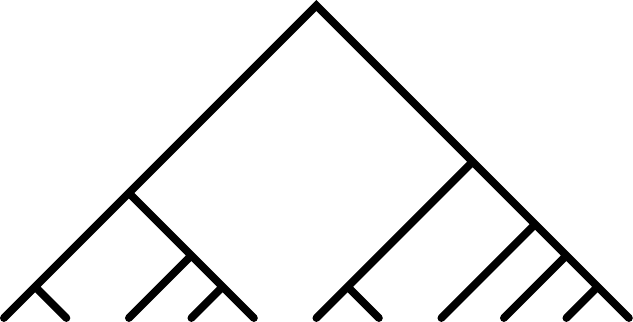}
        \caption{} \label{fig:B_1b}
    \end{subfigure}
    \caption{Counterexample for clade-number basis representations. These two trees of size $n = 11$ have identical clade numbers $\big(C_2(T), C_3(T), \ldots, C_{11}(T) \big) = (4, 2, 1, 1, 1, 0, 0, 0, 0, 1)$. However, they have different values for certain balance indices, found in Table~\ref{table:B_1}.}
    \label{fig:B_1}
\end{figure}

In Appendix~\ref{Appendix A}, we claimed that nine balance indices are represented by the daughter-number basis but not by the clade-number basis: the Colless index (\ref{colless}), corrected Colless index (\ref{corrected colless}), equal weights Colless ($I_2$) index (\ref{I2}), total $I$ index (\ref{total I index}), total $I'$ index (\ref{total I' index}), Rogers $J$ index (\ref{rogers j index}), rooted quartet index for binary trees (\ref{rqib}), stairs1 index (\ref{stairs1}), and the stairs2 index (\ref{stairs2}) To prove that these indices cannot be represented by the clade-number basis, we demonstrate a counterexample. We found two trees of size $n = 11$ with identical clade numbers but different values for each of these nine indices. The trees are depicted in Figure~\ref{fig:B_1}, and their values for the 9 indices appear in Table~\ref{table:B_1}. 

\begin{table}[htbp]
    \centering
    \renewcommand{\arraystretch}{1.2}
    \setlength{\tabcolsep}{6pt}
    \begin{tabular}{|l|c|c|}
        \hline
        \textbf{Index name} & \textbf{Tree in Fig.~\ref{fig:B_1a}} & \textbf{Tree in Fig.~\ref{fig:B_1b}} \\
        \hline
        Colless index & 6 & 8 \\
        Corrected Colless index & 6/45 & 8/45 \\
        Equal weights Colless ($I_2$) index & 28/81 & 71/162 \\
        Total $I$ index & 1 & 3/2 \\
        Total $I'$ index & 1 & 7/6 \\
        Rogers $J$ index & 4 & 6 \\
        Rooted quartet index for binary trees & 160 & 159 \\
        stairs1 index & 2/5 & 3/5 \\
        stairs2 index & 497/600 & 11/15 \\
        \hline
    \end{tabular}
    \caption{Values of a set of balance indices for the two trees in Figure~\ref{fig:B_1} with identical clade numbers. This counterexample proves that none of the listed statistics can be represented by the clade-number basis.} \label{table:B_1}
\end{table}

\subsection{Balance indices not represented by either basis} \label{B.2}
\begin{figure}[ptb]
    \centering
    \begin{subfigure}[t]{0.48\textwidth}
        \centering
        \includegraphics[height=1.2in]{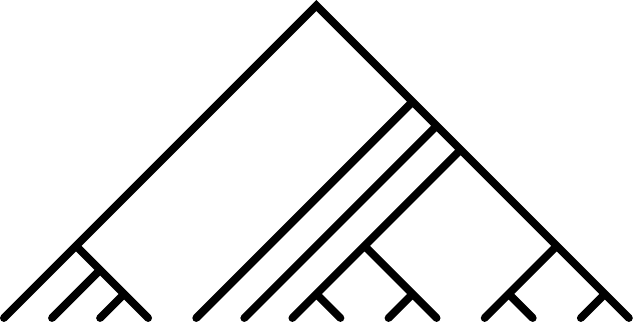}
        \caption{} \label{fig:B_2a}
    \end{subfigure}%
    ~ 
    \begin{subfigure}[t]{0.48\textwidth}
        \centering
        \includegraphics[height=1.2in]{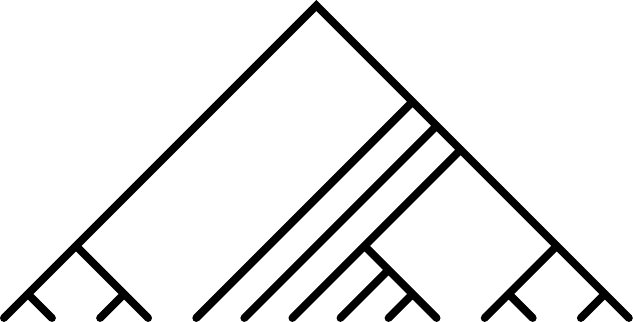}
        \caption{} \label{fig:B_2b}
    \end{subfigure}
    \caption{Counterexample for daughter-number basis representations. These two trees of size $n = 14$ have identical clade numbers $\big(C_2(T), \ldots, C_{14}(T)\big) = (5, 1, 3, 0, 0, 0, 1, 1, 1, 0, 0, 0, 1)$ and identical daughter numbers. The nonzero daughter numbers for both trees are $D_{1,1}(T) = 5$, $D_{1,2}(T) = 1$, $D_{1,3}(T) = 1$, $D_{1,8}(T) = 1$, $D_{1,9}(T) = 1$, $D_{2,2}(T) = 2$, $D_{4,4}(T) = 1$, $D_{4,10}(T) = 1$. They have different values for certain balance indices, computed in Table~\ref{table:B_3}.}
    \label{fig:B_2}
\end{figure}

Similarly, in Appendix~\ref{Appendix A} we claimed that twelve balance indices cannot be represented by the clade-number basis or the daughter-number basis: the variance in leaf depths (\ref{var leaf depths}), $B_1$ index (\ref{b1}), $B_2$ index (\ref{b2}), mean $I$ index (\ref{mean I index}), mean $I'$ index (\ref{mean I' index}), symmetry nodes index (\ref{symmetry nodes index}), Colijn-Plazzotta rank (\ref{CP rank}), Furnas rank (\ref{Furnas rank}), maximum depth (\ref{maximum depth}), maximum width (\ref{maximum width}), modified maximum difference in widths (\ref{modified max dif widths}), and maximum width over maximum depth (\ref{max width over max depth}). For all these indices except for the mean $I$ index and mean $I'$ index, we prove they cannot be represented in the two bases using the same argument as in Section~\ref{B.1}. In Figure~\ref{fig:B_2} and Table~\ref{table:B_3}, we demonstrate two trees of size $n = 14$ that have identical clade numbers and daughter numbers but different values for these indices. 

For the remaining two indices, the mean $I$ index ($\bar{I}(T)$) and mean $I'$ index ($\bar{I'}(T)$), we require another approach. These two indices are normalized versions of the total $I$ index ($\Sigma I (T)$) and total $I'$ index ($\Sigma I'(T)$), respectively. Both the total $I$ index and total $I'$ index can be represented by the daughter-number basis; moreover, the mean $I$ and mean $I'$ indices related to the total $I$ and total $I'$ indices by
\begin{align*}
    \bar{I}(T) &= \frac{1}{|\{v \in \inter, \, n_v \geq 4\}|} \Sigma I (T), \\
    \bar{I'}(T) &= \frac{1}{|\{v \in \inter, \, n_v \geq 4\}|} \Sigma I' (T).
\end{align*} 
Because 
$$|\{v \in \inter, \, n_v \geq 4\}| = \sum_{i = 1}^{\lfloor n/2 \rfloor} \sum_{j = i + 1}^{n-i} \mathbbm{1}_{\{i + j \geq 4\}} \, D_{i,j}(T), $$
the mean $I$ index and mean $I'$ index can both be written as nonlinear functions of the daughter-number basis. This observation has two implications: first, any trees with identical daughter numbers have identical mean $I$ and mean $I'$ indices. Second, they cannot be represented by the daughter-number basis because these functions are not linear functions of the daughter-number basis. To formally prove that these indices cannot be represented by the daughter-number basis, we present a method based on linear algebra in Appendix~\ref{Appendix C}.

\begin{table}[tpb]
    \centering
    \renewcommand{\arraystretch}{1.2}
    \setlength{\tabcolsep}{6pt}    
    \begin{tabular}{|l|c|c|}
        \hline
        \textbf{Index name} & \textbf{Tree in Fig.~\ref{fig:B_2a}} & \textbf{Tree in Fig.~\ref{fig:B_2b}} \\
        \hline
        Variance of leaf depths & 122/49 & 143/49 \\
        $B_1$ index & 457/60 & 149/20 \\
        $B_2$ index & 3 & 199/64 \\
        Symmetry nodes index & 5 & 6 \\
        Colijn-Plazzota rank & 1276009 & 16088633 \\
        Furnas rank & 1663 & 1760 \\
        Maximum depth & 6 & 7 \\
        Maximum width & 8 & 6 \\
        Modified maximum difference in widths & 4 & 2 \\
        Maximum width over maximum depth & 4/3 & 6/7 \\
        \hline
    \end{tabular}
    \caption{Values of a set of balance indices for the two trees in Figure~\ref{fig:B_2} with identical clade numbers and daughter numbers. This counterexample proves that none of the listed statistics can be represented by the clade-number basis or the daughter-number basis.} \label{table:B_3}
\end{table}

%% file: Appendix_C.tex
\section{Linear algebraic techniques for basis representations} \label{Appendix C}

A basis of statistics, such as the clade-number basis or the daughter-number basis, can be interpreted as a mapping from the space of trees $\mathcal{T}_n$ to a subset of $\mathbb{R}^k$, where $k$ is the number of basis elements. A summary statistic $A(T)$ that can be represented by the basis defines a linear map from $\mathbb{R}^k$ to $\mathbb{R}$. Therefore, we can study the representations of summary statistics by a particular basis by studying the possible linear maps between the basis and a given statistic. 

In this appendix, we use linear algebra to study possible linear mappings from a basis to a summary statistic. First, in Section~\ref{sec:c_1} we study exact basis representations of summary statistics by examining linear systems. We apply this approach to prove that two balance indices cannot be represented by the daughter-number basis, a fact that could not be shown by counterexample as discussed in Section~\ref{B.2}. Second, in Section~\ref{sec:c_2}, we study approximate basis representations of statistics by solving least squares problems for the linear system. This approach allows us to find approximate basis representations for indices that are not represented by a basis, to evaluate the fit of the approximations, and to identify how a statistic measures the importance of each component of the basis. 

\subsection{Linear system method} \label{sec:c_1}

In this section, we study a linear system associated with a basis and a summary statistic. Fix $n \geq 1$, the size of the trees. Let $\mathcal{B}(T)$ be a basis of statistics with $k$ components, and let $A(T)$ be any summary statistic. For a set $\mathcal{A} \subseteq \mathcal{T}_n$ containing $|\mathcal{A}| = m$ trees, compute $\mathcal{B}(T)$ and $A(T)$. Now consider the linear system
\begin{equation} \label{linear system basis statistic}
    \mathbf{B} \mathbf{x} = \mathbf{a},
\end{equation}
where $\mathbf{B} \in \mathbb{R}^{m \times k}$ and $\mathbf{a} \in \mathbb{R}^m$ are the values of $\mathcal{B}(T)$ and $A(T)$ for the $m$ trees. 

If the statistic $A(T)$ is represented by the basis $\mathcal{B}(T)$, then a solution $\mathbf{x} \in \mathbb{R}^k$ to eq.~\ref{linear system basis statistic} exists for any set of trees $\mathcal{A}$ for all $n$. If the statistic is not represented by the basis, then for some $n$, eq.~\ref{linear system basis statistic} has no solution. 

The study of solutions to eq.~\ref{linear system basis statistic} is a fundamental problem in linear algebra and many solution methods exist. For instance, by creating the augmented matrix $[\mathbf{B} \, | \, \mathbf{a}]$ and performing Gaussian elimination to recover the reduced-row echelon form, a solution can be obtained or the nonexistence of a solution can be proven (e.g.~\cite[Chapter 1]{Hill1986}). Finding a solution to eq.~\ref{linear system basis statistic} for many values of $n$ and sets of trees $\mathcal{A} \subset \mathcal{T}_n$ does not prove that a statistic can be represented by the basis, but it provides evidence that a representation may exist, and the particular value of $\mathbf{x}$ discovered may aid in finding a correct representation. Conversely, a single value of $n$ and a subset $\mathcal{A} \subset \mathcal{T}_n$ for which the associated linear system (eq.~\ref{linear system basis statistic}) does not have a solution suffice to provide proof that the statistic cannot be represented by the basis. 

Using this argument, we prove that the mean $I$ index (\ref{mean I index}) and mean $I'$ index (\ref{mean I' index}) are not represented by the daughter-number basis, as noted in Appendix~\ref{B.2}. If $n = 7$ and $\mathcal{A} = \mathcal{T}_7$, eq.~\ref{linear system basis statistic} has no solution for the mean $I$ index or mean $I'$ index, proving that these indices are not represented by the daughter-number basis (see Table~\ref{tab:mean I table}). By Proposition~\ref{clade to daughter}, they are also not represented by the clade-number basis. 

We comment that, unlike the basis representations found in Appendix~\ref{Appendix A}, this method only uses values of the summary statistic computed on specific trees and does not rely on any information about the functional form of the the summary statistic. Therefore, generating a set of trees $\mathcal{A}$ and searching for solutions to eq.~\ref{linear system basis statistic} provides a method of testing unknown statistics that have definitions that differ greatly from the tree properties used by a basis. 

\begin{table}[tb]
    \centering
    \setlength{\tabcolsep}{3pt}
    \renewcommand\arraystretch{1.5}
    \begin{tabular}{|c||c|c|c|c|c|c|c|c|c|c|c|c||c|c|}
    \hline
    & \multicolumn{12}{c||}{$\mathbf{B}$} & $\mathbf{a}_1$ & 
    $\mathbf{a}_2$ \\ \hline
    \textbf{Tree} & $D_{1,1}$ & $D_{1,2}$ & $D_{1,3}$ & $D_{1,4}$ & $D_{1,5}$ & $D_{1,6}$ & $D_{2,2}$ & $D_{2,3}$ & $D_{2,4}$ & $D_{2,5}$ & $D_{3,3}$ & $D_{3,4}$ & $\bar{I}(T)$ & $\bar{I'}(T)$ \\ \hline
    \includegraphics[width=1.2cm]{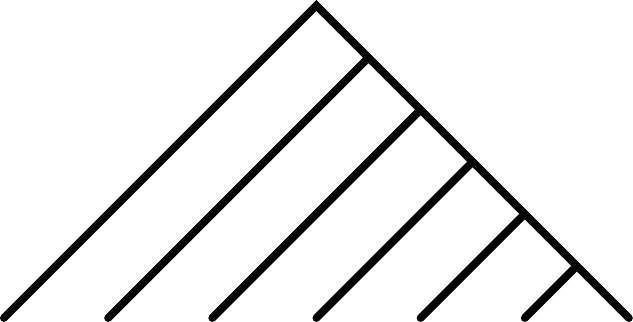} & 1 & 1 & 1 & 1 & 1 & 1 & 0 & 0 & 0 & 0 & 0 & 0 & 1 & $\frac{43}{48}$ \\ \hline
    \includegraphics[width=1.2cm]{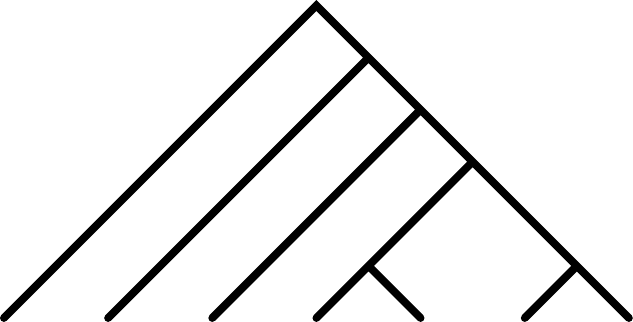} & 2 & 0 & 0 & 1 & 1 & 1 & 1 & 0 & 0 & 0 & 0 & 0 & $\frac{3}{4}$ & $\frac{17}{24}$ \\ \hline
    \includegraphics[width=1.2cm]{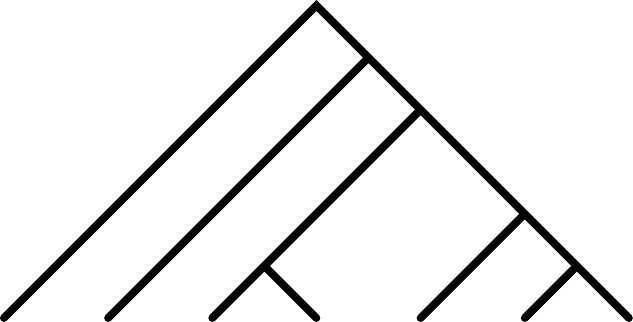} & 2 & 1 & 0 & 0 & 1 & 1 & 0 & 1 & 0 & 0 & 0 & 0 & $\frac{2}{3}$ & $\frac{11}{18}$\\ \hline
    \includegraphics[width=1.2cm]{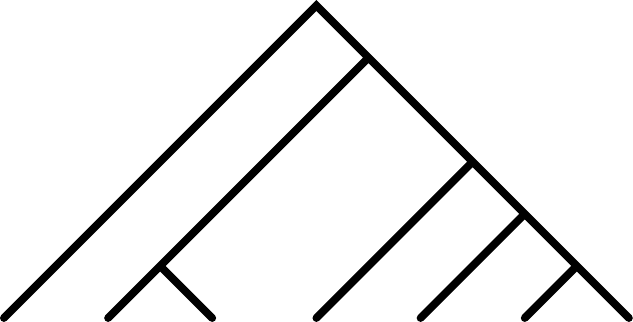} & 2 & 1 & 1 & 0 & 0 & 1 & 0 & 0 & 1 & 0 & 0 & 0 & $\frac{5}{6}$ & $\frac{13}{18}$ \\ \hline
    \includegraphics[width=1.2cm]{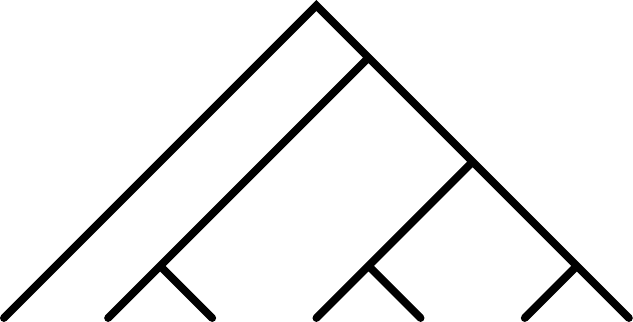} & 3 & 0 & 0 & 0 & 0 & 1 & 1 & 0 & 1 & 0 & 0 & 0 & $\frac{1}{2}$ & $\frac{17}{36}$ \\ \hline
    \includegraphics[width=1.2cm]{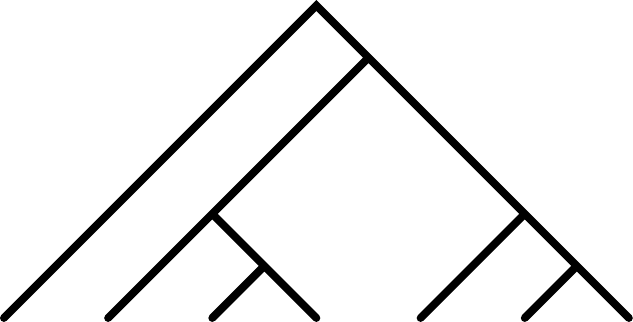} & 2 & 2 & 0 & 0 & 0 & 1 & 0 & 0 & 0 & 0 & 1 & 0 & $\frac{1}{2}$ & $\frac{1}{2}$ \\ \hline
    \includegraphics[width=1.2cm]{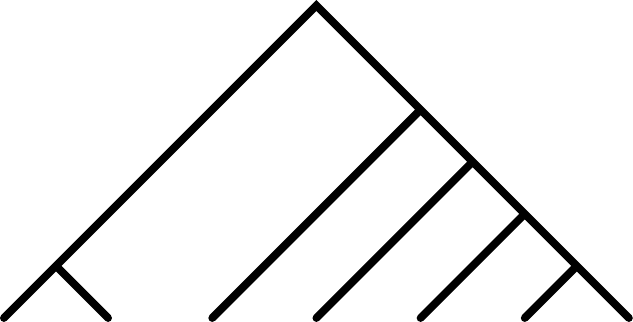} & 2 & 1 & 1 & 1 & 0 & 0 & 0 & 0 & 0 & 1 & 0 & 0 & $\frac{5}{6}$ & $\frac{3}{4}$ \\ \hline
    \includegraphics[width=1.2cm]{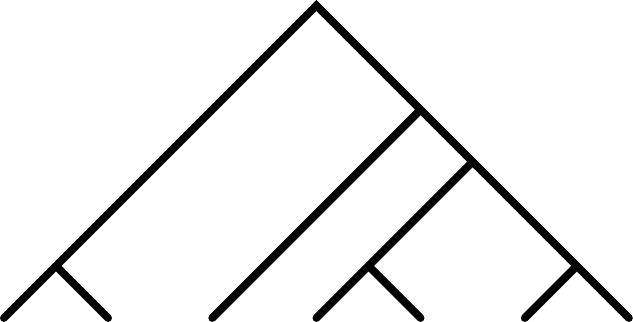} & 3 & 0 & 0 & 1 & 0 & 0 & 1 & 0 & 0 & 1 & 0 & 0 & $\frac{1}{2}$ & $\frac{1}{2}$ \\ \hline
    \includegraphics[width=1.2cm]{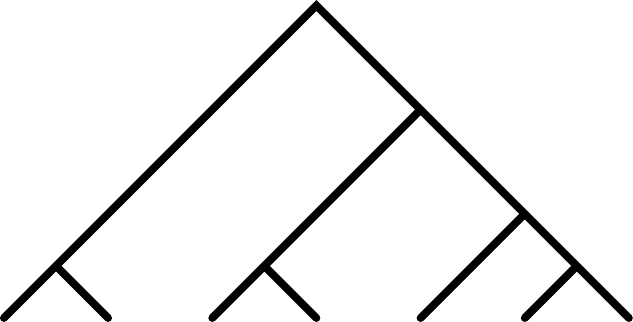} & 3 & 1 & 0 & 0 & 0 & 0 & 0 & 1 & 0 & 1 & 0 & 0 & $\frac{1}{4}$ & $\frac{1}{4}$ \\ \hline
    \includegraphics[width=1.2cm]{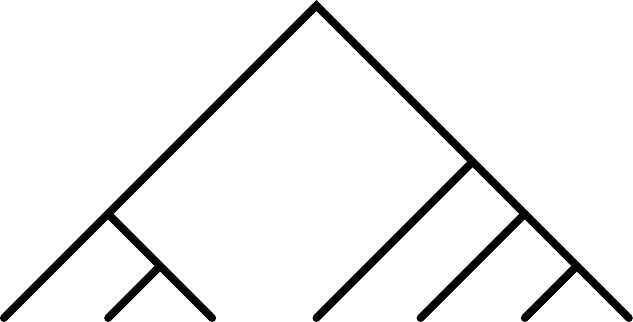} & 2 & 2 & 1 & 0 & 0 & 0 & 0 & 0 & 0 & 0 & 0 & 1 & $\frac{1}{2}$ & $\frac{3}{8}$\\ \hline
    \includegraphics[width=1.2cm]{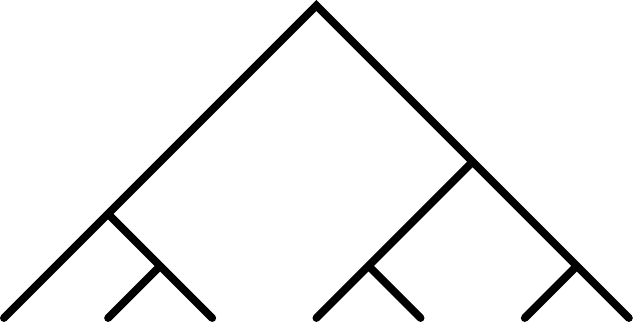} & 3 & 1 & 0 & 0 & 0 & 0 & 1 & 0 & 0 & 0 & 0 & 1 & 0 & 0 \\ \hline

    \end{tabular}

    \caption{The values $D_{i,j}$ of the daughter-number basis and the mean $I$ index $\bar{I}(T)$ and mean $I'$ index $\bar{I'}(T)$ for all 11 tree shapes of size $n = 7$. It can be verified by Gaussian elimination that the equation $\mathbf{B} \mathbf{x} = \mathbf{a}_i$ has no solution for $i \in \{1, 2\}$, where $\mathbf{B} \in \mathbb{R}^{11 \times 12}$ is the matrix of daughter-number basis values, $\mathbf{a}_1 \in \mathbb{R}^{11}$ is a vector of the values of $\bar{I}(T)$, $\mathbf{a}_2 \in \mathbb{R}^{11}$ is a vector of the values of $\bar{I'}(T)$, and $\mathbf{x} \in \mathbb{R}^{12}$ is a vector representing the possible coefficients of a daughter-number basis representation. This lack of solution to $\mathbf{B} \mathbf{x} = \mathbf{a}$ proves that the mean $I$ index and the mean $I'$ index cannot be represented by the daughter-number basis. } \label{tab:mean I table}
\end{table}

\subsection{Least squares method} \label{sec:c_2}

Even if the linear system in eq.~\ref{linear system basis statistic} has no solution, we can learn additional information about the statistic $A(T)$ by studying the associated least-squares problem,
\begin{equation} \label{least squares problem}
    \mathbf{x}^* = \arg\min_{\mathbf{x} \in \mathbb{R}^k} \sum_{i=1}^m \big(\mathbf{b}_{i}^t \mathbf{x} - a_i \big)^2 = \arg\min_{\mathbf{x} \in \mathbb{R}^k} \|\mathbf{B} \mathbf{x} - \mathbf{a}\|_2^2 ,
\end{equation}
where $\mathbf{b}_i^t$ is the $1 \times k$ $i$th row of $\mathbf{B}$ (corresponding to the transposed basis vector $\mathcal{B}(T_i) \in \mathbb{R}^k$) and $a_i$ is the $i$th element of the $m \times 1$ vector $\mathbf{a}$ (corresponding to $A(T_i)$). The residual of the solution, defined by 
\begin{equation} \label{residual eq}
    R(\mathbf{x}^*; \mathbf{B}, \mathbf{a}) := \sum_{i=1}^m \big( \mathbf{b}_i^t \mathbf{x}^* - a_i \big)^2,
\end{equation}
measures the error in the least-squares approximation across the set of trees that were considered. 

Assume now that all trees of size $n$ are considered, or $\mathcal{A} = \mathcal{T}_n$. In some contexts, the error in the approximation is not equally important for all trees. For instance, trees $T$ have different probabilities $\mathbb{P}[T]$ of being sampled under the Yule model (Theorem~\ref{Theorem 2}), and we may wish to find an approximate basis representation that minimizes the mean-squared error, 

\begin{figure}[tb]
    \centering
    \includegraphics[width = \linewidth]{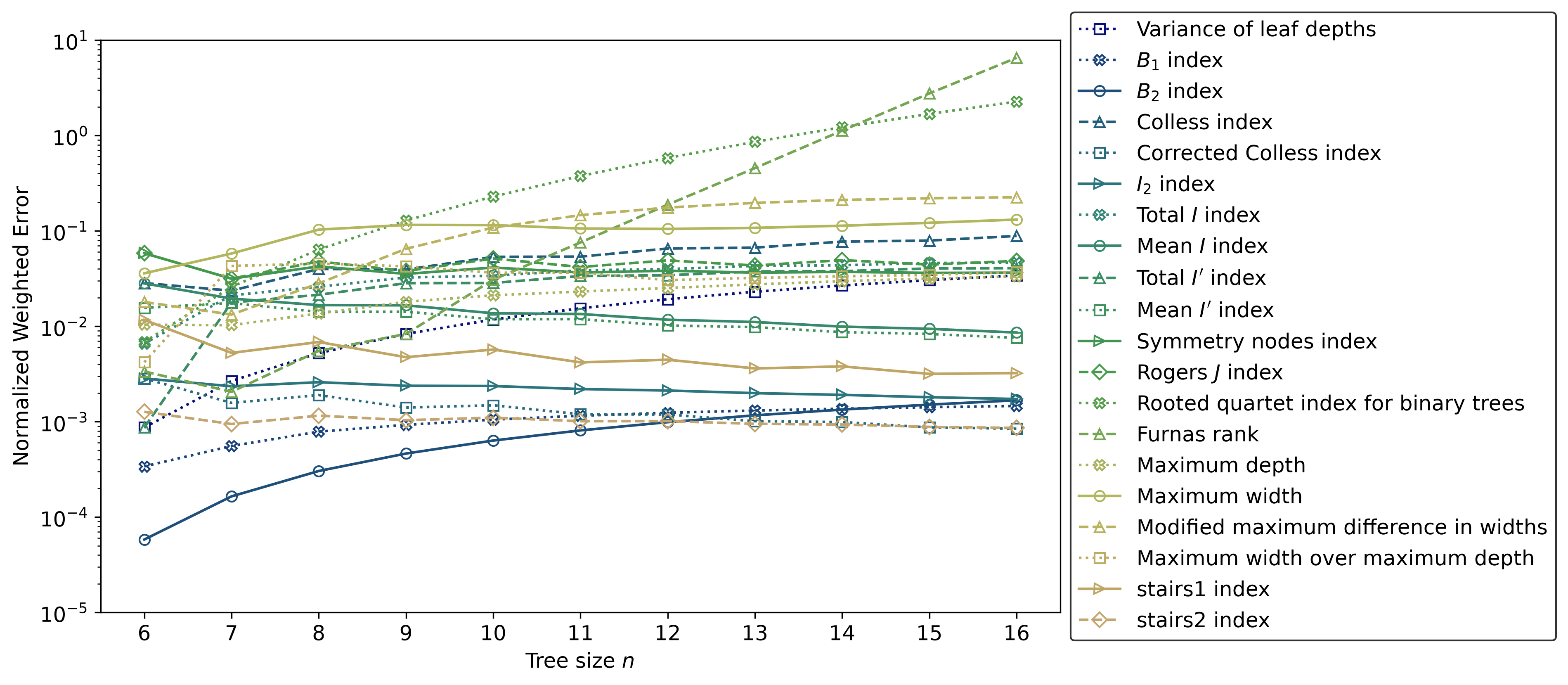}
    \caption{The error of the approximate representations of 20 balance indices by the clade-number basis, for all trees of size $n$ for each $n \in \{6, 7, \ldots, 16\}$. All 20 indices shown are not represented by the clade-number basis (Table~\ref{tab:index_representations}). The approximate representation is obtained by weighted least-squares (eq.~\ref{weighted least squares problem}), and it minimizes the mean-squared error of the approximation under the Yule model (eq.~\ref{Yule MSE}). The mean-squared error is normalized following eq.~\ref{weighted residual error}. All 8 statistics that are represented by the clade-number basis (Section~\ref{sec:4_1}) have error 0 up to numerical precision, and they are excluded from the plot. The magnitude of the error for the 20 indices shown provides a measurement of the relative fit of the index to the clade-number basis relative to other indices. } \label{fig:clade projection}
\end{figure}

\begin{equation} \label{Yule MSE}
    \textrm{MSE}(\mathbf{x}^*; \mathcal{B}, A) = \mathbb{E} \bigg[ \big(\mathcal{B}(T)^t \mathbf{x}^* - A(T)\big)^2 \bigg] := \sum_{T \in \mathcal{T}_n} \mathbb{P}[T] \, \big(\mathcal{B}(T)^t \mathbf{x}^* - A(T)\big)^2.
\end{equation}
Finding the value of $\mathbf{x}$ that minimizes this error is a special case of the weighted least-squares problem,
\begin{equation} \label{weighted least squares problem}
    \mathbf{x}^*_{\mathbf{W}} = \arg\min_{\mathbf{x} \in \mathbb{R}^k} \big\|\mathbf{W}^{1/2}(\mathbf{B} \mathbf{x} - \mathbf{a})\big\|_2^2 = \arg\min_{\mathbf{x} \in \mathbb{R}^k} \sum_{i=1}^m w_i \big( \mathbf{b}_i^t \mathbf{x} - a_i \big)^2,
\end{equation}
where $\mathbf{W}  \in \mathbb{R}^{m \times m}$ is a diagonal matrix of weights $w_{i} > 0$ assigned to trees $T_i \in \mathcal{A}$. If $\mathbf{W}$ is chosen to be the identity matrix, $\mathbf{W} = \mathbf{I}$, then eq.~\ref{weighted least squares problem} recovers eq.~\ref{least squares problem}. If the weight $w_i$ is chosen to be the probability that $T_i$ is generated under the Yule model, then eq.~\ref{weighted least squares problem} recovers the solution $\mathbf{x}^*_{\mathbf{W}}$ that minimizes the mean-squared error (eq.~\ref{Yule MSE}) of the approximate basis representation for a tree sampled under the Yule model. 

We use this approach to study approximate representations of tree balance indices by the clade-number basis $\mathcal{C}(T)$. We examined all balance indices listed in Appendix~\ref{Appendix A} but exclude the Colijn-Plazzotta rank because it has extremely large values for certain trees, leading to numerical difficulties. For $6 \leq n \leq 16$, we generated the set of all $m_n = |\mathcal{T}_n|$ tree shapes of size $n$. For each tree shape, we calculated the corresponding $n-1$ clade numbers to form a matrix $\mathbf{C} \in \mathbb{R}^{m_n \times (n-1)}$ with $i$th row $\mathbf{c}_i^t = \mathcal{C}(T_i)$. For each tree shape $T_i$, $1 \leq i \leq m_n$, we found the probability of generating $T_i$ under the Yule model using Theorem~\ref{Theorem 2} and set the weight $w_i$ to be this probability. For each balance index $A(T)$, we calculated a $m_n \times 1$ vector of values $\mathbf{a}$ of the index calculated for this set of $m_n$ trees. Finally, for each balance index, we numerically obtained the solution $\mathbf{x}^*_{\mathbf{W}; \, \mathcal{C}}$ to eq.~\ref{weighted least squares problem} that minimizes the mean-squared error of the approximate basis representation under the Yule model. We then repeated this process for the daughter-number basis $\mathcal{D}(T)$ to find an analogous solution $\mathbf{x}^*_{\mathbf{W}; \, \mathcal{D}}$ for this basis.

For each basis $\mathcal{B}(T) \in \{\mathcal{C}(T), \mathcal{D}(T)\}$, we analyzed the mean-squared error of the solutions $\mathbf{x}^*_{\mathbf{W}; \mathcal{B}}$ obtained for each index to evaluate the accuracy of the approximate representation. Because each index has a different magnitude, we normalized the mean-squared error (eq.~\ref{Yule MSE}) by the weighted magnitude of the index to find a standardized error in the approximation that could be compared across indices,
\begin{equation} \label{weighted residual error}
    \textrm{Error} \big(\mathbf{x}^*_{\mathbf{W}; \mathcal{B}}\big) := \frac{\sum_{i=1}^{m_n} w_i \big( \mathbf{b}_i^t \mathbf{x}^*_{\mathbf{W}; \mathcal{B}} - a_i \big)^2}{\big(\sum_{i=1}^{m_n} w_i a_i^2\big)^{1/2}}. 
\end{equation}
Larger values of this error indicate that the statistic is not well-approximated by the basis. 

\begin{figure}[tb]
    \centering
    \includegraphics[width = \linewidth]{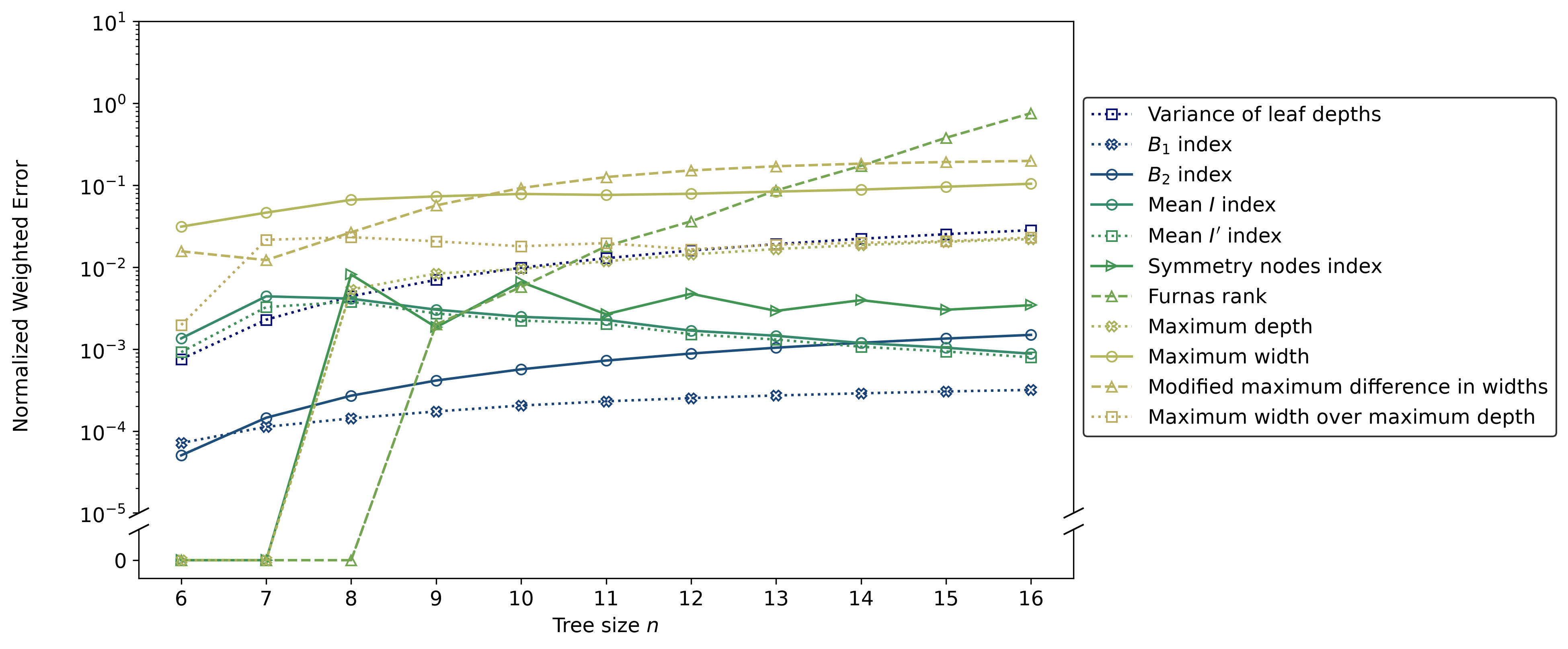}
    \caption{The error of the approximate representations of 11 balance indices by the daughter-number basis, for all trees of size $n$ for each $n \in \{6, 7, \ldots, 16\}$. All 11 indices shown are not represented by the daughter-number basis (Table~\ref{tab:index_representations}).The approximate representation is obtained by weighted least-squares (eq.~\ref{weighted least squares problem}), and it minimizes the mean-squared error of the approximation under the Yule model (eq.~\ref{Yule MSE}). The mean-squared error is normalized following eq.~\ref{weighted residual error}. All 17 statistics that are represented by the daughter-number basis (Section~\ref{sec:4_2}) have error 0 up to numerical precision, and they are excluded from the plot. The magnitude of the error for the 11 indices shown provides a measurement of the relative fit of the index to the daughter-number basis relative to other indices. } \label{fig:daughter projection}
\end{figure}

We plot the results for the clade-number basis in Figure~\ref{fig:clade projection}. As expected, the 8 indices that are represented by the clade-number basis (Table~\ref{tab:index_representations}) have an error of 0 (up to numerical precision), and we exclude them from the plot. All statistics that cannot be represented by the clade-number basis have an error that is nonzero for $n \geq 6$. The relative ordering between statistics helps identify the indices that are closest and furthest from being represented by the clade-number basis. Interestingly, although the error for several statistics grows, as $n$ increases, some statistics appear to have nearly constant or even shrinking error in the approximate basis representation. 

We plot corresponding results for the daughter-number basis in Figure~\ref{fig:daughter projection}. For two statistics, the symmetry nodes index and the Furnas rank, the error for small trees is 0 (up to numerical precision). This result indicates that for small trees, these indices are able to be represented by the daughter-number basis. In such cases, the number of trees $\mathcal{T}_n$ is comparable to the number of basis elements in the daughter-number basis $\mathcal{D}(T)$ for small $n$, so that the existence of an exact solution to the linear system (eq.~\ref{linear system basis statistic}) is not surprising. 

By comparing Figure~\ref{fig:daughter projection} with Figure~\ref{fig:clade projection}, we notice that for each index, the error is smaller in the daughter-number basis than in the clade-number basis. This property can be explained by the fact that the daughter-number basis is a refinement of the clade-number basis (Section~\ref{sec:3_4}). The daughter-number basis contains more information about the tree than does the clade-number basis, so the approximate basis representations can only improve in accuracy. 

Beyond exploring the error of each approximate representation obtained, we can also examine the coefficients of the solutions themselves. For the clade-number basis $\mathcal{C}$, the least-squares solution $\mathbf{x}^*_{\mathbf{W}; \mathcal{C}}$ for a particular statistic $A(T)$ provides an approximate representation of that statistic by the clade-number basis. Write this solution as a vector of coefficients $\mathbf{x}^*_{\mathbf{W}; \mathcal{C}} = (a_2, a_3 \ldots, a_n)^t$. Then for trees $T \in \mathcal{T}_n$,
\begin{equation}
    A(T) \approx \sum_{r=2}^n a_r \, C_r(T),
\end{equation}
where the mean-squared error (eq.~\ref{Yule MSE}) indicates the accuracy of the approximation. Therefore, we can interpret coefficient $a_r$ as an indication of how the statistic $A(T)$ measures clades in $T$ of size $r$. This result enables an interpretation of the statistic $A(T)$ in terms of the features encoded by the clade-number basis. A similar analysis can be done with the daughter-number basis. 

\begin{figure}[tb]
    \centering
    \includegraphics[width = \linewidth]{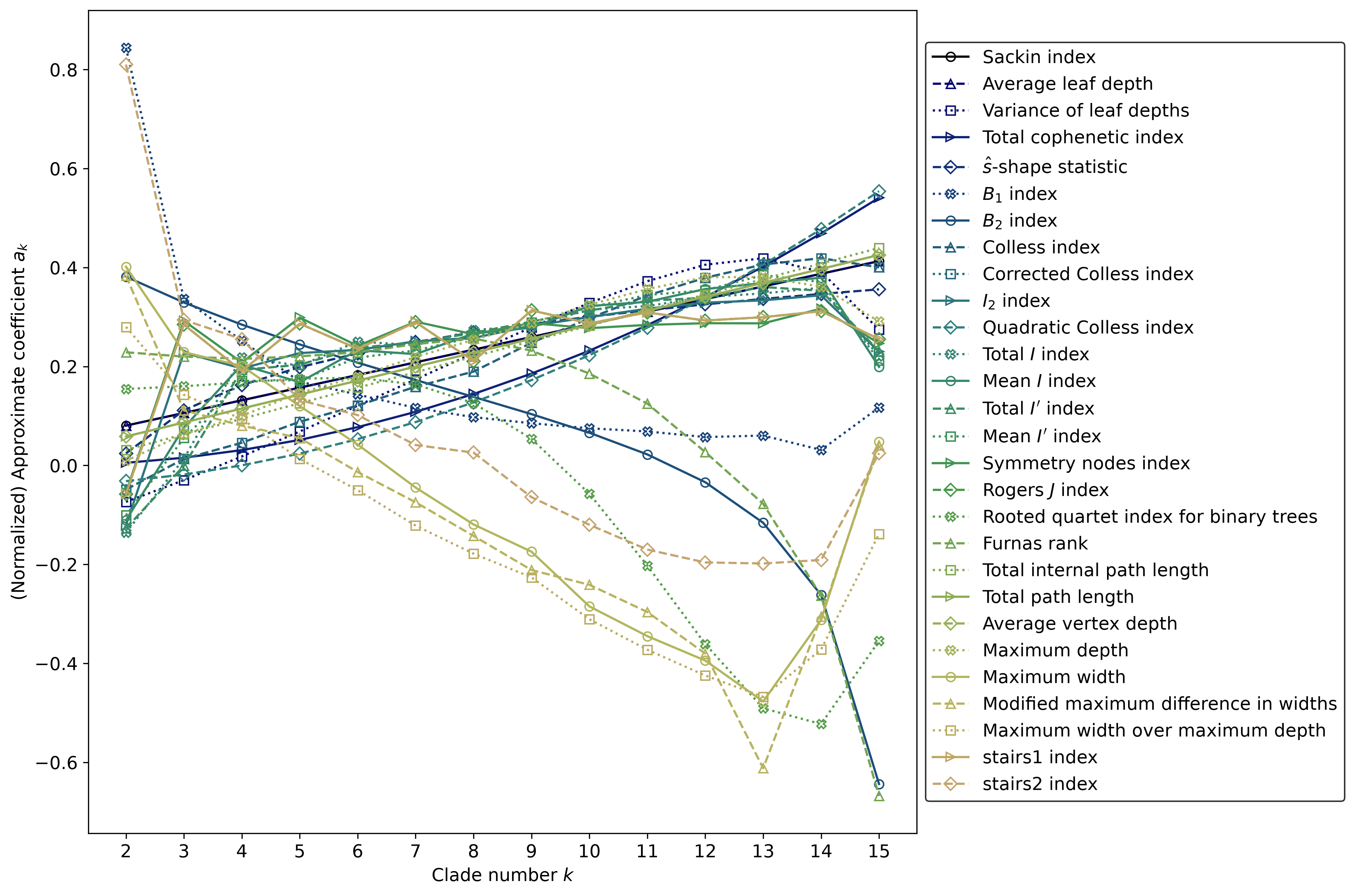}
    \caption{Approximate coefficients for representations of 28 balance indices by the clade-number basis for trees of size $n = 16$. The approximate representation was obtained by weighted-least squares (eq.~\ref{weighted least squares problem}) and minimizes the mean-squared error of the approximation under the Yule model (eq.~\ref{Yule MSE}). For each statistic, the coefficient corresponding to the clade number $C_{16}(T)$ has been set to 0. Each vector of coefficients $(a_2, a_3, \ldots, a_{15})$ was normalized to facilitate direct comparison of the relative shape of each set of coefficients. } \label{fig:coefficients}
\end{figure}

Studying the relative shapes of these coefficient vectors provides a tool to compare statistics. For $n =16$, we examined the solutions $\mathbf{x}^*_{\mathbf{W}; \mathcal{C}}$ to the least-squares problem for the clade-number basis for all balance indices. To ensure uniqueness in the coefficient vectors due to the fact that $C_n(T)$ can be written as a linear combination of $C_2(T), C_3(T), \ldots, C_{n-1}(T)$ (eq.~\ref{remove Cn}), we removed this column from the matrix $\mathbf{C}$ of clade numbers when we calculated the least-squares solution $\mathbf{x}^*_{\mathbf{W}; \mathcal{C}}$. For each resulting vector of coefficients $\mathbf{x}^*_{\mathbf{W}; \mathcal{C}} = (a_2, a_3, \ldots, a_{15})^t$, we normalized it by its magnitude $\|\mathbf{x}^*_{\mathbf{W}; \mathcal{C}}\|_2$ to directly compare the shapes of these vectors. 

In Figure~\ref{fig:coefficients}, we plot these normalized coefficients as a function of $r$, the size of the clade. We observe a broad range of shapes of coefficient vectors across the indices. Many statistics appear to have entirely monotonically increasing or decreasing coefficients, whereas others are not monotonic. Some statistics have entirely positive values, whereas others have both positive and negative values. The range of shapes indicates that across balance indices, the different statistics behave differently in the way that they take into account the relative importance of clades of different sizes.

%% file: Appendix_D.tex
\section{Identity relating the quadratic Colless, total cophenetic, and Sackin indices} \label{Appendix D}

In this appendix, we present a proof of Proposition~\ref{cophenetic sackin qc theorem}, which states that for binary trees $T \in \mathcal{T}_n$, the Sackin index $S(T)$, quadratic Colless index $QC(T)$, and total cophenetic index $\Phi(T)$, are related by
\begin{equation} \label{focal equation}
    QC(T) = 2\Phi(T) + S(T) - n(n-1). 
\end{equation}

For convenience, we restate the definitions of these three balance indices from Appendix~\ref{Appendix A}: 
\begin{align}
    S(T) &= \sum_{\inter} n_v, \label{eq:D.1} \\
    QC(T) &= \sum_{\inter} (n_{v_L} - n_{v_R})^2, \label{eq:D.2} \\
    \Phi(T) &= \sum_{v \in \inter \setminus \rho} \binom{n_v}{2}. \label{eq:D.3}
\end{align}
Recall that for a vertex $v \in \inter$, $n_v$ denotes the size of the subtree induced by $v$, and $n_{v_L}, n_{v_R}$ denote the sizes of the subtrees induced by its daughter vertices $v_L, v_R$, with $n_{v_L} \leq n_{v_R}$.

First, we prove the following lemma. 
\begin{lemmaA} \label{lemma 2}
    Let $T$ be a binary tree with $n$ leaves. Then
    \begin{equation} \label{eq:lemmaA}
        \sum_{\inter} \big( n_{v_L}^2 + n_{v_R}^2 \big)  = n +  \sum_{v \in \inter \setminus \rho} n_v^2.
    \end{equation}
\end{lemmaA}

\begin{proof}

We prove the statement by defining a correspondence between terms on both sides of eq.~\ref{eq:lemmaA}. For each internal vertex $v \in \inter$, both of its daughter nodes $v_L, v_R$ are non-root vertices in the tree. Correspondingly, each vertex $v \in V(T) \setminus \{\rho\}$ has a unique parent node that is an internal vertex. Therefore, 
\begin{equation} \label{eq:D.4}
    \sum_{\inter} \big( n_{v_L}^2 + n_{v_R}^2 \big) = \sum_{v \in V(T) \setminus \{\rho\}} n_v^2.
\end{equation}

Next, the equation
\begin{equation} \label{eq:D.5}
    \sum_{v \in V(T) \setminus \{\rho\}} n_v^2 = n + \sum_{v \in \inter \setminus \{\rho\}} n_v^2
\end{equation}
holds because each of the $n$ leaves $v \in V(T) \setminus \{\rho\}$ has $n_v = 1$. Combining eqs.~\ref{eq:D.4} and \ref{eq:D.5} proves Lemma~\ref{lemma 2}.
\end{proof}

Using this lemma, we now prove the main proposition. 

\begin{proof}[Proof of Proposition~\ref{cophenetic sackin qc theorem}]

Expanding the squared term in eq.~\ref{eq:D.2}, we find that
\begin{align}
    QC(T) &= \sum_{\inter} (n_{v_L} - n_{v_R})^2 \nonumber \\
    &=\bigg[\sum_{\inter} (n_{v_L}^2 + n_{v_R}^2)\bigg] -  \bigg[2 \sum_{\inter}n_{v_L}n_{v_R} \bigg]\nonumber \\
    &= n + \bigg[\sum_{v \in \inter \setminus \{\rho\}} n_v^2 \bigg]-  \bigg[2 \sum_{\inter} n_{v_L}n_{v_R}\bigg], \label{eq:D.6}
\end{align}
where we used lemma~\ref{lemma 2} in the final step. Next, because $n_v = n_{v_L} + n_{v_R}$, we can write 
\begin{align}
    \sum_{\inter} n_{v_L}n_{v_R} &= \sum_{\inter} (n_v - n_{v_R}) \, n_{v_R} \nonumber \\
    &= \sum_{\inter} n_v n_{v_R} - \sum_{\inter} n_{v_R}^2, \label{eq:D.7}
\end{align}
and by symmetry, 
\begin{align}
    \sum_{\inter} n_{v_L}n_{v_R} &= \sum_{\inter} n_v n_{v_L} - \sum_{\inter} n_{v_L}^2. \label{eq:D.8}
\end{align}
Adding eq.~\ref{eq:D.7} to eq.~\ref{eq:D.8} gives
\begin{align}
    2 \sum_{\inter} n_{v_L}n_{v_R} &= \bigg[\sum_{\inter} n_v (n_{v_R} + n_{v_L})\bigg] - \bigg[\sum_{\inter} (n_{v_L}^2 + n_{v_R}^2) \bigg] \nonumber \\
    &= \sum_{\inter} n_v^2 - \bigg[n +  \sum_{v \in \inter \setminus \{\rho\}} n_v^2 \bigg] \nonumber \\
    &= n_{\rho}^2 - n = n(n-1), \label{eq:D.9}
\end{align}
where we applied Lemma~\ref{lemma 2} in the second step and used the fact that $n_{\rho} = n$ in the last step. By rearrangement and substituting for definitions using eqs.~\ref{eq:D.1} and \ref{eq:D.3}, we find that
\begin{align}
    \sum_{v \in \inter \setminus \{\rho\}} n_v^2 &= \bigg[2\sum_{v \in \inter \setminus \{\rho\}} \binom{n_v}{2}\bigg] + \bigg[\sum_{v \in \inter \setminus \{\rho\}} n_v\bigg] \nonumber \\
    &= 2 \Phi(T) + S(T) - n_{\rho} \nonumber \\
    &= 2\Phi(T) + S(T) - n. \label{eq:D.10}
\end{align}
Therefore, combining eqs.~\ref{eq:D.6}, \ref{eq:D.9}, and \ref{eq:D.10} gives the desired formula 
\begin{align}
    QC(T) &= 2\Phi(T) + S(T) - n(n-1),
\end{align}
completing the proof of Proposition~\ref{cophenetic sackin qc theorem}.
\end{proof}